\pdfoutput=1
\documentclass[letterpaper,12pt]{article}

\usepackage[top=0.95in,bottom=0.95in,left=0.95in,right=0.95in]{geometry}
\usepackage{amssymb,amsmath,amsthm,mathtools}
\usepackage[utf8]{inputenc}
\usepackage[T1]{fontenc}
\usepackage{lmodern}
\usepackage{crimson}
\usepackage[kerning]{microtype}
\usepackage[english]{babel}
\usepackage{setspace}
\usepackage{graphicx,booktabs,array,tabularx,longtable,xltabular,multirow,makecell,ragged2e}
\usepackage{threeparttable}
\usepackage{caption}
\usepackage[section]{placeins}
\usepackage{enumitem}
\setlist{nosep,leftmargin=*}
\usepackage{pdflscape,needspace}
\usepackage[dvipsnames]{xcolor}
\usepackage[authoryear]{natbib}
\usepackage{etoc}
\usepackage[colorlinks,urlcolor=NavyBlue,linkcolor=NavyBlue,citecolor=NavyBlue,breaklinks=true]{hyperref}
\hypersetup{pdftitle={That Time of the Year Again: Government Pay Revisions in the Aggregate Economy},
  pdfauthor={Faisal Quaiyyum},
  pdfkeywords={public sector pay, pay commissions, constitutional courts, indexation, fiscal stress, dynamic discrete choice}}

\newcommand{\sym}[1]{\ifmmode^{#1}\else\(^{#1}\)\fi}
\newcolumntype{P}[1]{>{\RaggedRight\arraybackslash}p{#1}}
\newcolumntype{C}[1]{>{\centering\arraybackslash}p{#1}}

\theoremstyle{plain}
\newtheorem{proposition}{Proposition}
\newtheorem{lemma}{Lemma}
\newtheorem{corollary}{Corollary}
\newtheorem{definition}{Definition}
\newtheorem{assumption}{Assumption}
\theoremstyle{remark}

\newcommand{\E}{\mathbb E}
\newcommand{\ind}{\mathbf 1}

\newcommand{\dd}{\,\mathrm d}

\newcommand{\sd}[1]{(s.e.\ #1)}

\graphicspath{{figures/}}
\newcommand{\RauditGen}{31}
\newcommand{\RauditN}{100}
\newcommand{\RbihsPreLow}{83}
\newcommand{\RdocGdpRaiseHtwo}{3.30}
\newcommand{\RdocGdpRaiseHzero}{\ensuremath{-}0.85}
\newcommand{\RdocGdpRaiseNttwo}{11}
\newcommand{\RdocNstartRaise}{27}
\newcommand{\RhtmPubTwentySixteen}{47.6}
\newcommand{\RnEventCountries}{50}
\newcommand{\RnLedger}{151}
\newcommand{\RnLedgerCut}{28}
\newcommand{\RnLedgerRaise}{123}
\newcommand{\RprovIndex}{30}
\newcommand{\RprovLedger}{21}
\newcommand{\RprovNctry}{51}
\newcommand{\RprovNewGen}{292}
\newcommand{\RprovNoEvent}{250}
\newcommand{\RprovNyears}{830}
\newcommand{\RprovSector}{54}
\newcommand{\RprovUnknown}{157}
\newcommand{\RqBgdNPostW}{nine}
\newcommand{\RqBgdNPreW}{nine}
\newcommand{\RqBgdPost}{176}
\newcommand{\RqBgdPostsPost}{2,198}
\newcommand{\RqBgdPostsPre}{2,222}
\newcommand{\RqBgdPre}{83}
\newcommand{\RqHBgdQ}{7.5}
\newcommand{\RqHBgdQSE}{26.0}
\newcommand{\RqHIndQ}{\ensuremath{-}10.9}
\newcommand{\RqHIndQSE}{10.5}
\newcommand{\RqLowRailApps}{19}
\newcommand{\RqLowRailPosts}{62,907}

\newcommand{\RfzNimpl}{19}
\newcommand{\RfzNmeas}{6}

\newcommand{\RvAttainDirections}{20}
\newcommand{\RvAttainGainIRL}{0.11}
\newcommand{\RvAttainGainUSA}{0.04}
\newcommand{\RvAttainImprovingList}{Ireland and the United States}
\newcommand{\RvAttainMaxGain}{0.11}

\newcommand{\RvBgdDocCycle}{54.6}
\newcommand{\RvBgdDocInterval}{6.33}
\newcommand{\RvBgdLRDetCycle}{52.0}
\newcommand{\RvBgdLRDetInterval}{5.80}
\newcommand{\RvBgdLRDetKI}{60}
\newcommand{\RvBgdLRDetSw}{1.00}
\newcommand{\RvCFbgdFitDocmeanscale}{60.4}
\newcommand{\RvCFbgdFitDocscalesperOneZeroZeroy}{11.1}
\newcommand{\RvCFbgdFitDocstatlossOne}{17.22}
\newcommand{\RvCFbgdFitIdxstatlossOne}{5.14}
\newcommand{\RvCFbgdLRDocstatlossOne}{16.58}
\newcommand{\RvCFbgdLRIdxstatlossOne}{7.62}
\newcommand{\RvCFprtNoRevlosstotalOne}{16.87}
\newcommand{\RvCFprtNoRevtotalfinancing}{4.78}
\newcommand{\RvCFprtObslosstotalOne}{22.32}
\newcommand{\RvCFprtObstotalfinancing}{11.70}
\newcommand{\RvCFprtRevObsNoRevOne}{0.21}
\newcommand{\RvCFsensBaseObsMinusNoRev}{5.44}
\newcommand{\RvCFsensInDoubleObsMinusNoRev}{5.72}
\newcommand{\RvCFsensInHalfObsMinusNoRev}{4.34}
\newcommand{\RvCFsensLamOneObsMinusNoRev}{5.95}
\newcommand{\RvCFsensLamZeroObsMinusNoRev}{5.32}
\newcommand{\RvCFsensRestLateObsMinusNoRev}{5.52}
\newcommand{\RvCFsensRestNowObsMinusNoRev}{5.90}
\newcommand{\RvCFsensTolFixedObsMinusNoRev}{3.80}
\newcommand{\RvHeldoutMiss}{27}
\newcommand{\RvHeldoutN}{11}
\newcommand{\RvHeldoutTotConst}{$-$189.1}
\newcommand{\RvHeldoutTotModel}{$-$224.9}
\newcommand{\RvHeldoutTotMyopic}{$-$216.4}
\newcommand{\RvHeldoutWins}{3}
\newcommand{\RvHeldoutYears}{194}
\newcommand{\RvMu}{0.092}
\newcommand{\RvMyopicDiffESP}{1.20}
\newcommand{\RvMyopicDiffPRT}{6.88}
\newcommand{\RvNestCases}{12}
\newcommand{\RvNestMaxDiff}{0.0014}
\newcommand{\RvOnD}{0.329}
\newcommand{\RvOnDeltaPp}{1.6}
\newcommand{\RvOnGreeceD}{0.310}
\newcommand{\RvOnHinge}{$-$0.021}
\newcommand{\RvOnObjD}{0.254}
\newcommand{\RvProfileBEhi}{$-$0.01}
\newcommand{\RvProfileBElo}{$-$0.19}
\newcommand{\RvProfileKhi}{$-$19.86}
\newcommand{\RvProfileKloInfeasible}{Hungary}
\newcommand{\RvProfileKloRest}{2.47}
\newcommand{\RvRecPRTchiA}{0.58}
\newcommand{\RvRecPRTchiB}{0.22}
\newcommand{\RvRecPRTchiTrue}{0.34}
\newcommand{\RvRecPRTgamFA}{0.21}
\newcommand{\RvRecPRTgamFB}{0.42}
\newcommand{\RvRecPRTgamFTrue}{0.39}
\newcommand{\RvRecoveryN}{10}
\newcommand{\RvRecoveryReached}{8}
\newcommand{\RvRobMu}{0.051}
\newcommand{\RvRobMuSe}{0.017}
\newcommand{\RvRobOnD}{0.355}
\newcommand{\RvRobOnDse}{0.141}
\newcommand{\RvRobOnEvents}{7}
\newcommand{\RvRoneCalibLR}{20.2}
\newcommand{\RvRoneFitFullstatZero}{0.34}
\newcommand{\RvRoneFitHalfstatZero}{0.99}
\newcommand{\RvRoneGovFitHalf}{2.46}
\newcommand{\RvRoneGovLRHalf}{2.15}
\newcommand{\RvRoneLRFullstatZero}{0.44}
\newcommand{\RvRoneLRHalfstatZero}{1.12}
\newcommand{\RvRoneReqFixHalf}{1.09}
\newcommand{\RvRoneReqHalf}{0.99}
\newcommand{\RvRoneReqRobustFull}{0.13}
\newcommand{\RvRoneReqRobustHalf}{0.77}
\newcommand{\RvRoneUOneKiHi}{150}
\newcommand{\RvRoneUOneKiLo}{80}
\newcommand{\RvRoneUOneSwHi}{0.8}
\newcommand{\RvRoneUOneSwLo}{0.0}
\newcommand{\RvTransferLeadRest}{11.2}
\newcommand{\RvTransferN}{9}
\newcommand{\RvTransferShortHL}{18.5}
\newcommand{\RvTransferTotConstant}{$-$126.0}
\newcommand{\RvTransferTotFiscal}{$-$92.2}
\newcommand{\RvTransferTotLegal}{$-$141.7}
\newcommand{\RvTransferTotModel}{$-$99.5}
\newcommand{\RvTransferTotState}{$-$92.3}
\newcommand{\RvTransferWinsConst}{6}
\newcommand{\RvTransferWinsOut}{4}
\newcommand{\RvUchiEstPRT}{0.34}
\newcommand{\RvUchiPRT}{1.11}
\newcommand{\RvUdiffPRT}{$-$1.84}
\newcommand{\RvVFourB}{19}
\newcommand{\RvVFourCutChi}{14.61}
\newcommand{\RvVFourCutGamF}{8.59}
\newcommand{\RvVFourCutX}{2.57}
\newcommand{\RvVFourHiCsqChi}{0.63}
\newcommand{\RvVFourHiCsqGamF}{0.447}
\newcommand{\RvVFourHiGamF}{0.519}
\newcommand{\RvVFourLoChi}{0.17}
\newcommand{\RvVFourLoCsqChi}{0.24}
\newcommand{\RvVFourLoCsqGamF}{0.351}
\newcommand{\RvVFourLoGamF}{0.306}
\newcommand{\RvVFourMedChi}{0.08}
\newcommand{\RvVFourMedGamF}{1.10}
\newcommand{\RvVFourMedX}{0.24}
\newcommand{\RvphiFBGD}{0.271}

\newcommand{\RsBgdEndEabs}{121}
\newcommand{\RsBgdEndP}{15}
\newcommand{\RsBgdErodeEabs}{10.1}
\newcommand{\RsBgdErodePabs}{5.8}
\newcommand{\RsBgdRevMean}{41.0}
\newcommand{\RsBgdScaleGapW}{six}
\newcommand{\RsBgdScaleNW}{four}
\newcommand{\RsBrBgdJumpData}{34.0}
\newcommand{\RsBrBgdJumpHi}{61.3}
\newcommand{\RsBrBgdJumpLo}{29.3}
\newcommand{\RsBrBgdJumpMed}{41.3}
\newcommand{\RsBrBgdYearsData}{6.0}
\newcommand{\RsBrBgdYearsHi}{12.0}
\newcommand{\RsBrBgdYearsLo}{6.0}
\newcommand{\RsBrBgdYearsMed}{8.0}
\newcommand{\RsBrIndJumpData}{25.4}
\newcommand{\RsBrIndJumpHi}{39.0}
\newcommand{\RsBrIndJumpLo}{21.0}
\newcommand{\RsBrIndJumpMed}{29.4}
\newcommand{\RsBrIndScaleData}{25.6}
\newcommand{\RsBrPrtAllInside}{every}
\newcommand{\RsBrPrtEndData}{$-$15.9}
\newcommand{\RsBrPrtEndMed}{$-$14.5}
\newcommand{\RsBrPrtPersistData}{0.57}
\newcommand{\RsBrPrtPersistMed}{0.57}
\newcommand{\RsBrPrtTroughData}{$-$24.7}
\newcommand{\RsBrPrtTroughMed}{$-$23.7}
\newcommand{\RsBrUsaRaiseData}{2.90}
\newcommand{\RsBrUsaRaiseHi}{3.27}
\newcommand{\RsBrUsaRaiseLo}{2.71}
\newcommand{\RsBrUsaRaiseMed}{2.92}
\newcommand{\RsCoverPassIn}{31}
\newcommand{\RsCoverPassTot}{34}
\newcommand{\RsCoverRpEndFailPhrase}{neither of the two}
\newcommand{\RsCoverRpEndPassMiss}{Spain}
\newcommand{\RsCoverRpEndPassMissGap}{0.6}
\newcommand{\RsCoverRpEndPassMissSide}{above the model's 95th percentile}
\newcommand{\RsCoverRpEndPassPhrase}{six of the seven}
\newcommand{\RsCtBeforeCumEx}{6}
\newcommand{\RsCtBeforeCumIn}{41}
\newcommand{\RsCtBeforeLossOne}{29.06}
\newcommand{\RsCtBeforePayCrisis}{86.8}
\newcommand{\RsCtBeforeRest}{6.2}
\newcommand{\RsCtCertainBeforeCumEx}{0}
\newcommand{\RsCtCertainBeforeCumIn}{30}
\newcommand{\RsCtCertainBeforeDebt}{95.4}
\newcommand{\RsCtCertainBeforePayCrisis}{87.7}
\newcommand{\RsCtCertainBeforeTermTreasury}{18.96}
\newcommand{\RsCtEstCumEx}{25}
\newcommand{\RsCtEstCumIn}{62}
\newcommand{\RsCtEstDebt}{93.1}
\newcommand{\RsCtEstImposed}{19.5}
\newcommand{\RsCtEstLossOne}{37.04}
\newcommand{\RsCtEstPayCrisis}{83.8}
\newcommand{\RsCtEstRest}{20.0}
\newcommand{\RsCtEstTemp}{12.9}
\newcommand{\RsCtEstTermFinancing}{17.16}
\newcommand{\RsCtEstTermTreasury}{9.93}
\newcommand{\RsCtFinMax}{25.08}
\newcommand{\RsCtFinMin}{1.89}
\newcommand{\RsCtMinAboveNone}{2.29}
\newcommand{\RsCtNEnv}{21}
\newcommand{\RsCtNoneCumEx}{11}
\newcommand{\RsCtNoneCumIn}{121}
\newcommand{\RsCtNoneImposed}{4.0}
\newcommand{\RsCtNoneLossOne}{25.83}
\newcommand{\RsCtNonePayCrisis}{85.9}
\newcommand{\RsCtNoneTermFinancing}{1.92}
\newcommand{\RsCtNoneTermTreasury}{14.39}
\newcommand{\RsCtStabMax}{5.04}
\newcommand{\RsCtStabMin}{4.62}
\newcommand{\RsCtTreasMax}{18.96}
\newcommand{\RsCtTreasMin}{7.04}
\newcommand{\RsCutCtryW}{seven}
\newcommand{\RsCutDef}{7.3}
\newcommand{\RsCutN}{15}
\newcommand{\RsCutNom}{1.2}
\newcommand{\RsCutNomFall}{80}
\newcommand{\RsEroBetween}{0.80}
\newcommand{\RsEroCtryHi}{0.73}
\newcommand{\RsEroCtryLo}{0.02}
\newcommand{\RsEroInt}{$-$0.11}
\newcommand{\RsEroIntSE}{0.09}
\newcommand{\RsEroLevel}{$-$0.09}
\newcommand{\RsEroLevelSE}{0.20}
\newcommand{\RsEroLongN}{24}
\newcommand{\RsEroLooHi}{0.37}
\newcommand{\RsEroLooLo}{0.32}
\newcommand{\RsEroMedInfl}{5.3}
\newcommand{\RsEroMedRaise}{7.0}
\newcommand{\RsEroN}{185}
\newcommand{\RsEroOneYear}{82}
\newcommand{\RsEroPassOne}{0.36}
\newcommand{\RsEroPassOneSE}{0.12}
\newcommand{\RsEroSFmax}{0.38}
\newcommand{\RsEroSize}{0.35}
\newcommand{\RsEroSizeSE}{0.06}
\newcommand{\RsFailList}{Hungary and Lithuania}
\newcommand{\RsForwardBetterList}{Portugal and Spain}
\newcommand{\RsIdxFitDocLossOne}{17.22}
\newcommand{\RsIdxFitDocLossZero}{5.89}
\newcommand{\RsIdxFitDocPninety}{34}
\newcommand{\RsIdxFitFullFinLossOne}{13.76}
\newcommand{\RsIdxFitFullFinLossZero}{11.58}
\newcommand{\RsIdxFitFullFinPay}{85.7}
\newcommand{\RsIdxFitFullFinPninety}{27}
\newcommand{\RsIdxFitFullFreePay}{88.1}
\newcommand{\RsIdxFitFullFreePninety}{24}
\newcommand{\RsIdxFitHalfFinLossOne}{14.06}
\newcommand{\RsIdxFitHalfFinLossZero}{5.91}
\newcommand{\RsIdxLRDocLossOne}{16.58}
\newcommand{\RsIdxLRFullFinLossOne}{16.91}
\newcommand{\RsIdxLRHalfFinLossOne}{14.59}
\newcommand{\RsIndErodeEabs}{4.5}
\newcommand{\RsIndErodeP}{0.1}
\newcommand{\RsLedgerCutCtry}{17}
\newcommand{\RsLedgerCutGrZeroAbs}{5.2}
\newcommand{\RsLedgerCutN}{27}
\newcommand{\RsLedgerRaiseCtry}{42}
\newcommand{\RsLedgerRaiseGrZeroAbs}{0.7}
\newcommand{\RsLedgerRaiseN}{122}
\newcommand{\RsMatchOtherHi}{0.93}
\newcommand{\RsMatchOtherHiName}{the United States}
\newcommand{\RsMatchOtherLo}{0.67}
\newcommand{\RsMatchOtherLoName}{Romania}
\newcommand{\RsMechBgdNoFinAllow}{10}
\newcommand{\RsMechBgdNoFixAllow}{17}
\newcommand{\RsMechBgdNoFixScales}{6}
\newcommand{\RsMechBgdNoFixSize}{20.9}
\newcommand{\RsMechEstNone}{0.35}
\newcommand{\RsMechEstPayEighteen}{$-$8.5}
\newcommand{\RsMechEstPayTwelve}{$-$23.7}
\newcommand{\RsMechNoCourtCutsW}{two}
\newcommand{\RsMechNoCourtPayEighteen}{$-$13.0}
\newcommand{\RsMechNoOverNone}{0.06}
\newcommand{\RsMechNoOverPayEighteen}{$-$3.6}
\newcommand{\RsMechNoReqCutsW}{no}
\newcommand{\RsMechNoReqPayTwelve}{$-$2.2}
\newcommand{\RsMissYearsUSA}{2011 and 2014}
\newcommand{\RsNoneCrisis}{26}
\newcommand{\RsNoneDef}{4.8}
\newcommand{\RsNoneN}{34}
\newcommand{\RsOnEventsW}{nine}
\newcommand{\RsPassList}{Portugal, Ireland, Spain, Cyprus, Italy, Romania, the United States, India and Bangladesh}
\newcommand{\RsPrtCutEleven}{5.1}
\newcommand{\RsPrtCutFourteen}{5.0}
\newcommand{\RsPrtCutTwelve}{15.1}
\newcommand{\RsPrtEndPabs}{16}
\newcommand{\RsPrtInforceFourteen}{2.1}
\newcommand{\RsPrtPaidBelowStanding}{0.6}
\newcommand{\RsPrtPaidFourteen}{5.7}
\newcommand{\RsPrtPaidTwelve}{20.2}
\newcommand{\RsRaiseCrisis}{4}
\newcommand{\RsRaiseDef}{2.2}
\newcommand{\RsRaiseN}{56}
\newcommand{\RsRaiseNom}{7.1}
\newcommand{\RsRecDec}{194}
\newcommand{\RsRecRest}{14}
\newcommand{\RsRecRestricted}{19}
\newcommand{\RsRecUnknown}{3}
\newcommand{\RsRecUnscored}{11}
\newcommand{\RsRecYears}{233}
\newcommand{\RsRightUSAW}{27}
\newcommand{\RsSelCutBalAbs}{1.35}
\newcommand{\RsSelCutGrAbs}{1.34}
\newcommand{\RsSelRaiseGr}{2.04}
\newcommand{\RsSizeSpacingN}{33}
\newcommand{\RsSizeSpacingP}{0.01}
\newcommand{\RsSizeSpacingRho}{0.44}
\newcommand{\RsSizeSpacingSlope}{4.2}
\newcommand{\RsSizeSpacingSlopeSE}{2.7}
\newcommand{\RsTransferGainUSA}{11.3}
\newcommand{\RsUsaBaseAtW}{sixteen}
\newcommand{\RsUsaBaseBelowW}{thirteen}
\newcommand{\RsUsaFormulaYearsW}{29}
\newcommand{\RsUsaRevMean}{2.9}
\newcommand{\RsWrongLTUW}{seven}
\newcommand{\RsWrongROUW}{four}
\newcommand{\RsYearsLTUW}{seventeen}
\newcommand{\RsYearsROUW}{twelve}
\newcommand{\RsYearsUSAW}{29}

\newcommand{\RwChiFive}{0.33}
\newcommand{\RwChiZero}{0.34}
\newcommand{\RwCrossNWord}{nine}
\newcommand{\RwDecExcessBefore}{1.9}
\newcommand{\RwEdgeBeforeGain}{4.5}
\newcommand{\RwEdgeCalMinusNone}{8.6}
\newcommand{\RwEdgeFinCal}{12.5}
\newcommand{\RwEdgeHoldWord}{sixteen}
\newcommand{\RwEdgeMinAboveNone}{3.3}
\newcommand{\RwEdgeNWord}{seventeen}
\newcommand{\RwEdgeTempCal}{11.1}
\newcommand{\RwGridIn}{0.05}
\newcommand{\RwGridOut}{0.1}
\newcommand{\RwKdnFive}{1.22}
\newcommand{\RwKdnZero}{0.22}
\newcommand{\RwLROne}{9.02}
\newcommand{\RwLamLFive}{2.35}
\newcommand{\RwLamLZero}{3.15}
\newcommand{\RwPOne}{0.003}
\newcommand{\RwPhiFive}{1.30}
\newcommand{\RwPhiOne}{0.04}
\newcommand{\RwPhiZero}{1.28}
\newcommand{\RwRaisesLossOne}{4.0}
\newcommand{\RwSlopeDrop}{0.018}
\newcommand{\RwSlopeW}{0.001}
\newcommand{\RwTwelveLossFive}{10.3}
\newcommand{\RwUpper}{0.066}
\newcommand{\RwUpperRound}{0.07}
\newcommand{\RwZeroBeforeGain}{8.0}
\newcommand{\RwZeroCalMinusNone}{11.2}
\newcommand{\RwZeroExcessBefore}{2.7}
\newcommand{\RwZeroFinCal}{17.2}
\newcommand{\RwZeroTempCal}{12.9}

\newcommand{\pending}[1]{{\color{red!70!black}[Pending: #1]}}
\newcommand{\inputif}[2]{\IfFileExists{#1}{\input{#1}}{\pending{#2}}}

\newcommand{\figwidth}{\textwidth}%
\newcommand{\figifat}[5]{\IfFileExists{figures/#2}{\begin{figure}[#1]\centering\includegraphics[width=\figwidth]{#2}\caption[#3]{#3. #5}\label{#4}\end{figure}}{\pending{figure: #3}}}
\newcommand{\figif}{\figifat{t}}


\makeatletter\@clubpenalty=10000 \clubpenalty=10000 \makeatother
\makeatletter
\renewcommand{\@maketitle}{%
  \newpage
  \begin{center}%
    {\fontsize{15.1}{18.1}\selectfont\bfseries \@title \par}%
    \vskip 1.4em%
    {\fontsize{14}{17}\selectfont \begin{tabular}[t]{c}\@author\end{tabular}\par}%
    \vskip 0.5em%
    {\normalsize \@date}%
  \end{center}%
  \par\vskip 10pt}
\renewcommand\section{\@startsection{section}{1}{\z@}{-2.4ex plus -0.2ex}%
  {1.2ex}{\normalfont\fontsize{14.1}{17}\selectfont\bfseries}}
\renewcommand\subsection{\@startsection{subsection}{2}{\z@}{-2.1ex plus -0.8ex}%
  {0.8ex}{\normalfont\fontsize{13.1}{15.7}\selectfont\bfseries}}
\renewcommand\paragraph{\@startsection{paragraph}{4}{\z@}{1.4ex plus 0.2ex}%
  {-1em}{\normalfont\normalsize\bfseries}}
\renewenvironment{abstract}{%
  \small\par\centerline{\bfseries\abstractname}\par\vskip 0.35em%
  \list{}{\leftmargin=1.2em\rightmargin=1.2em\listparindent=1.5em\itemindent=0pt%
          \parsep=0pt\topsep=0pt}\item\relax}{\endlist}
\def\@spart#1{{\parindent\z@\raggedright\interlinepenalty\@M\normalfont\fontsize{15.1}{18.1}\selectfont\bfseries #1\par}\nobreak\vskip 3ex\@afterheading}
\makeatother
\makeatletter
\newcommand{\condbreak}[1]{\par\begingroup\dimen@=#1\dimen@ii\pagegoal\advance\dimen@ii-\pagetotal
  \ifdim\dimen@>\dimen@ii\penalty-\@M\fi\endgroup}
\makeatother
\begin{document}

\title{That Time of the Year Again:\\ Government Pay Revisions in the Aggregate Economy}
\author{Faisal Quaiyyum\thanks{faisalquaiyyum@uchicago.edu. The usual disclaimers apply.}\\[2pt]{\normalsize University of Chicago}}
\date{November 2026}
\maketitle
\thispagestyle{empty}

\begin{abstract}
\noindent
Public pay is set by law in nominal terms, so inflation erodes its real value between legal acts until the government
revises it. I model each revision as the choice of a forward-looking government, shaped by two costs that push in
opposite directions. A fixed cost of acting makes revisions large and rare, whereas a convex cost of financing the
increase at once makes them small and frequent. Through the budget, each revision also moves debt and the risk of fiscal
stress under which the next one is made. In the documented pay decisions of eleven countries, pay set at discretion
erodes against prices and private earnings and is restored in discrete steps, every cut was decided during market stress
or an adjustment program, and court rulings drove a wedge between the cuts legislated, the cuts paid and the cuts that
lasted. Estimated on the legal acts, the model reproduces the documented paths of pay in most countries when data and
simulations are measured the same way. In Portugal's estimated model, the savings that the Treasury requires once spreads
rise produce the cuts, the debt overhang holds pay down after the crisis, and a court that strikes cuts after they take
effect invites cuts it will strike. Review of cuts before they took effect would have lowered the social cost of
Portugal's crisis relative to review after the fact, and partial indexation lowers Bangladesh's social loss unless
its automatic increases are costly to finance.


\end{abstract}
\vspace{0.5pt}

\onehalfspacing\advance\baselineskip by 0pt plus 1pt\relax

\section{Introduction}
\label{sec:intro}

Public pay is one of the largest items of government spending, about a tenth of GDP in the euro area
\citep{eurostat2026}, and governments pay it by schedules that the law sets in nominal terms. Between one legal act and
the next the schedule stays fixed while prices and private earnings rise, so inflation lowers its real value until the
government acts again. Every government therefore faces a recurring decision about when to restore public pay and by how
much, and, when public finances come under pressure, about whether to cut it instead. This decision matters well beyond
the public sector. Cuts in the wage bill are a central margin of fiscal consolidations
\citep{alesina1997}, whose output costs depend on their composition \citep{alesina2015}. The wage bill moves
procyclically in many countries \citep{lane2003}, and public wages feed into private wages, profits and investment
\citep{alesina2002,bradley2017}.

\looseness=-1 Macroeconomic research on public pay has asked what level it should take and what it does to the economy.
\citet{gomes2015} derives the optimal public-sector wage in a general-equilibrium model in which pay set too high draws
the unemployed into queues for public jobs, and \citet{chang2021} estimate a New Keynesian model in which cutting public
wages affects output and the fiscal balance about as much as cutting government purchases. Public wages and public
employment have distinct macroeconomic effects \citep{quadrini2007,bermperoglou2017}, and public and private wages move
together \citep{lamo2012,afonso2014}. In these papers the public wage is a price, or a policy process, that the model
sets every period. In practice, however, it moves only when a legal act moves it, so public pay responds to inflation and
to fiscal pressure through acts whose timing, size and sign the government chooses. These acts have rarely been the
object of study.

This paper studies them, and its account is simple. Because pay is fixed in nominal terms, inflation does the adjusting
between acts, and governments restore pay in occasional steps whose size reflects what it costs to act and to finance a
raise at once. When spreads rise, the Treasury needs savings faster than inflation delivers them, and the same decision
turns into a legislated cut. Debt that stays high after the crisis can then hold pay down. Courts and indexation rules do
not remove this mechanism. Instead, they change when savings and raises arrive and what they cost.

The central idea is that public pay is a nominal price set by law, and that the budget makes it costly to change in two
distinct ways. The first is the cost of acting at all. A legislative or bargaining round, or a pay commission, costs much
the same whatever the size of the change. Like a firm that faces a menu cost, a government that faces this cost
tolerates erosion and revises in discrete steps, and the higher the cost, the larger and rarer the steps
\citep{sheshinski1977,dixit1991,golosov2007,caballero2007,stokey2009}. The second cost has no counterpart in menu-cost pricing. A
raise is paid from the day it takes effect, so the government must finance the whole increase in its payroll at once,
and borrowing within a year costs more at the margin the more is borrowed. This convex cost of financing pushes the other
way, toward smaller and more frequent revisions, for the same reason that convex costs of taxation lead governments to
smooth taxes \citep{barro1979}. The size and spacing of revisions alone therefore cannot tell the two costs apart, but
the response of size to the speed of erosion can (Proposition~\ref{prop:band} and Corollary~\ref{cor:costs}), and
inflation varies that speed across years and countries. In the record that response is about a third, as under a
fixed cost alone, but most raises come a year apart, so the estimation separates the two costs with the yearly
decisions.

The budget also links each decision to the next, which ties public pay to sovereign debt and its pricing. The payroll
enters the deficit and the deficit enters debt, the dynamics that shape fiscal policy in sovereign-debt models
\citep{arellano2008}, where the stabilization benefit of spending and the exposure to a debt crisis are determined
jointly \citep{bianchi2023}. A larger deficit raises the risk of fiscal stress, and under stress the Treasury's required
savings on the payroll rise with the spread. The cuts that deliver them lower the deficit on which the next year's risk
of stress depends.

I document the decision with the legal acts themselves (Section~\ref{sec:patterns}). For eleven countries, eight in
Europe, the United States, Bangladesh and India, I read the general pay decision of each budget cycle from the act or
the official notice that reports it, and I date each change by the year in which it was first paid. Unlike the aggregate
wage bill, which mixes general revisions with employment, promotions and arrears, the acts isolate the policy decision,
just as narrative records isolate fiscal policy actions from the endogenous movements of the budget
\citep{romer2010,ramey2011}.

Three patterns emerge, one for each part of the mechanism. First, where governments revise pay at discretion, legal pay
erodes between acts and is restored in discrete steps, and raises are larger where they are rarer. Second, almost all raises come outside crises, whereas every one of the \RsCutN{} cuts in the decision histories of the eight European countries came
during market stress or an adjustment program. Third, where a court reviews cuts, the cut that is legislated, the cut
that is paid and the cut that lasts can differ for years, and so can the savings that the Treasury receives. Portugal's
Constitutional Court struck three of the four crisis cuts, two of them after they had been in force.

One dynamic program covers every country (Section~\ref{sec:model}). Each year the government chooses the change in pay
that maximizes the expected discounted value of a payoff that weighs public services against the payroll's cost, given
the budget constraint and, in the European countries, a spread that rises with debt and a risk of fiscal stress that
rises with the deficit. The documents of each country determine which actions its
government has, whether an annual raise or cut, a court that can strike a cut, or a commission that must report before
a new scale is set. Because the decisions are discrete, the problem is a dynamic discrete choice in the tradition of
\citet{rust1987}, surveyed by \citet{aguirregabiria2010}, and I estimate it country by country on the legal acts
(Section~\ref{sec:quant}). The fiscal relations are estimated on fiscal data alone, before any pay decision is used, and
in them the deficit, a flow, predicts the onset of stress while a higher stock of debt does not.

Although the estimation fits the decisions one at a time, the decisions that the model generates add up to the
documented paths of pay in most countries. When the same definitions are applied to
the documented and the simulated paths, the documented value lies inside the model's 90 percent interval for
\RsCoverPassIn{} of \RsCoverPassTot{} statistics in the annual countries that pass an adequacy test fixed in advance.
Portugal's real pay lies inside the interval in \RsBrPrtAllInside{} year of 2008--2024, through the cuts of its sovereign
debt crisis, the Court's strikes and the restorations. In Bangladesh the simulated scales are larger and rarer than the
documented ones, although the documented scales lie within the simulated range.

Removing one feature at a time, at fixed parameters, shows which feature produces each pattern inside the model. The
cuts disappear with the Treasury's requirement, because inflation alone does not erode pay fast enough to deliver the
saving that the Treasury requires once spreads rise. With the Court gone, every cut is permanent, and with the debt
overhang gone, pay recovers years earlier after the crisis. Bangladesh's erosion-and-reset cycle needs both costs. The
fixed cost of appointing a commission keeps new scales rare, while the cost of financing the interim allowances paid
between commissions makes a new scale the cheaper way to restore pay once pay has fallen far behind.

I then use the estimated model to evaluate two institutions that shape how public pay adjusts to prices and to fiscal
pressure (Section~\ref{sec:design}). The first is the judicial review of pay cuts in Portugal. A court that strikes a cut
after it has taken effect cannot undo the saving that the cut made while it was in force, so a government facing such a
court has a reason to propose cuts it expects to lose. This saving disappears only if review is both certain and comes
before the cut takes effect (Proposition~\ref{prop:court}), so the speed and the certainty of review are complements. In
the estimated model, a court that reverses cuts shortens how long cuts last, but it does not raise pay during the
crisis, and it raises the crisis's social cost, mainly because the restorations that the court orders must be financed
at once. Review before a cut takes effect removes the temporary savings and most of the restorations, and its social
cost is lower than that of review after the fact.

The second institution is the indexation of pay between commissions in Bangladesh, a spending rule that replaces a
decision with an automatic adjustment, as automatic stabilizers do for taxes and transfers \citep{mckay2016}. A rule's
value against discretion turns on the flexibility a government gives up \citep{kydland1977,amador2006,halac2014}, and
\citet{campolieti2025} finds that Canadian public-sector settlements trade indexation against smaller base increases and
longer contracts, the joint choice that \citet{gray1978} analyzed. In the estimated model, price indexation would have
lowered Bangladesh's social loss if its automatic increases cost nothing to finance. Once their financing enters the
government's problem, however, the government lets pay fall further behind, because a lower level of pay makes each
automatic increase cheaper to finance. Since the cost of financing rises with the square of an increase, indexing half
of inflation keeps half of the protection against prices at a quarter of the cost, and partial indexation therefore
survives a much wider range of financing costs than full indexation.

These results rest on two features of the design. First, the model accounts for the histories it fits but does not
forecast them. Re-estimated without a block of years, it predicts the block worse than constant probabilities, and
outside \RsForwardBetterList{} a government that does not look ahead fits the decisions as well. Second, the model is
closed through the budget rather than through output. What a revision does to output enters the welfare comparisons
through a calibrated, conditional transmission block, because the panel of legislated acts does not identify the
aggregate response. In the Portuguese experiments the alternative paths of pay differ little in their output effects but
widely in the cost of financing them. The block draws on research showing that the response of output to government spending
depends on the exchange-rate regime \citep{ilzetzki2013,nakamura2014,farhi2016}, that whether households spend
anticipated income gains depends on the setting \citep{parker1999,hsieh2003} and on liquidity constraints
\citep{zeldes1989,kaplan2014}, and that where domestic banks hold the government's debt, its borrowing and its default risk reach private
credit \citep{gennaioli2014,becker2018}, which matters most for industries that depend on external finance
\citep{rajan1998}.

The rest of the paper proceeds as follows. Section~\ref{sec:patterns} describes the institutions, the data and the three
patterns. Section~\ref{sec:model} presents the government's dynamic program, and Section~\ref{sec:quant} its estimation
and the comparison of model and data. Section~\ref{sec:design} evaluates the two institutions, and
Section~\ref{sec:conclusion} concludes.

\section{Institutions, data and empirical patterns}
\label{sec:patterns}

This section documents three regularities in how governments revise public pay, which the model of
Section~\ref{sec:model} is built to explain. Legal pay set at discretion erodes between legal acts and is restored in
discrete steps. Raises and cuts are decided in different fiscal circumstances. Finally, where courts review cuts, the cut
that is paid and the cut that lasts can differ from the cut that is legislated. All three are measured with observable
series alone, namely the documented legal acts, consumer prices, private earnings and the fiscal accounts, so no
parameter of the model enters any of them. Section~\ref{sec:limits_patterns} explains why these descriptive
regularities cannot identify what a revision does to the economy.

\subsection{Pay-setting institutions and the documented decision histories}
\label{sec:institutions}
\label{sec:regimes}
\label{sec:histories}
\label{sec:chronology}

A public pay scale lists the pay of every grade and step in currency units. Employees move up the scale through
seniority increments, but the schedule itself stays fixed in nominal terms until a legal act changes it. A revision
passes through several dates. It is announced, enacted and made effective, often retroactively, and it is first paid
later still, with arrears when the entitlement precedes the payment. The documents keep these dates apart. The annual
series below date each change by the year in which it is first paid, because that is when employees' income and the
Treasury's cash change.

The eleven countries studied most closely, eight in Europe, the United States, Bangladesh and India, belong to three
institutional families. Pay is decided in the annual budget, adjusted by a statutory formula, or set by commissions. In most
of the eight European countries the general change of pay is decided in the annual budget. Portugal's state budget
updates, freezes or cuts its single pay table, and the budgets of Spain, Lithuania, Hungary and Romania set their pay
tables, base amounts or pay bases. In the other three the budget works alongside other instruments. Ireland's national
agreements fixed general increases for several years at a time, and its crisis cuts were imposed by emergency
legislation. Italy suspended its national contracts by decree from 2010 to 2015, and Cyprus negotiates general increases
and indexes pay to prices. In the United States a statutory formula sets the default January adjustment of General Schedule pay,
which an alternative plan of the President or an act of Congress can replace \citep{crs2022}. In
Bangladesh and India pay commissions set new scales several years apart. Between commissions Bangladesh pays interim
allowances, while India's dearness allowance indexes pay to consumer prices. Finally, constitutional courts constrained
the crisis measures of three of the European countries. Portugal's court struck several cuts and ordered the rest reversed.
Italy's struck the 2010 levy on salaries above EUR 90,000, a measure that left general pay unchanged, and Lithuania's
held the crisis reductions unconstitutional in 2013.

The structural model is estimated on a record built for this purpose, which contains the general pay decision of each
budget cycle, read from the act or the official notice that reports it. From the first documented decision to the last,
the record covers the budgets of 2007--2024 in the European countries (2008--2024 in Portugal, 2007--2023 in Spain and
Lithuania, and 2007--2012 in Hungary, whose general pay base stayed frozen afterward), 1995--2024 in the United States,
and 1989--2024 and 1987--2024 in Bangladesh and India. Not every year in which pay changes is a year in which the
government chose the change. Before any estimation, each year is therefore classified from the documents into one of
four kinds. In a \emph{decision year} the government chooses the general change. In an \emph{implementation year} the
change was fixed by an instrument adopted two or more years earlier, so the year records a change but no decision. In a \emph{restricted decision year} the government chose while a court ruling still bound pay. In a \emph{measured year} the decision is not
documented but the average change of pay is, so the year carries the pay path without being treated as a decision. Of
the \RsRecYears{} country-years in the eleven records, \RsRecDec{} are decision years (\RsRecRestricted{} of them
restricted), \RfzNimpl{} are implementation years and \RfzNmeas{} are measured years. The remaining \RsRecRest{} enter
no decision. In \RsRecUnknown{} of them no document was found, and in \RsRecUnscored{} a commission country's earlier
measure of unknown size leaves the gap unknown until the next scale. A year for which no document was found is unknown,
and it is never coded as a year without a change.

The patterns draw on three further sources. A ledger of \RnLedger{} legislated changes in public pay in
\RnEventCountries{} countries between 1977 and 2024, \RnLedgerRaise{} raises and \RnLedgerCut{} cuts, records each act
with its source and a confidence grade. A year without a recorded act is evidence of no change only if someone looked.
An audit therefore went through the years that the ledger leaves without an act and checked whether a general revision
had in fact been made (Section~\ref{sec:limits_patterns}). The fiscal data are Eurostat's government finance statistics
for the European countries \citep{eurostat2026} and the IMF's Fiscal Monitor and World Economic Outlook for the other
three \citep{imf2026}. Each European budget's fiscal state is dated from its circumstances. A year is normal unless
market access was lost or threatened without a program (market stress), an adjustment program under IMF or EU
conditionality was in force (a program), or a crisis had ended while debt remained above 85 percent of GDP (the
aftermath). A year of market stress or a program is a crisis year. Appendix~\ref{app:data} defines the series and cites their sources.

\subsection{Pattern 1: legal pay erodes between acts and is restored in discrete steps}
\label{sec:pattern_erosion}

\begin{figure}[t]
\centering
\includegraphics[width=\textwidth]{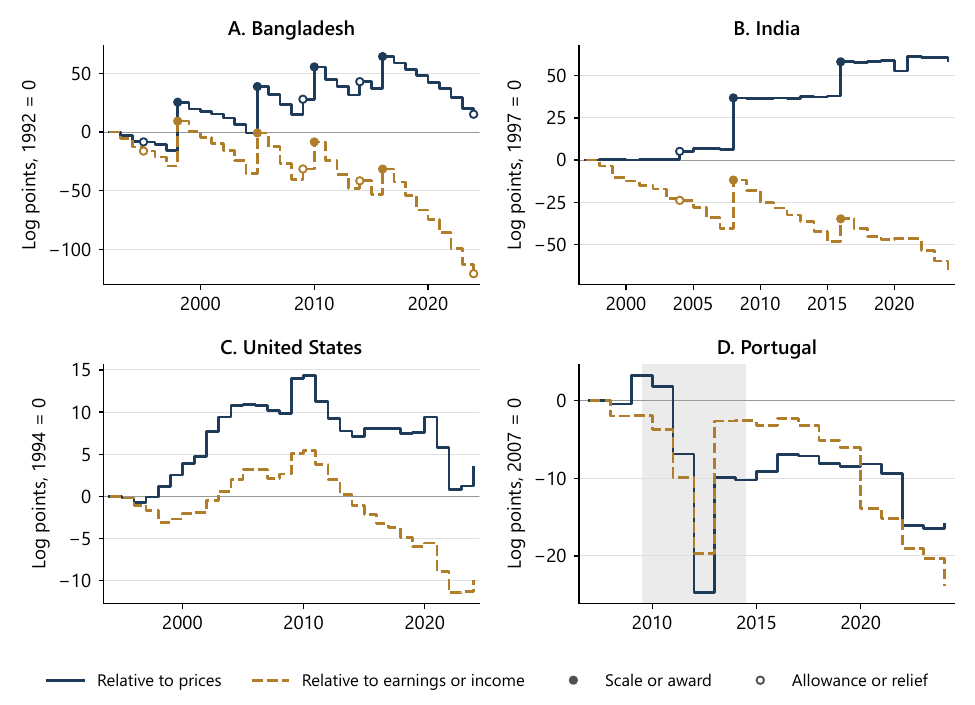}
\caption[Legal pay against prices and private earnings]{Legal pay against prices and private earnings. Each panel
cumulates the documented general changes of legal pay, as paid in each year, less consumer-price inflation (solid) and
less the growth of private earnings (dashed), in log points relative to the base year. Private earnings are
business-economy hourly wages in Portugal (Eurostat's Labour Cost Index) and the employment cost index of private wages and
salaries in the United States. In Bangladesh and India the dashed line is relative to income per head, the growth of
real GDP per head plus consumer-price inflation. Filled markers are new pay scales or commission awards, and open
markers are interim allowances or reliefs. Portugal's changes are annual averages, so a cut struck by the Court counts
for the part of the year it was in force, and the shaded years are its years of market stress or an adjustment program.
Within-grade increments and promotions are excluded.}
\label{fig:pay_erosion}
\end{figure}

Figure~\ref{fig:pay_erosion} follows the legal pay of a fixed position in the general schedule, cumulating the
documented general changes in the years in which they were paid, and measures it against two benchmarks. Against
consumer prices it shows what pay buys, and against private earnings it shows how pay keeps up with the labor market.
Private earnings are business-economy hourly wages in Portugal \citep{eurostat2026} and the employment cost index of
private wages and salaries in the United States \citep{blseci2026}. Bangladesh and India have no long series of private
earnings, and their nominal GDP series break where the national accounts were rebased. Their second benchmark is
therefore income per head, constructed by cumulating the growth of real GDP per head and consumer-price inflation.

\begin{figure}[!b]
\centering
\includegraphics[width=\textwidth]{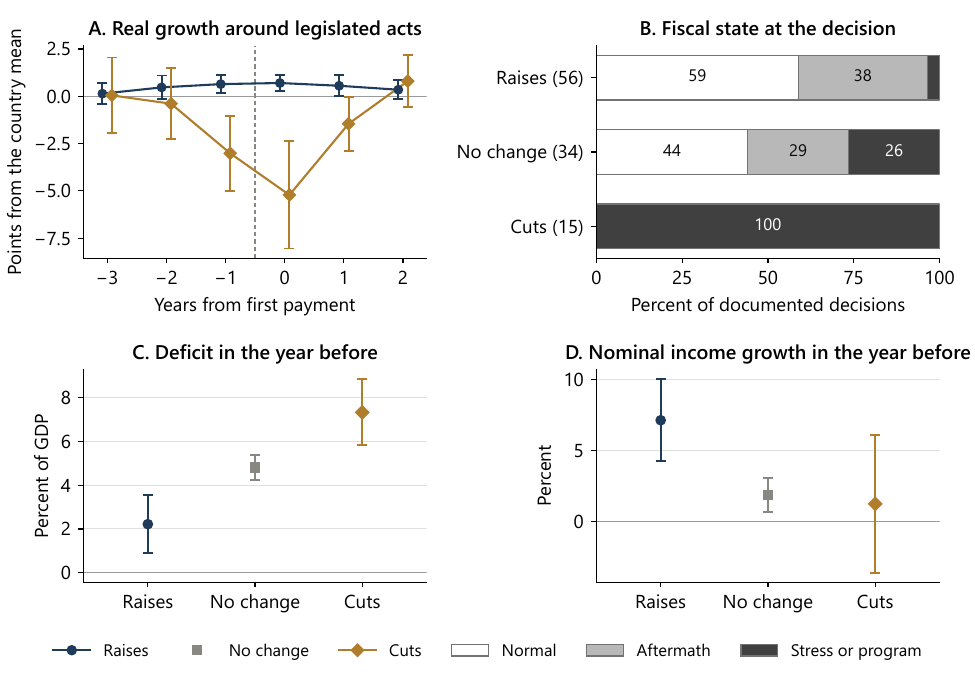}
\caption[The circumstances of raises and cuts]{The circumstances of raises and cuts. Panel~A plots real GDP growth
around the ledger's legislated raises and cuts, in points from the country's mean over the years researched for the ledger, by years
from first payment. A year with both is a cut year, and the dashed vertical line separates the years before the first
payment. Panels~B--D use the decision histories of Portugal, Ireland, Spain, Cyprus, Italy, Romania, Hungary and
Lithuania, one observation per decision, with the number of decisions in parentheses, and show decisions of no change
for comparison. They plot the fiscal state at the decision (B), and the general government deficit in percent of GDP (C)
and the growth of nominal GDP per head in percent (D) in the year before it. The intervals are 95 percent intervals
clustered by country.}
\label{fig:circumstances}
\end{figure}

Where pay is revised at discretion, it erodes between acts and is restored in discrete steps. In Bangladesh real legal
pay fell by \RsBgdErodePabs{} log points a year between revisions. The \RsBgdScaleNW{} national pay scales since 1992 raised legal pay by \RsBgdRevMean{} log points on average, or by \RsBrBgdJumpData{} in real terms, about one scale
every \RsBgdScaleGapW{} years. Against income per head the erosion was faster, \RsBgdErodeEabs{} log points a year.
Thus the scales restored purchasing power without restoring pay's position relative to incomes. By 2024 legal pay was
\RsBgdEndP{} log points above its 1992 level in real terms and \RsBgdEndEabs{} log points below it relative to income per
head. India shows what indexation changes. Its dearness allowance held legal pay level against prices between
commission awards (an average change of \RsIndErodeP{} log points a year), but not against incomes. Relative to incomes,
pay fell by \RsIndErodeEabs{} log points a year until an award restored part of the loss, and the two awards raised nominal pay by \RsBrIndScaleData{} log points on average.

Where a formula is the default, revisions are small and frequent. The General Schedule moved every January except during the
statutory freeze of 2011--2013, by \RsUsaRevMean{} log points on average, so its value relative to prices and to
private wages drifts without visible steps. Portugal's annual budgets kept legal pay close to prices until the crisis,
when the cuts lowered its real value by as much as 25 log points. In 2024 it was still \RsPrtEndPabs{} log points below
its 2007 level in real terms.

Across countries, raises are larger where they are rarer. In the \RsSizeSpacingN{} countries with at least three
raises, in the ledger or found by the audit, and two documented spells between them, the rank correlation between the
mean interval and the mean raise is \RsSizeSpacingRho{} ($p=\RsSizeSpacingP$). A linear fit adds
\RsSizeSpacingSlope{} log points to the mean raise for each additional year between raises \sd{\RsSizeSpacingSlopeSE}.
The correlation is unconditional, in that it does not hold inflation fixed. Erosion punctuated by discrete steps is what
a fixed cost of revising predicts. Moreover, Section~\ref{sec:benchmark} shows that a positive association between size
and rarity is what differences in the costs of revising would produce across countries, whereas differences in the
speed of erosion would make larger raises more frequent.

Within a country, raises are larger when inflation has been faster, but far less than in proportion. Across
\RsEroN{} documented spells between raises, the elasticity of a raise's size to mean inflation over the spell is
\RsEroSize{} \sd{\RsEroSizeSE} with country fixed effects (Appendix Table~\ref{tab:size_erosion}). The interval is no
shorter. Its elasticity is \RsEroInt{} \sd{\RsEroIntSE}, and \RsEroOneYear{} percent of the spells last a single year.
Both are associations, not effects of inflation.

\subsection{Pattern 2: raises and cuts arise in different circumstances}
\label{sec:pattern_circumstances}

Almost all raises come outside crises, while cuts follow a deterioration of the fiscal position (Figure~\ref{fig:circumstances}).
Panel~A uses the ledger's acts whose surrounding years have growth data, \RsLedgerRaiseN{} raises in
\RsLedgerRaiseCtry{} countries and \RsLedgerCutN{} cuts in \RsLedgerCutCtry{}. Around the raises, real growth is close
to its country mean, and slightly above it, by \RsLedgerRaiseGrZeroAbs{} points, in the year of first payment. Around
the cuts, in contrast, growth falls in the years before and is \RsLedgerCutGrZeroAbs{} points below the mean in the
year the cut is first paid.

\suppressfloats[t]
\begin{table}[!ht]\centering
\begin{threeparttable}
\caption{Selection into legislated raises and cuts}
\label{tab:selection}
\footnotesize\setlength{\tabcolsep}{2pt}
\begin{tabular*}{\textwidth}{@{\extracolsep{\fill}}l*{4}{C{1.9cm}}}
\toprule
& \multicolumn{4}{c}{Legislated act first paid in year $t$} \\
\cmidrule(lr){2-5}
& Raise & Cut & Raise & Cut \\
& (1) & (2) & (3) & (4) \\
\midrule
GDP growth, $t-1$ & 2.04\sym{***} & $-1.34$\sym{***} &  &  \\
 & (0.44) & (0.41) &  &  \\
GDP growth, $t-2$ & 2.02\sym{***} & $-0.38$ &  &  \\
 & (0.42) & (0.23) &  &  \\
Fiscal balance, \% of GDP, $t-1$ &  &  & 1.08 & $-1.35$\sym{***} \\
 &  &  & (0.85) & (0.44) \\
Public debt, \% of GDP, $t-1$ &  &  & $-0.01$ & $-0.11$\sym{*} \\
 &  &  & (0.19) & (0.06) \\
\midrule
$p$-value, reported predictors & $<0.001$ & 0.005 & 0.429 & 0.009 \\
Mean of the dependent variable & 51.5 & 4.9 & 47.7 & 5.2 \\
Observations & 728 & 728 & 539 & 539 \\
Countries & 51 & 51 & 44 & 44 \\
Lagged growth and inflation & Yes & Yes & Yes & Yes \\
Country fixed effects & Yes & Yes & Yes & Yes \\
Year fixed effects & Yes & Yes & Yes & Yes \\
\bottomrule
\end{tabular*}
\begin{tablenotes}[flushleft]\footnotesize
\item Notes: Each column is a linear-probability regression of the indicator that a legislated raise or cut is first paid in year $t$ on documented country-years. Coefficients are in percentage points of probability per percentage point of growth or of GDP, so coefficients and standard errors are multiplied by 100. Standard errors clustered by country are in parentheses, and stars use $p$-values from $t$ tests with one fewer degree of freedom than countries. The $p$-value row tests the reported predictors jointly. Every column includes two lags of real GDP growth, which columns 1 and 2 report, and lagged inflation. The events are the ledger's acts and the general acts that the year-by-year audit found, and country-years of unknown status are excluded. \mbox{\sym{*} $p<0.10$}, \mbox{\sym{**} $p<0.05$}, \mbox{\sym{***} $p<0.01$}.
\end{tablenotes}
\end{threeparttable}
\end{table}

The decision histories of the eight European countries sharpen the contrast. All \RsCutN{} of their cuts, in
\RsCutCtryW{} countries, were decided in market stress or an adjustment program, against \RsRaiseCrisis{} percent of the
\RsRaiseN{} raises and \RsNoneCrisis{} percent of the \RsNoneN{} decisions to leave pay unchanged (Panel~B). In the year
before the decision the deficit averaged \RsCutDef{} percent of GDP before cuts, \RsNoneDef{} percent before decisions
of no change and \RsRaiseDef{} percent before raises (Panel~C). Nominal income had grown by \RsRaiseNom{} percent in
the year before a raise and by \RsCutNom{} percent before a cut (Panel~D), and it fell in the year of the cut or the
year before in \RsCutNomFall{} percent of cuts.

Table~\ref{tab:selection} shows that the contrast survives country and year fixed effects. On documented
country-years, a percentage point more growth in the previous year is associated with a probability of a raise that is
higher by \RsSelRaiseGr{} percentage points and a probability of a cut that is lower by \RsSelCutGrAbs{} points. Neither
the fiscal balance nor debt predicts raises, but a balance weaker by one point of GDP is associated with a probability of
a cut that is higher by \RsSelCutBalAbs{} percentage points. Raises are thus associated with growth alone, and cuts with
both growth and the fiscal position. The model accordingly treats raises and cuts as distinct actions with different
triggers, not as positive and negative values of one policy.

\suppressfloats[t]
\begin{figure}[!ht]
\centering
\includegraphics[width=\textwidth]{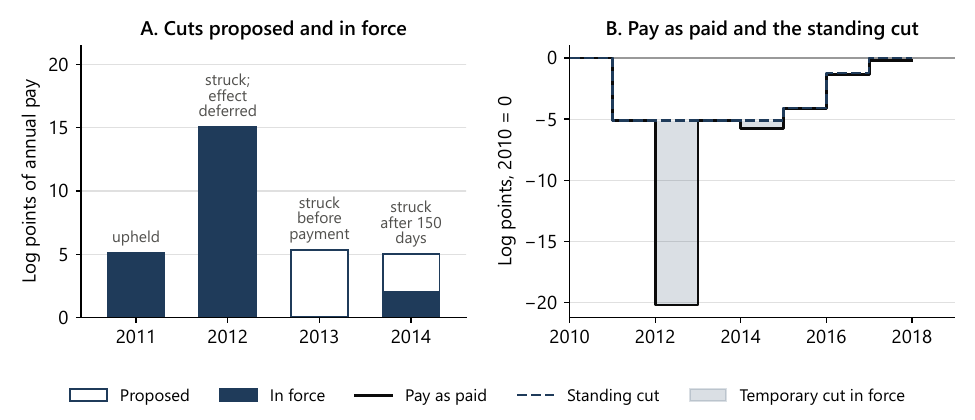}
\caption[Portugal's cuts proposed, in force and standing]{Portugal's cuts proposed, in force and standing. Panel~A plots
the cut that each budget of 2011--2014 legislated (outlined) and the part of it in force during the year (solid), in log
points of annual pay, and the labels give the Constitutional Court's ruling. Panel~B plots annual-average pay as paid
(solid) and the standing cut (dashed), in log points relative to 2010. The standing cut is the 2011 reductions that the
Court upheld less their legislated restoration, a fifth from January 2015 and the rest in quarterly tranches in 2016,
and the shaded area is the part of struck cuts in force. In 2014 the 2011 reductions were not applied between the ruling
of 30 May and 13 September, so pay as paid fell only \RsPrtPaidBelowStanding{} log points below the standing cut,
although \RsPrtInforceFourteen{} log points of the 2014 reductions were in force. Sources are the budget laws and the Court's rulings 396/2011, 353/2012, 187/2013 and 413/2014.}
\label{fig:prt_payment}
\end{figure}

\subsection{Pattern 3: legal reversals create payment paths}
\label{sec:pattern_payment}

Where a court can strike a cut, the cut that is legislated, the cut that is paid and the cut that lasts can differ for
years, and Portugal shows all three (Figure~\ref{fig:prt_payment}). The 2011 budget reduced base pay by
\RsPrtCutEleven{} log points, and the Constitutional Court upheld the reduction for the duration of the emergency. The
2012 budget added the suspension of the holiday and Christmas payments, worth \RsPrtCutTwelve{} log points of annual
pay. The Court struck the suspension in July 2012 but deferred the ruling's effect to 2013, so the suspension remained in
force for the whole year, and pay as paid fell \RsPrtPaidTwelve{} log points below its 2010 level. The 2013 budget
suspended the holiday payment again, and the Court struck it before it was due. The 2014 budget deepened the base
reductions by \RsPrtCutFourteen{} log points, and the Court struck them after 150 days. From that ruling until
September the 2011 reductions were not applied either, so pay as paid in 2014 was only \RsPrtPaidFourteen{} log points
below its 2010 level. Once the emergency had passed, the Court held that the 2011 reductions could not continue, and the
legislature restored them in 2015 and 2016.

These paths have three features that a model of revisions must reproduce. First, a struck cut is a temporary saving,
because it saves pay until the court strikes it. Portugal thus kept the saving from the 2012 suspension for a year and from part of the 2014 reductions for five
months. Second, a cut that stands is durable. The 2011 reductions, for example, lasted four years
before the legislature reversed them in tranches. Third, the Treasury may later
repay what employees lost, so the same legal act produces one path for employees' pay and another for the Treasury's
cash. Spain refunded its 2012 suppression of the December extra payment, worth a fourteenth of annual pay, as back pay in
2015 and 2016. Similarly, Lithuania paid compensation in 2016--2020 for the cuts of 2009--2013 that its Constitutional
Court had held disproportionate.

\subsection{What the patterns measure}
\label{sec:limits_patterns}

The patterns describe the decision rather than its consequences. They cannot identify what a revision does to the
economy, for two reasons. First, raises follow growth (Table~\ref{tab:selection}), so a comparison of countries that
raise pay with countries that do not confounds the raise with the conditions that selected it. Second, the comparison
group is contaminated, because many years that look like years without a revision were not. The natural design
compares output around the start of a raise episode with output in years without an act (Appendix~\ref{app:data}). The
audit applied the ledger's definition of an act, year by year, to the \RprovNyears{} comparison and pre-episode years of
the \RprovNctry{} countries that enter this design, the ledger's \RnEventCountries{} and the United States. It found \RprovNewGen{} general revisions that the ledger
omits, many of them annual adjustments decided in each year's budget, and in a random sample of \RauditN{} years outside the windows researched for the ledger it found a general revision in \RauditGen{}. Once these years are removed, too few raise
episodes are preceded by three years without an act to estimate the response of output with useful precision
(Appendix~\ref{app:data}). Section~\ref{sec:transmission} therefore treats the transmission of a revision through a
calibrated, conditional model, and the estimation of Section~\ref{sec:quant} rests on the decisions themselves.

\section{Model}
\label{sec:model}

This section builds the government's dynamic program, which can generate all three patterns. A government revises a nominal pay
schedule with the instruments its institutions provide, and the budget links each decision to the next. The payroll
moves the deficit and debt, which move the spread, the risk of fiscal stress and the length of a crisis's aftermath, and
these in turn shape the next decision. Section~\ref{sec:benchmark} isolates the trade-off behind Pattern~1 in an
analytical benchmark. Section~\ref{sec:component} adds the fiscal state and the courts behind Patterns~2 and~3 and
characterizes what a court's review changes. Section~\ref{sec:transmission_model} follows a revision through the
economy, and Appendix Table~\ref{tab:notation} collects the notation.

\subsection{Environment and timing}
\label{sec:environment}
\label{sec:gap}

Time is annual, and the government decides at each budget. Let $W_t$ be the legal pay of a grade step and $W^\ast_t$ a
reference schedule in the same currency units, defined as the pay that keeps the quality of public services at its
efficient level. Quality depends both on what pay buys and on pay relative to private earnings, so the reference grows
with prices and with a share $s_w\in[0,1]$ of the real growth of private wages (Appendix~\ref{app:model}). In Bangladesh
and India, which have no long series of private wages, nominal income per head stands in for them. The reference is not
observed. It is a construct of the model, and $s_w$ is estimated from the decisions. The state of the government's
problem is the gap
\begin{equation}
 x_t=100\,(\ln W^\ast_t-\ln W_t),
 \label{eq:gap}
\end{equation}
the distance in log points by which legal pay falls short of the reference. A raise from $W$ to $W'$ lowers the gap by
$100\ln(W'/W)$, and a cut raises it. While the schedule is fixed, the gap grows by
\begin{equation}
 \Delta x_{t+1}=(1-\kappa_P)\,\pi_t+s_w\,(\omega_t-\pi_t),
 \label{eq:drift}
\end{equation}
where $\pi_t$ is inflation and $\omega_t$ the growth of nominal private wages over budget year $t$, which the government
does not yet know when it decides, and $\kappa_P$ is the share of inflation that a rule indexes (India's dearness allowance has $\kappa_P=1$). Erosion is thus the default. Without a legal act, the gap widens
by the unindexed part of inflation and a share of real wage growth. A government that weighs the taxpayers who finance
the payroll more heavily than the employees who receive it prefers to hold pay somewhat below the reference, at a
positive gap $x^\star$ defined in Section~\ref{sec:benchmark}.

\looseness=-1 The government proposes the general change of pay in each budget, and other agents constrain it. The Treasury requires
savings on the payroll when market access is threatened. The employees and their union tolerate erosion only up to a
point, beyond which a dispute is costly. And where they exist, a court may strike cuts and a pay commission must report
before a new scale can be set. Households, firms and banks respond to the resulting path of pay in the transmission block
of Section~\ref{sec:transmission_model}. Within a budget year the state is observed, the government proposes a change, a
court rules on any cut, pay is paid and the budget executed, and the fiscal state, debt and the macroeconomic state move
to the next year.

\subsection{The reset problem: an analytical benchmark}
\label{sec:benchmark}

A government that leaves its schedule unchanged while the reference rises lets the gap grow. A small correction would
reduce employees' discontent, but a revision carries a political cost even when it is small. Waiting saves that cost
while making the eventual adjustment larger and harder to finance. The government therefore tolerates some erosion and
revises only once waiting has become costly enough.

The benchmark keeps only this trade-off. The flow payoff is $-\tfrac a2(x-x^\star)^2$, where $a>0$ weighs misalignment.
The preferred gap $x^\star=(\eta_F-\eta_W)/a$ is where the net political value of a smaller payroll equals the marginal
cost of holding pay that far below the reference, $a\,x^\star$. That net value is the extra weight $\eta_F$ that the
government gives the taxpayers who finance the payroll less the extra weight $\eta_W$ it gives the employees who receive
it. A raise from $x$ to $x'<x$ incurs a fixed cost $K^+$ and a convex financing cost $\phi_F(x-x')^2$. Between revisions
the gap follows a Brownian motion with drift $\mu>0$ and variance $\sigma_d^2$ per year, and the government discounts at
the rate $\delta_I=-\ln\beta_I$. Its value $V$ solves
\begin{equation}
 \max\Big\{-\tfrac a2(x-x^\star)^2+\mu V'(x)+\tfrac{\sigma_d^2}{2}V''(x)-\delta_IV(x),\ \max_{x'<x}\big[V(x')-K^+-\phi_F(x-x')^2\big]-V(x)\Big\}=0 ,
 \label{eq:qvi}
\end{equation}
whose first term is the return to waiting and whose second is the gain from revising now \citep{stokey2009}. Both are
at most zero, and one of them is zero at every gap. Thus the government waits while revising would lose value, and it
revises as soon as revising is worth as much as waiting.

\begin{proposition}[The reset band]
\label{prop:band}
(i) Without uncertainty, $\sigma_d=0$, and among policies that only raise pay, keep the gap bounded and do not accumulate
intervention times, the highest long-run average payoff is attained by resetting the gap from $x^\star+J^\ast/2$ to
$x^\star-J^\ast/2$ every $T^\ast=J^\ast/\mu$ years, where $J^\ast$ is the unique positive root of
\[
 \frac{a}{12}J^3+\phi_F\mu J^2=K^+\mu .
\]
(ii) With uncertainty, $\sigma_d>0$, if the optimal policy raises pay to a reset point $\hat x$ whenever the gap reaches a
trigger $\bar x>\hat x$, then below the trigger
\[
 V(x)=v_0+v_1x+v_2x^2+v_3\exp\Big(\frac{\sqrt{\mu^2+2\delta_I\sigma_d^2}-\mu}{\sigma_d^2}\,x\Big),
\]
where $v_0+v_1x+v_2x^2$ sets the return to waiting to zero, and $(v_3,\bar x,\hat x)$ solve value matching, smooth
pasting and optimality of the reset point. The time between raises has an inverse Gaussian distribution with mean
$(\bar x-\hat x)/\mu$ and variance $(\bar x-\hat x)\sigma_d^2/\mu^3$.
\end{proposition}

\begin{corollary}[Two costs of a revision]
\label{cor:costs}
In part (i), let $s_F=\phi_FJ^{\ast2}/K^+\in[0,1)$, the financing cost of the optimal revision relative to its fixed cost.
The elasticities of the size $J^\ast$ are $1/(3-s_F)$ with respect to $K^+$, $-s_F/(3-s_F)$ with respect to $\phi_F$ and
$(1-s_F)/(3-s_F)$ with respect to $\mu$. The interval $T^\ast$ has the same elasticities with respect to $K^+$ and
$\phi_F$, and $-2/(3-s_F)$ with respect to $\mu$. Hence (a) a higher fixed cost makes revisions larger and rarer, and a
higher financing cost makes them smaller and more frequent, (b) faster erosion makes them larger and more frequent, and
(c) size and interval alone do not identify $s_F$, but the elasticity $e$ of size with respect to $\mu$ does, with
$s_F=(1-3e)/(1-e)$.
\end{corollary}

The proposition applies the economics of inaction \citep{sheshinski1977,dixit1991,stokey2009} to a price that the
government sets by law, and part~(i) shows its logic in closed form. A policy that raises pay by $J$ whenever erosion
has accumulated $J$ revises every $J/\mu$ years, so its average loss per year is
\begin{equation}
 C(J)=\frac{K^+\mu}{J}+\phi_F\,\mu J+\frac{a}{24}\,J^2 .
 \label{eq:avgloss}
\end{equation}
The first term is the fixed cost of one revision spread over the years until the next, and the second is the financing
cost $\phi_FJ^2$ of one revision spread in the same way. The third is the average misalignment of a gap that runs evenly
from $x^\star-J/2$ to $x^\star+J/2$. A larger revision spreads the fixed cost over more years but raises both the
financing cost per year and the misalignment, and the cubic of part~(i) is the first-order condition that balances the
three. The government also overshoots its preferred gap, resetting to $x^\star-J^\ast/2$, because the new schedule will
stay in force for several years.

What the government's problem adds to the menu-cost model of a price is the second cost, which comes from the
budget. A fixed cost makes the government economize on the number of revisions, and a convex cost of financing an
increase at once makes it economize on their size, as fixed and convex costs of adjustment do for a firm's investment
\citep{abel1994,cooper2006}. The two costs therefore
push the size of revisions in opposite directions (Corollary~\ref{cor:costs}(a)). Faster erosion makes revisions both
larger and more frequent (part~(b)), and how much larger reveals which cost dominates (part~(c)). With a fixed cost
alone, $s_F=0$, the size of a revision rises with the cube root of the speed of erosion. As the financing cost comes to
dominate, $s_F\to1$, the elasticity falls toward zero, so the size of a revision stops responding to erosion and only its
frequency adjusts. Inflation, which varies the speed of erosion across years and countries, can therefore separate two
costs that the size and spacing of revisions alone cannot. The documented spells of
Section~\ref{sec:pattern_erosion} do not make that separation. The size of a raise responds to inflation with an
elasticity of about a third, the value under a fixed cost alone, but the interval does not shorten as the benchmark
requires, because most raises already come a year apart. The estimate therefore does not measure $s_F$
(Appendix~\ref{app:size_erosion}), and Section~\ref{sec:quant} separates the two costs with the yearly decisions.

Pattern~1 is the band of part~(ii), erosion between discrete resets. The band also says what the cross-country
association of Section~\ref{sec:pattern_erosion} reflects. A reset undoes the erosion accumulated since the last one,
$J^\ast=\mu T^\ast$. At a given speed of erosion, any difference in the costs of revising, fixed or financing, therefore
moves size and interval together, while a difference in the speed of erosion moves them apart. That larger raises come
with longer intervals is what differences in the costs of revising would produce, and not what differences in erosion
would, and part~(c) says what would tell the two costs apart. Appendix~\ref{app:proofs} gives the proofs and checks part~(ii)
numerically.

\subsection{The government's decision with fiscal feedback}
\label{sec:component}
\label{sec:incumbent}

The benchmark gives the government one instrument and no budget. The estimated problem keeps its economics and adds each
institution where it operates.

\paragraph{State.} The state is $s_t=(x_t,b_t,E_t,F_t,m_t,L_t,r_t,\ind^C_t)$. It contains the gap, general government
debt at the start of the budget year in percent of GDP, the election phase, which counts the years to the next general election, and the fiscal state. It also contains the
macroeconomic state $m_t$, the outstanding cuts $L_t$ that a court may strike or order reversed, whether last year's
change was a raise ($r_t$) and whether a pay commission is sitting ($\ind^C_t$). The macroeconomic state describes the
country's inflation and real growth in the year before the budget year, the latest known when the government decides.
Those of budget year $t$ are drawn after the decision, as part of $m_{t+1}$, and they enter the drift~\eqref{eq:drift},
the deficit and debt of~\eqref{eq:deficit} and the onset of stress in~\eqref{eq:onset} together.

\paragraph{The budget.} In percent of GDP, the deficit and next year's debt are
\begin{equation}
 d_t=p(F_t)+s_g\,e^{-x'_t/100}+\frac{i(b_t,F_t)}{100}\,\frac{b_t}{1+n_t},\qquad b_{t+1}=\frac{b_t}{1+n_t}+d_t ,
 \label{eq:deficit}
\end{equation}
where $x'_t$ is the gap after the year's decision, $s_g$ the payroll at the reference, $p(F)$ the primary deficit
excluding the payroll in fiscal state $F$, $n_t$ nominal growth and $i$ the effective interest rate in percent. This is the
government budget constraint in ratio to GDP, in which inflation and real growth erode the debt inherited from earlier
years \citep{hall2011}. A raise therefore
widens the deficit in the year it is paid and adds to debt from then on. The spread over Germany and the effective rate
follow
\begin{equation}
 \zeta(b,F)=\zeta_0(F)+\zeta_1(F)\,(b-60)^+,\qquad i(b,F)=i_c+\iota\,\zeta(b,F).
 \label{eq:spread}
\end{equation}
Here $\zeta_0(F)$ is the spread at debt of 60 percent of GDP in fiscal state $F$, $\zeta_1(F)$ its slope in debt above 60 percent, and $i_c$ the effective rate at a zero spread.

\paragraph{Fiscal states.} The fiscal state is normal ($N$), market stress ($M$), an adjustment program under
conditionality ($C$) or an aftermath ($A$), and normal and aftermath years are further split by whether an excessive
deficit procedure of the European Union is open. From a normal or aftermath year, stress or a program begins with
probability
\begin{equation}
 \Pr(F_{t+1}\in\{M,C\}\mid F_t\in\{N,A\})=\Lambda\big(c_0+c_g\,g_t+c_d\,d_t\big),
 \label{eq:onset}
\end{equation}
where $\Lambda$ is the logistic function, $g_t$ real growth and $d_t$ the deficit of~\eqref{eq:deficit}. This is
the onset of fiscal stress. It begins with a budget of market stress, except in Hungary and Romania, whose crises began
with a program. A raise therefore increases the risk of stress in the following year through the deficit it adds. After a crisis the aftermath
lasts while debt is above 85 percent of GDP, so a cut that reduces debt also shortens the aftermath. After its first
budget, stress continues with probability 0.6 and escalates into a program with probability 0.2, and in later budgets it
continues with probability one half and escalates with probability 0.4. A program persists with probability two
thirds.

\paragraph{The flow payoff.} Let $S(x)=100\,(1-e^{-x/100})$ be the saving on the payroll at the gap $x$, in percent of
the payroll at the reference. The flow payoff at the gap $x$ after the year's decision is
\begin{equation}
\begin{split}
 u(x;b,E,F)={}&-\lambda_Fx-\tfrac{a_S}{2}x^2+\big[\lambda_F+\eta_F(b)-\eta_W(E)\big]S(x)\\
 &-c_U\big(x-x_s(b)\big)^+-\eta_T\big[Q(b,F)-S(x)\big]^+ .
\end{split}
 \label{eq:support}
\end{equation}
The first two terms are the value of public services. Their slope at the reference equals the excess burden $\lambda_F$ of
the revenue that pays for the services, which makes the reference efficient pay. The third term is the saving on the
payroll, which the government values at the excess burden plus the extra political weight $\eta_F$ of taxpayers, less the
weight $\eta_W$ of employees. That weight is calibrated at 0.3 in every election phase, so elections enter the estimated
problem only through the value of a visible raise below. The fourth is the cost of conflict with the union once the gap exceeds its tolerance
$x_s$, and the last is the Treasury's penalty when the saving falls short of the saving it requires. The curvature
$a_S=a-(\lambda_F+\eta_F-\eta_W)/100$ of the service value makes the curvature of the payoff at the reference equal to
the benchmark's $a$. In the aftermath, debt above its pre-crisis level $\bar b$ raises both the political cost of the
payroll and the union's tolerance, $\eta_F(b)=\eta_F+a\chi(b-\bar b)^+$ and $x_s(b)=x_s+\chi(b-\bar b)^+$. This is the
debt overhang, and because debt responds to pay through~\eqref{eq:deficit}, the overhang responds to pay as well.

\paragraph{The financing requirement.} In stress or a program the Treasury requires a saving on the payroll of
\begin{equation}
 Q(b,F)=\gamma_F\,\frac{\zeta\,b}{s_g}+\gamma_D\,\frac{100\,(d^{\mathrm{ref}}-3)^+}{s_g}
 \label{eq:requirement}
\end{equation}
percent of the payroll at the reference, where $\zeta b/100$ is the extra interest bill at the current spread, in percent of GDP,
and $d^{\mathrm{ref}}$ the deficit at the reference payroll and at mean nominal growth. The payroll is asked to bear the share $\gamma_F$ of the extra
interest bill and the share $\gamma_D$ of the deficit above three percent of GDP. Under an open excessive deficit
procedure outside a crisis, the requirement consists of the deficit term alone. Appendix~\ref{app:model}
derives~\eqref{eq:support} and~\eqref{eq:requirement} from the problems of employees, their union, taxpayers and the
Treasury's creditors.

\paragraph{Actions and their values.} Each country's documents determine which actions are available to its government.
Every action ends the year's decision at a post-decision gap $x'$, with outstanding cuts $L'$ and a record $r'$ of
whether pay was raised, and its value is the value of that position net of the action's cost. The value of a position is
the year's flow payoff plus the discounted expected value of the next year,
\begin{equation}
 \mathcal W(x',L',r';s)=u(x';b,E,F)+\beta_I\,\E\big[V(s_{t+1})\mid x',L',r',s\big].
 \label{eq:postvalue}
\end{equation}
The expectation runs over next year's state. The gap and the outstanding cuts move to
\begin{equation}
 x_{t+1}=x'+\Delta x_{t+1}-L^{\mathrm{rev}}_{t+1},\qquad L_{t+1}=L'-L^{\mathrm{rev}}_{t+1},
 \label{eq:lawofmotion}
\end{equation}
where $\Delta x_{t+1}$ is the drift~\eqref{eq:drift} and $L^{\mathrm{rev}}_{t+1}$ the cuts that a court reverses at the
start of the year, which are zero without a court. Next year's debt follows~\eqref{eq:deficit} at the payroll that $x'$
implies, and the fiscal state follows~\eqref{eq:onset} at the deficit that this payroll produces, together with the
durations of stress, programs and the aftermath. The macroeconomic state follows the country's Markov chain, and its
draw $m_{t+1}$ sets the year's drift, nominal growth and real growth. In the European countries the draw carries the
documented mean differences in inflation and real growth between crisis and normal years when next year is a second or later
budget of stress or a year of a program, and the private wage carries its documented mean deviation in crisis and aftermath years
(Appendix~\ref{app:components}). The election phase advances by
a year, and $r_{t+1}=r'$. The decision therefore moves next year's debt and, through the deficit, next
year's risk of stress.

\paragraph{Annual pay setting.} The government waits, raises pay by $j$ log points or cuts it by $c$ log points. A raise
from $x$ to $x-j$ costs $\phi_F[S(x)-S(x-j)]^2$, the square of the payroll increase that must be financed at once, plus a
restart cost $K_F$ if the previous year had no raise ($r=0$). A visible raise in an election year is worth $v_E$ per log
point \citep{nordhaus1975}. A cut of $c$ log points costs a fixed $K^-$ plus $\lambda_L$ per log point of pay actually
lost. The values of the three directions are therefore
\begin{equation}
\begin{aligned}
 v(s,\mathrm{wait})&=\mathcal W(x,L,0;s),\\
 v(s,\mathrm{raise},j)&=\mathcal W(x-j,L,1;s)-\phi_F\big[S(x)-S(x-j)\big]^2-K_F\,(1-r)+v_E\,\ind^E\,j,\\
 v(s,\mathrm{cut},c)&=\mathcal W(x+c,L+c,0;s)-K^--\lambda_L\,c,
\end{aligned}
 \label{eq:values_annual}
\end{equation}
where $\ind^E$ equals one in an election year. The outstanding cuts matter only where a court can strike or reverse
them.

\paragraph{A court that strikes cuts (Portugal and Italy).} In stress or a program, a cut that takes the outstanding cuts
above an admissible burden $\bar B$ is struck with probability $\rho_S$, after it has been in force for the share $\tau$
of the year. That share is the whole year in the budgets of stress and the first budget of a program, and a fifth of
the year in later program years. A struck
cut leaves the gap and the outstanding cuts where they were, and only the year's flow payoff is taken at the year's
average gap $x+\tau c$. The model takes next year's debt and risk of stress to be those of the gap $x$, so the saving
made while a struck cut is in force does not lower the deficit. A cut with $L+c>\bar B$ is thus worth
\begin{equation}
\begin{split}
 v(s,\mathrm{cut},c)={}&(1-\rho_S)\,\mathcal W(x+c,L+c,0;s)+\rho_S\big[\mathcal W(x,L,0;s)+u(x+\tau c)-u(x)\big]\\
 &-K^--\lambda_L\,\xi\,c ,
\end{split}
 \label{eq:values_court}
\end{equation}
where $u$ is evaluated at the year's $(b,E,F)$ and $\xi=1-\rho_S(1-\tau)$ is the expected share of the cut that is in
force during its year (Proposition~\ref{prop:court}). The court also reverses cuts. At the start of a later program year
it reverses the excess of the outstanding cuts over $\bar B$ with probability 0.9, and at the start of an aftermath year
it reverses all of them with probability $r_A$. Thus $L^{\mathrm{rev}}_{t+1}$ in~\eqref{eq:lawofmotion} is
$(L'-\bar B)^+$ or $L'$ with these probabilities, and zero otherwise. These probabilities and shares are calibrated to
Portugal's rulings. Lithuania's court orders reversals but does not strike cuts, and it reverses all outstanding cuts
with probability 0.25 at the start of every year outside stress or a program.

\paragraph{Implementation years.} A change fixed by an instrument adopted two or more years earlier is applied as
documented and is not a decision.

\paragraph{Pay commissions (Bangladesh and India).} The government may wait, pay an interim allowance at the fixed cost
$K_A$ plus the financing cost, cut, or appoint a commission at the fixed cost $K_I$ when none is sitting. A sitting
commission reports with the documented yearly probability $\rho_R$, and the government then sets the new scale at the
financing cost $\phi_C[S(x)-S(x')]^2$. In these countries the record $\ind^C$ of whether a commission is sitting takes
the place of $(L,r)$ in~\eqref{eq:postvalue}. Waiting and cutting are valued as in~\eqref{eq:values_annual}, with
$\ind^C$ unchanged, and the other actions are worth
\begin{equation}
\begin{aligned}
 v(s,\mathrm{allowance},j)&=\mathcal W(x-j,\ind^C;s)-K_A-\phi_F\big[S(x)-S(x-j)\big]^2+v_E\,\ind^E\,j,\\
 v(s,\mathrm{appoint})&=\rho_R\,V^C(x;s)+(1-\rho_R)\,\mathcal W(x,1;s)-K_I,\\
 v(s,\mathrm{scale},j)&=\mathcal W(x-j,0;s)-\phi_C\big[S(x)-S(x-j)\big]^2+v_E\,\ind^E\,j .
\end{aligned}
 \label{eq:values_commission}
\end{equation}
Here $V^C(x;s)$ is the value of a year in which the report arrives. In that year the government chooses between leaving
pay unchanged, worth $\mathcal W(x,0;s)$, and a new scale of size $j$, after which no commission sits. A commission
appointed in a year reports within that year with probability $\rho_R$, and otherwise it sits. While it sits, its report
arrives each year with probability $\rho_R$, and in the other years the government waits, pays an allowance or cuts.

The European countries have the fiscal states and the budget of~\eqref{eq:deficit}--\eqref{eq:onset}. No documented
episode of fiscal stress changed general pay in the United States, Bangladesh or India, so their problems have a single
fiscal state.

\paragraph{Decisions.} A decision is a direction $k$ (wait, raise, cut, or appoint a commission, where a raise in a
commission country is an interim allowance or a new scale) and, for a raise or a cut, a size $j$, written $c$ for a cut
in~\eqref{eq:values_annual} and~\eqref{eq:values_court}. The documented state
does not determine the documented decision exactly, and the model attributes what the state leaves unexplained to taste
shocks. Each year the government's valuation of every option is perturbed by shocks $\varepsilon_{kj}$ that it observes
when it decides and anticipates for future years. Its value function therefore solves the Bellman equation
\begin{equation}
 V(s)=\E_\varepsilon\Big[\max_{(k,j)\in\mathcal A(s)}\big\{v(s,k,j)+\varepsilon_{kj}\big\}\Big],
 \label{eq:bellman_max}
\end{equation}
where $\mathcal A(s)$ is the set of actions that the country's documents allow in the state $s$ and $v(s,k,j)$ is the
value of an action in~\eqref{eq:values_annual}--\eqref{eq:values_commission}. While a commission sits, the value is
$V(s)=\rho_RV^C(x;s)+(1-\rho_R)\,\E_\varepsilon[\max\{\cdot\}]$, with the maximum over waiting, allowances and cuts, and
$V^C$ is the same expected maximum over leaving pay unchanged and the new scales. Without the shocks,
\eqref{eq:bellman_max} is the deterministic Bellman equation $V(s)=\max_{(k,j)\in\mathcal A(s)}v(s,k,j)$, in which each
value combines the year's payoff, the action's cost and the discounted continuation value of~\eqref{eq:postvalue}. The
shocks have the generalized extreme value distribution of a nested logit, with scale $\sigma$ across the directions and
scale $\sigma_J\le\sigma$ across the sizes within a direction. With the shocks centered at zero, their expected maximum has a closed form, and with the sizes
listed on a grid, \eqref{eq:bellman_max} becomes
\begin{equation}
 V(s)=\sigma\ln\sum_k\exp\big(I_k(s)/\sigma\big),\qquad
 I_k(s)=\sigma_J\ln\sum_{j}h_j\exp\big(v(s,k,j)/\sigma_J\big),
 \label{eq:bellman}
\end{equation}
where $h_j$ is the width in log points of the size $j$, so that $I_k$ approximates an integral over a continuum of
sizes, and $I_k=v(s,k)$ for waiting and appointing, which have no size. The probability of a direction is
$\exp(I_k/\sigma)/\sum_{k'}\exp(I_{k'}/\sigma)$, and that of a size within it is proportional to $h_j\exp(v/\sigma_J)$.
This is a nested logit, and $\sigma_J\le\sigma$ makes it consistent with random utility \citep{mcfadden1978}.

\begin{assumption}[Separation from output]
\label{ass:separation}
Inflation and real growth follow the country's chain, estimated on its normal years, with the documented shifts of
crisis years (Appendix~\ref{app:components}), and they respond to pay only through the fiscal state. The fiscal state
responds to the pay decisions only through the deficit and debt of~\eqref{eq:deficit}. In the documented histories the
spread is the documented October spread, and its deviation from the schedule~\eqref{eq:spread} is held at its
documented value in counterfactuals. The onset of stress and the end of the aftermath follow the deficit and debt.
\end{assumption}

\begin{definition}[Solution]
\label{def:solution}
Given the macroeconomic chain, the fiscal law of motion of~\eqref{eq:deficit}--\eqref{eq:onset}, the court's rules and
the drift~\eqref{eq:drift}, a solution of a country's problem is a value function $V$ that satisfies~\eqref{eq:bellman_max}
at every state, together with the choice probabilities it implies and the law of motion of the state that they induce.
\end{definition}

The problem is the same for every country. Countries differ in their action sets, which the documents determine, and in
their parameters, which are estimated country by country. The benchmark of Section~\ref{sec:benchmark} is the special
case of the commission countries' problem in which every other feature is shut down.

\begin{lemma}[Nesting]
\label{lem:nesting}
With one fiscal state, no court, no electoral or conflict term, no indexation, a constant erosion $\mu$ per year, no
interim allowances or cuts, a commission that reports in the year of its appointment, a quadratic payoff and financing
cost and $\sigma\to0$, problem~\eqref{eq:bellman} is the reset problem of Section~\ref{sec:benchmark} with decisions once a
year, $K^+=K_I$ and $\phi_F=\phi_C$. Without discounting, the average loss of a policy whose cycles all have one length
differs from that of Proposition~\ref{prop:band}(i) by a constant, so the best such policy has one of the two whole
numbers of years around $T^\ast$. The optimum over all policies may alternate between the two.
\end{lemma}

\paragraph{What a court's review changes.} A court in this problem does not forbid cuts. Instead, it changes what a cut
beyond the admissible burden is worth to the government. Such a cut is struck with probability $\rho_S$, but only after
it has been in force for the share $\tau$ of the year, and with probability $1-\rho_S$ it stands.

\begin{proposition}[Temporary savings]
\label{prop:court}
In a crisis year, let a cut of $c$ log points take the outstanding cuts above the admissible burden $\bar B$.
(i) The expected part of the cut that is in force during its year is $\xi(\rho_S,\tau)\,c$ log points of pay, where
\[
 \xi(\rho_S,\tau)=1-\rho_S\,(1-\tau),
\]
and this is also the expected pay that employees lose, which the government weighs at $\lambda_L$ per log point.
(ii) If review is certain and comes before the cut takes effect, $\rho_S=1$ and $\tau=0$, the cut is never in force, and
its value falls short of the value of waiting by the fixed cost $K^-$, so proposing it is dominated.
(iii) Otherwise $\xi(\rho_S,\tau)>0$, and the speed and the certainty of review are complements in lowering it. The
effect of certainty, $\partial\xi/\partial\rho_S=-(1-\tau)$, is larger in magnitude the earlier review comes, and the
effect of speed, $\partial\xi/\partial\tau=\rho_S$, is larger the more certain review is.
\end{proposition}

A struck cut is thus a temporary saving, the feature of Portugal's history in Pattern~3, and the proposition says when a
government can use it. Under the calibrated court, $\rho_S=0.9$, a cut beyond the burden is in force in the budgets of stress
and the first budget of a program for as much of the year in expectation as a cut that stands, $\xi=1$, because a struck cut stays in force for the
whole year. In later program years, when it stays in force for a fifth of the year, $\xi=0.28$. Review before a cut
takes effect lowers $\xi$ to $0.1$, the chance that the cut stands, and only review that is also certain lowers it to
zero. Whether a government proposes such a cut when $\xi>0$ depends on what the saving is worth to it, which in a crisis
is governed mainly by the Treasury's requirement. Section~\ref{sec:cf_prt} measures what the incentives of each court
cost.

\subsection{Predictions}
\label{sec:predictions}

The problem predicts each of the three patterns, together with a feedback that links them.

\paragraph{Erosion and discrete revisions.} Where the scale changes only through a commission, the fixed cost of
appointing one and the costs of financing a new scale and the interim allowances produce a band. The gap widens until a scale resets it in a large
step, and a higher fixed cost makes the steps larger and rarer. Where an annual budget sets pay, the restart cost $K_F$
makes freezes persist and raises resume in steps. Indexation removes the part of the erosion that comes from prices.

\paragraph{Raises and cuts in different states.} Outside stress the government rarely needs to cut, because inflation
erodes real pay without a law. Cuts come when the requirement~\eqref{eq:requirement} exceeds the saving at the current
gap, which ties them to market pressure and the deficit. Because weak growth raises the risk of stress
in~\eqref{eq:onset}, it ties them to the business cycle as well.

\paragraph{Payment paths under a court.} A struck cut still saves pay for the part of the year it is in force, so a
government facing a court may propose cuts it expects to lose (Proposition~\ref{prop:court}). Each struck cut is a
temporary saving that is later undone, and each cut that stands lasts until a reversal is ordered or legislated.

\paragraph{Fiscal feedback.} A cut lowers the deficit and debt, and with them the risk of stress, the next
requirement, the overhang and the length of the aftermath, whereas a raise does the opposite. The overhang keeps pay
low after a crisis for as long as debt remains high.

\subsection{The transmission of a revision}
\label{sec:transmission_model}

A raise pays public employees more and must be financed by the Treasury. It reaches firms as demand and, where the
Treasury borrows the deposits that fund their loans, as a reduction in credit. The block that follows this path serves
the welfare comparisons of Section~\ref{sec:design}, which need the output path that each path of pay produces. It runs
one way, from the decision to the economy. Output does not feed back into the pay decision, so the block is a
conditional transmission model rather than a completed general equilibrium (Section~\ref{sec:transmission}). It is a
log-linear small open economy of households, firms, banks and a government that finances a revision by taxes, bank
credit, retail bonds and the market, in shares $\theta_T,\theta_B,\theta_R,\theta_F$ set by each country's institutions, and Appendix~\ref{app:transmission} states its equations and calibration. Each country's exchange-rate regime is part of its calibration, since under a fixed exchange
rate or in a monetary union monetary policy does not offset a fiscal expansion \citep{gali2008,nakamura2014,farhi2016}.
The fiscal size of a raise, its dose $\Omega$, is the increase in pay times the payroll it covers, in points of GDP.

\paragraph{Households.} Households value consumption and leisure and discount the future at the factor $\beta$. They hold
deposits, which pay the gross rate $R^D$, and retail government bonds, and they can borrow up to a limit that rises with documented income $\hat y$,
$\tilde b'\le \tilde b_0+\theta_C\,\hat y$.

\begin{lemma}[Recipients' consumption]
\label{lem:mpc}
Let a revision raise the disposable income of public employee $i$ by $\Delta y_{i,t}=\rho_y^t\Delta y_i$ relative to its
path without the revision, with persistence $\rho_y$. (i) If $i$'s borrowing limit binds before and after the increment,
consumption rises on impact by $(1+\theta_C\,\partial\hat y/\partial y)\Delta y_{i,0}$, and it rises by
$\Delta y_{i,t}$ in every period if the limit does not respond to the increment
($\partial\hat y/\partial y=0$). (ii) If $i$ is
unconstrained, faces no income risk and $\beta R^D=1$, consumption rises by $m^u\Delta y_i$ in every period from the date
of the news, with $m^u=(1-\beta)/(1-\beta\rho_y)$. (iii) If recipients in case (i) receive the share $s_c$ of the
increment and their limits do not respond to it, the impact propensity of recipients is $\bar m=s_c+(1-s_c)m^u$.
\end{lemma}

Unconstrained recipients adjust their consumption when a revision is announced, and constrained recipients when it is
paid. This is why the dates of news, payment and arrears are kept apart, as in the narrative dating of anticipated tax
changes \citep{mertens2012}. When the 2013 shutdown delayed the pay of US federal employees, their spending fell and
recovered with the back pay \citep{baker2017,gelman2020}. How much an announcement moves spending also depends on how
recipients revise their expectations of future pay. The lemma assumes rational expectations, whereas individual
professional forecasts of macroeconomic variables overreact to news \citep{bordalo2020}.

\paragraph{Banks and credit.} Banks hold loans $L^p$, government securities and reserves, funded by deposits, equity and
foreign funding.

\begin{proposition}[Credit displacement: a conditional accounting decomposition]
\label{prop:captive}
Suppose that within the year equity, foreign funding and excess reserves are fixed, required reserves are a fraction
$r_D$ of deposits, banks absorb the bank-financed share, taxes and a share $s_R$ of retail purchases are paid from
deposits, and a share $s_D$ of the payroll returns to deposits within the year. Then
\begin{equation}
 \Delta L^p=-\kappa_L\,\Omega,\qquad \kappa_L\equiv\theta_B+(1-r_D)\,(\theta_T+s_R\theta_R-s_D) .
 \label{eq:captive}
\end{equation}
\end{proposition}

Each unit of the dose that the Treasury takes from banks displaces a unit of loans. Each unit it draws from deposits,
through taxes or retail bonds, displaces the part of a unit that banks would have lent, net of the pay that returns to
deposits. The decomposition is an accounting identity that holds when banks' lending constraint binds, and it does not
describe an equilibrium. The constraint binds for banks that cannot replace the deposits they lose, the premise of the
bank lending channel, through which a monetary tightening cuts lending most at banks with less liquid balance sheets
\citep{kashyap2000}. Where domestic banks hold the government's debt, its borrowing and its default risk reach
private credit \citep{gennaioli2014,acharya2013,bocola2016,becker2018}.

\paragraph{Firms.} Firms feel a revision through the recipients' demand, through credit and, if public pay pulls up
private wages, through their costs. Firms in the nontraded and export sectors hire labor at the private wage, set prices
subject to quadratic adjustment costs, invest subject to adjustment costs, and finance part of their working capital and
investment with bank loans up to a collateral limit. The limit is on collateral rather than on earnings, the form of
borrowing on which small firms rely more than large ones \citep{lian2021}, and what a firm can borrow against its fixed
assets depends on what they would fetch in liquidation \citep{kermani2023}. A revision therefore moves four margins of the firm over time. Labor
demand rises with output and falls with the private wage. Prices follow a Phillips curve in marginal cost, which contains
the private wage. Investment follows an accelerator in output and, where the lending rate is administered and loans are
rationed, falls with credit at the rate $\eta_K$. Borrowing is the residual of investment and working capital over
internal funds, so a firm whose credit is displaced cuts investment first and labor second. A cut in one bank's lending
lowers output and employment both at the firms that borrow from it and, through local demand and agglomeration
spillovers, at other firms nearby
\citep{huber2018}. The private wage responds
to public pay through a direct elasticity $\psi$ and to output through $\psi_Y$.

\paragraph{The private labor market.} A raise also changes who wants a public job. Applicants for public posts at skill
$z$ pay $c_a$. When applications exceed the posts on offer, each applicant is hired with probability equal to the ratio
of posts to applications, and otherwise every applicant is hired. The direct elasticity of the private wage to public
pay, holding private labor demand fixed, is $\psi(z)\equiv\partial\ln w_p(z)/\partial\ln W$.

\begin{proposition}[The direct recruitment channel]
\label{prop:psi}
Suppose applicants can queue while employed, a raise does not change the skill composition of applicants, and private
labor supply responds to public pay only through recruitment. (i) If applications exceed posts at skill $z$, then
$\psi(z)=0$. (ii) If posts exceed applications, then $\psi(z)\ge0$, with equality when hires come from an informal margin
that is elastic at a reservation wage the raise does not affect.
\end{proposition}

When the queue is long, a raise lengthens it without drawing anyone away from private firms, because the posts on offer
set the number hired. Only where posts outnumber applicants can a raise draw workers out of private firms and so push
up private wages.

\paragraph{The impact decomposition.} Holding the monetary response, the exchange rate and expectations fixed, to first
order in the dose the impact response of output is
\begin{equation}
 \frac{\Delta Y}{Y}=\Omega\Big[\underbrace{(\bar m-m_T\theta_T)(1-\alpha_M)\kappa_D}_{\text{demand}}
 -\underbrace{\eta_K\kappa_L}_{\text{financing}}
 -\underbrace{\varepsilon_W\psi/s_g}_{\text{wage cost}}\Big],
 \label{eq:formula}
\end{equation}
where $m_T$ is taxpayers' propensity to consume, $\alpha_M$ the import content of spending, $\kappa_D$ the nontraded demand
multiplier and $\varepsilon_W$ the output elasticity to the private wage. A raise adds demand to the extent that
recipients' spending on domestic goods exceeds the spending that taxpayers give up. It subtracts investment where the
Treasury takes credit from firms that cannot borrow elsewhere, and it raises costs if public pay pulls up private wages.
The decomposition connects the impact effect to the dynamic system, but for a persistent revision it is not a sufficient
statistic \citep{auclert2024}.

\section{Estimation and validation}
\label{sec:quant}
\label{sec:estimation}

This section estimates the decision problem country by country and asks three questions of the estimates, namely what
in the data pins down each parameter, whether the estimated model reproduces the paths of pay when model and data are
measured the same way, and which of its features produce the three patterns. A plan written before any result reported here
was computed fixed the specification, the criterion, the estimator, the tests of Section~\ref{sec:validation} and the
counterfactual environments (Appendix~\ref{app:estimation}). The comparison of Section~\ref{sec:bridge} and the mechanism exercise of
Section~\ref{sec:mechanism} were added after the estimates and tests, and the alternative courts and financed rules of
Section~\ref{sec:design} were specified after them but before they were computed. Two coding errors found after the
tests had run, one in the legislated reversals of the documented pay paths and one in Bangladesh's series of income per
head, were corrected, and every result they affect was recomputed (Appendix~\ref{app:estimation}).

\subsection{Parameters and their discipline}
\label{sec:parameters}
\label{sec:fiscal_data}

Table~\ref{tab:params} sorts the parameters by the evidence that disciplines them. Panel~A collects the parameters
estimated from the documented decisions, Panel~B those common to all countries and Panel~C those estimated on fiscal data
alone, and the notes list the main calibrated constants, with the others in Appendix~\ref{app:components}. The fiscal relations are estimated first, on fiscal data alone, so
that the pay decisions identify the government's response to the fiscal position rather than the response of the fiscal
position to pay. On the documented fiscal histories of the eight European countries and Latvia, the deficit predicts the
onset of fiscal stress. A deficit larger by one point of GDP raises the yearly probability of stress by about
\RvOnDeltaPp{} percentage points at the sample's mean rate. Higher debt, in contrast, does not raise the risk of onset. Put in
place of the deficit, it enters with a negative coefficient, because the onsets of 2009 came at low debt. The spread
rises gently with debt above 60 percent of GDP, and a point of spread raises the effective interest rate by \RvMu{}
points within a year outside programs (Appendix Table~\ref{tab:fiscal_block}). The court's probabilities are calibrated
to Portugal's rulings, and the misalignment weight $a$ normalizes the units in which every cost is measured.

The remaining parameters are estimated from the decisions, and different features of the decisions pin down different
parameters. The financing cost $\phi_F$ governs the size of raises given the gap, so a government whose raises stay
small although pay has fallen far behind faces a high financing cost. The restart cost $K_F$ governs how long freezes
persist, because only the first raise after a year without one pays it. The costs of a cut, $K^-$ and $\lambda_L$,
govern how often and by how much governments cut once the Treasury's requirement binds. The requirement's shares,
$\gamma_F$ and $\gamma_D$, scale crisis cuts to the interest pressure and the excess deficit. The overhang $\chi$ governs
how long pay stays low in the aftermath as debt falls, although its size is not separately identified from the union's
tolerance (Appendix~\ref{app:experiments}). The reference share $s_w$ determines how far pay may fall behind real private
wages before a revision, and in the commission countries the cost of appointing a commission, $K_I$, determines the
interval between scales. Bangladesh's decisions pin down its reference share only loosely. The values that its decisions
do not reject at the 5 percent level, with every other parameter re-estimated at each, span reference shares from
\RvRoneUOneSwLo{} to \RvRoneUOneSwHi{} (Appendix~\ref{app:experiments}). By Corollary~\ref{cor:costs}(c), the size and
spacing of revisions alone identify only a locus of fixed and convex costs. What separates the two costs is the response
of the size of revisions to the speed of erosion, which inflation varies across years and countries, together with the
variation that the fiscal states, the court and the commission calendar add.

\begin{table}[p]\centering
\begin{threeparttable}
\caption{The parameters and what disciplines them}
\label{tab:params}
\footnotesize\setlength{\tabcolsep}{2pt}
\begin{tabularx}{\textwidth}{>{\RaggedRight\arraybackslash}p{0.205\textwidth}>{\RaggedRight\arraybackslash}X ccc>{\RaggedRight\arraybackslash}p{0.19\textwidth}}
\toprule
Parameter & Meaning and units & Portugal & \makecell{Set or\\s.e.} & Bangladesh & Disciplined by \\
\midrule
\multicolumn{6}{@{}l}{\textit{Panel A. Estimated from the documented decisions}} \\
\addlinespace[2pt]
Financing cost of a raise ($\phi_F$) & Cost per squared percent of the payroll raised at once & 1.28 &  & 0.27 & Size of raises given the gap \\
Restart cost ($K_F$) & Fixed cost of the first raise after a year without one & 1.87 &  & 60.0\textsuperscript{a} & Persistence of decisions of no change \\
Financing cost of a scale ($\phi_C$) & As $\phi_F$, for a new scale &  &  & 0.0002\textsuperscript{a} & Size of new scales \\
Fixed cost of a cut ($K^-$) & Fixed cost of legislating a cut & 0.22 &  & 50.0\textsuperscript{a} & Cuts when the requirement binds \\
Cost of pay lost ($\lambda_L$) & Cost per log point of pay cut & 3.15 &  & 15.0\textsuperscript{a} & Size of cuts \\
Required share of interest pressure ($\gamma_F$) & Share of the extra interest bill required from the payroll & 0.39 & [0.31, 0.52] &  & Size of crisis cuts \\
Required share of the excess deficit ($\gamma_D$) & Share of the deficit above 3 percent of GDP required & 0.03 &  &  & Cuts under a deficit procedure \\
Debt overhang ($\chi$) & Log points of preferred gap per point of debt above its pre-crisis level & 0.34 & [0.17, 0.86]\textsuperscript{b,c} &  & Pay restraint in the aftermath \\
Reference share ($s_w$) & Share of real private wage growth in the reference & 0.00\textsuperscript{a} &  & 0.22 & Erosion tolerated before a revision \\
Cost of a commission ($K_I$) & Fixed cost of appointing a pay commission &  &  & 101.2 & Interval between new scales \\
Scale of taste shocks ($\sigma$) & Scale of the taste shocks to the directions & 0.46 &  & 14.4 & Dispersion of directions \\
Size shocks ($\sigma_J/\sigma$) & Relative scale of size shocks & 1.00\textsuperscript{a} &  & 0.94 & Dispersion of sizes \\
\midrule
\multicolumn{6}{@{}l}{\textit{Panel B. Common to all countries}} \\
\addlinespace[2pt]
Preferred gap ($x^\star$) & Gap the government prefers, in log points & 7.50 & [5, 10]\textsuperscript{b} & 7.50 & Pooled criterion, left at its starting value \\
Treasury's penalty ($\eta_T/a$) & Penalty per percent of payroll short of the requirement & 300 &  & 300 & Left at its starting value \\
\midrule
\multicolumn{6}{@{}l}{\textit{Panel C. Estimated on the pooled fiscal data of the eight European countries and Latvia}} \\
\addlinespace[2pt]
Spread slope ($\zeta_1$) & Points of spread per point of debt above 60 percent of GDP & 0.017 & (0.002) &  & Spreads \\
Pass-through ($\iota$) & Points of effective interest per point of spread & 0.09 & (0.05) &  & Interest payable \\
Onset on the deficit ($c_d$) & Logit coefficient per point of deficit & 0.33 & (0.14) &  & Onsets of stress \\
\bottomrule
\end{tabularx}
\begin{tablenotes}[flushleft]\footnotesize
\item Notes: Panel A gives each country's estimates at the best fit of its documented decisions, and Panel B the two searched ratios common to all countries. Panel C gives relations estimated on fiscal data alone, pooled over the eight European countries of the estimation and Latvia and used for every country, with standard errors in parentheses (Appendix Table~\ref{tab:fiscal_block}). Square brackets give the 95 percent confidence sets of Portugal's parametric bootstrap (19 replicates). Costs are in the normalized loss units of Section~\ref{sec:objective}, and the Treasury's penalty is given as its ratio to $a$. In Bangladesh the costs of a raise are those of an interim allowance, whose fixed cost $K_F$ is the $K_A$ of Section~\ref{sec:component}, and the costs of a cut apply to cuts between commissions. The calibrated constants are the misalignment weight $a=0.02$, which normalizes the units, the government's discount factor $\beta_I=0.9$ and the excess burden of taxation $\lambda_F=0.3$, with 0 and 1 in the sensitivities. The court's probabilities come from Portugal's rulings, 0.9 of striking a cut beyond the admissible burden and 0.5 a year of reversing the outstanding cuts in the aftermath. \textsuperscript{a}At a bound of the search range. \textsuperscript{b}At an end of the range considered. \textsuperscript{c}An inner bound, since a longer search could extend the set downward (Appendix~\ref{app:estimation}). Appendix Table~\ref{tab:v2_estimates} reports every country's estimates.
\end{tablenotes}
\end{threeparttable}
\end{table}

\subsection{Estimation}
\label{sec:criterion}
\label{sec:pooling}

The criterion is the log likelihood of the documented decisions under the nested logit of~\eqref{eq:bellman}, each
evaluated at its documented state. That state consists of the gap traced along the documented pay path (in the
commission countries, from the gap that the model would choose after each documented scale), the debt at the start of
the budget year, the October spread before it, the fiscal state and the outstanding cuts. A raise or a cut of
documented size contributes the log probability of its direction and of a size within one log point of the documented
one, and a decision of undocumented size contributes the probability of its direction. Because the government
anticipates its own taste shocks, their scales enter the continuation values and are estimated with the other
parameters. The criterion is not concave, and every evaluation solves the problem anew. Each country's parameters are
therefore found by a pattern search, which needs no derivatives \citep{hooke1961}, started from many points spread over
the parameter space \citep{mckay1979}. The preferred gap $x^\star$ and the Treasury's penalty are common to all
countries and are searched on the sum of all countries' criteria. The pooled search left both at their starting values,
so it is not established that the histories identify them jointly with the countries' parameters. Similarly, the value
of a visible raise in an election year is not identified, and the cost of conflict with the union is bounded above but
not below (Appendix~\ref{app:estimation}).

\looseness=1 An adequacy test fixed in advance asks whether the model describes each country's decisions at all. A country fails if
the model's most likely direction is wrong in more than a third of its decision years, if it is wrong in most of its
crisis budgets, or if it is right no more often than the rule that pay never changes. A country that fails remains in
every table. \RsPassList{} pass. In Portugal the model's most likely direction is the documented one in every one of
the seventeen decision years, the five crisis budgets included. In the other passing countries it matches between
\RsMatchOtherLo{} (\RsMatchOtherLoName) and \RsMatchOtherHi{} (\RsMatchOtherHiName) of the decision years. Romania
passes at the margin, since the most likely direction is wrong in \RsWrongROUW{} of its \RsYearsROUW{} decision years,
the most that the test allows. Each of them is a year without a general change in which the model expects a raise.
\RsFailList{} fail. Hungary's general pay base stayed frozen after 2012 while sector-specific raises lifted average
pay, so its general decisions describe only part of how pay is set. Lithuania's \RsWrongLTUW{} misses in
\RsYearsLTUW{} decision years are the years of the freeze that followed its cut of 2009, which lasted until 2017 apart
from a rounding increase of 0.13 log points at the euro changeover in 2015. In each of them the model expects pay to move, with a cut in 2010 and a raise
thereafter.

\paragraph{The United States.} \looseness=1 In the United States a statutory formula is meant to set the January raise of General
Schedule pay. The formula combines a base increase that follows the growth of private wages with locality increases that
close measured gaps with private pay. It was not paid in full in any decision year. An alternative plan of the President
or an act of Congress set the locality increase every year, and in \RsUsaBaseBelowW{} years it held the base increase
below the formula as well (Appendix~\ref{app:data}). Each raise is therefore a decision, and because no episode of fiscal
stress changed general pay, the country's problem has a single fiscal state. The model's most likely direction is the
documented one in \RsRightUSAW{} of the \RsYearsUSAW{} decision years. The two it misses, \RsMissYearsUSA, are the first
year of the freeze of 2011--2013 and the year after the freeze ended. With a single fiscal state the model has nothing
that signals the freeze, so it predicts a raise for 2011. It also predicts that the freeze continues into 2014, when pay
rose by one percent, because only the first raise after a freeze pays the restart cost and the decisions put that cost
high. The financing cost lies at the upper end of its range, which keeps the raises small. Simulated along the
documented history, the mean raise is \RsBrUsaRaiseMed{} log points [\RsBrUsaRaiseLo, \RsBrUsaRaiseHi] against a
documented \RsBrUsaRaiseData{}, and every statistic of the data--model comparison lies within the model's interval
(Section~\ref{sec:bridge}). Predicted from the other eight annual countries' estimates alone, the United States is where
the model gains most over constant probabilities, by \RsTransferGainUSA{} points of log score. Re-estimated without a
block of years, however, the model predicts that block slightly worse than the country's own direction frequencies
(Appendix Table~\ref{tab:country}).

\paragraph{Identification and inference.} Three checks show what the decisions reveal about the parameters. First, in
\RvRecoveryReached{} of \RvRecoveryN{} histories simulated from known parameters, the search from random starts attains
at least the likelihood of the true parameters, and it does so in both histories simulated from Portugal's estimate.
Attaining the likelihood does not pin down each parameter, however, because a history of seventeen budgets identifies
the fit more sharply than the separate costs. Second, Portugal's confidence sets therefore take their critical values
from a parametric bootstrap, in which \RvVFourB{} histories are simulated at the estimate and each is re-estimated by the
full search. The requirement's share of interest pressure lies between \RvVFourLoGamF{} and \RvVFourHiGamF{}, and the
overhang's set runs from \RvVFourLoChi{} to the end of the range considered. The restricted searches at the overhang's
lowest values stopped at their maximum number of rounds, so this lower end is an inner bound, and the data do not firmly
exclude a zero overhang. The preferred gap is not pinned down between 5 and 10 log points. Third, steps along random
directions from each country's estimate raise its criterion by at most \RvAttainMaxGain{} points of log likelihood, so
the estimates are optima up to about a tenth of a point.

\paragraph{Year-by-year estimates and the long-run calibration.}
\label{sec:fit_commission}
The commission countries need a second discipline, because the year-by-year criterion conditions on the documented
timing of the commissions and so does not discipline the cycle that the government's own policy would produce.
Bangladesh's year-by-year estimates fit the timing of its scales given the documented states. Once the taste shocks are
switched off, however, the policy they imply, the government's systematic rule, revises pay
\RvCFbgdFitDocscalesperOneZeroZeroy{} times a century, by \RvCFbgdFitDocmeanscale{} log points. Over the documented
commissions since 1977, which reach back before the yearly record of decisions, scales came \RvBgdDocInterval{} years apart on average, and a cycle, a new scale together with
the allowances paid since the previous one, raised pay by \RvBgdDocCycle{} log points. The long-run calibration sets the reference share and the cost of appointment so that the policy without taste shocks matches this interval and cycle as closely as the range of the reference share allows, with scales \RvBgdLRDetInterval{} years apart and cycles of \RvBgdLRDetCycle{} log points ($s_w=\RvBgdLRDetSw$, the upper end of its range, and $K_I=\RvBgdLRDetKI$). The estimates and the calibration rest on
different evidence. The calibration targets the long-run moments, and the year-by-year decisions reject it, with a
likelihood-ratio statistic of \RvRoneCalibLR{} against a 5 percent critical value of 5.99 for its two restrictions.
Section~\ref{sec:cf_bgd} reports the indexation experiment at both.

\subsection{Data and model on the same objects}
\label{sec:bridge}
\label{sec:fit}

\looseness=-1 The estimation fits decisions one at a time. A sharper test is whether the decisions the model generates add up to the
documented paths of pay, which Figure~\ref{fig:bridge} and Table~\ref{tab:moments} examine. In the annual countries that
pass the adequacy test, the documented value lies within the model's 90 percent interval for \RsCoverPassIn{} of their
\RsCoverPassTot{} statistics, and for 12 of the 13 that the criterion does not target (Table~\ref{tab:moments},
Panel~D). In Bangladesh the simulated scales are larger and rarer than the documented ones but
bracket them. Each country's model is simulated 500 times along the documented history. In every year the government's
decision is drawn from the estimated model at the gap that its own earlier simulated decisions produced, while the
macroeconomic and fiscal history, the implementation years, the legislated restorations, indexation and the Court's
calendar follow the documents. Every statistic is then computed on the documented history and on each simulated history
with the same definitions, dates, normalization and event selection. The spread of the model's statistic across
histories therefore measures how much a statistic varies across histories of the documented length.

\suppressfloats[t]
{\renewcommand{\figwidth}{\textwidth}\figifat{!ht}{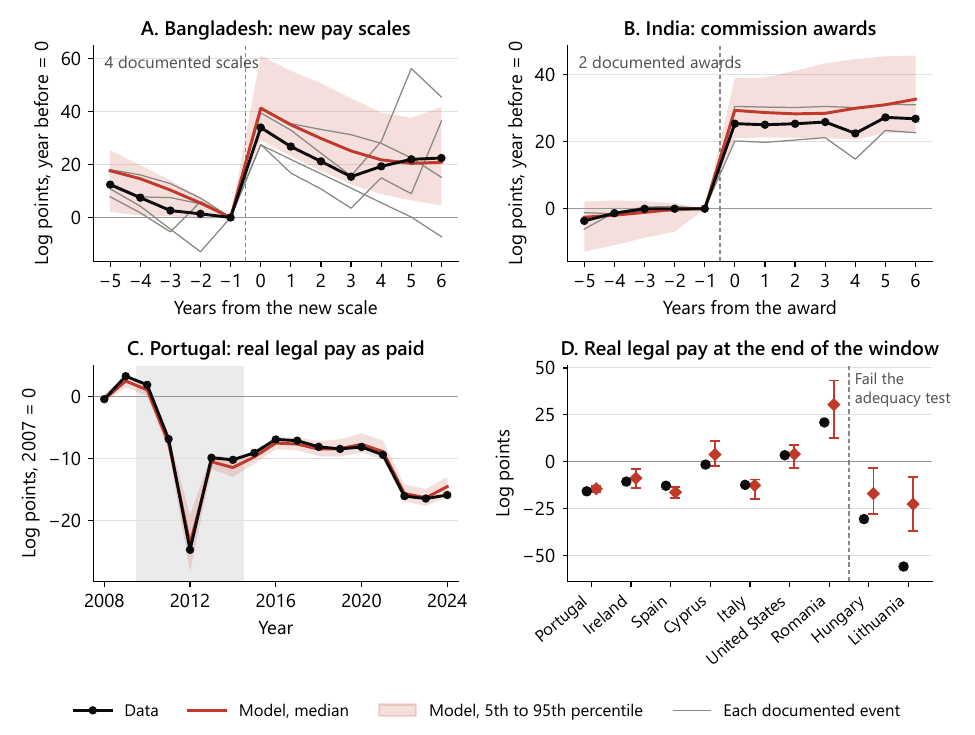}{Data and model on the same objects}{fig:bridge}{Each panel compares a documented statistic
(black) with the same statistic on 500 histories simulated from the estimated model along the documented history (red),
shown as the median and the 5th to 95th percentiles across histories, shaded in Panels~A--C and as diamonds with
whiskers in Panel~D. Panels~A and~B plot real legal pay, in log points relative to the year before a new scale or award,
by years from it, for Bangladesh's four scales of 1998--2016 and India's two awards of 2008 and 2016, with each documented event in gray. A simulated history contributes the mean over its own scales. Panel~C plots Portugal's real
legal pay as paid, in annual averages relative to 2007, and the shaded years are years of stress or a program. Panel~D
plots each annual country's real legal pay at the end of its window relative to the year before the window (Portugal
from 2007, the United States from 1994, the others from 2006), with the countries that fail the adequacy test to the
right of the dashed line. Real pay is legal pay less consumer-price inflation, and the parameters are the year-by-year
estimates.}}

Three comparisons stand out. First, Portugal's simulated path follows the documented one through the crisis and its
aftermath. Real pay as paid lies inside the model's 90 percent interval in \RsBrPrtAllInside{} year of 2008--2024,
including the trough of 2012 (\RsBrPrtTroughData{} log points against a median of \RsBrPrtTroughMed{}) and 2024
(\RsBrPrtEndData{} against \RsBrPrtEndMed{}). The decisions behind this path are the ones the criterion fits, so the
panel shows that the fitted decisions are consistent with the path rather than validating them out of
sample. The path itself is not part of the criterion, however, and neither is the persistence of decisions of no
change, of which \RsBrPrtPersistData{} are followed by another, against a simulated median of \RsBrPrtPersistMed{}.
Second, India's model reproduces the shape of pay under indexation. Real pay is level between awards in both, and in the
year of an award the documented real pay rises by \RsBrIndJumpData{} log points, against a simulated \RsBrIndJumpMed{}
[\RsBrIndJumpLo, \RsBrIndJumpHi] (Figure~\ref{fig:bridge}, Panel~B). Finally, across the nine annual countries, the
documented level of real pay at the end of the window lies within the model's interval in \RsCoverRpEndPassPhrase{}
countries that pass the adequacy test and in \RsCoverRpEndFailPhrase{} that fail it (Panel~D). The exception among the
passing countries is \RsCoverRpEndPassMiss, whose documented pay ends \RsCoverRpEndPassMissGap{} log points
\RsCoverRpEndPassMissSide.

\par\condbreak{110pt}
\begingroup\setstretch{1}\footnotesize\setlength{\tabcolsep}{2pt}
\setlength{\LTleft}{0pt}\setlength{\LTright}{0pt}
\setlength{\LTpre}{\intextsep}\setlength{\LTpost}{0pt}
\begin{longtable}{@{\extracolsep{\fill}}l*{4}{c}cl}
\multicolumn{7}{@{}c@{}}{\small\tablename~\thetable. Targeted and untargeted moments\label{tab:moments}}\\[6pt]
\toprule
& & \multicolumn{3}{c}{Simulated histories} & & \\
\cmidrule(lr){3-5}
Moment & Data & Median & 5th pct. & 95th pct. & Targeted & Parameters \\
\midrule
\endfirsthead
\multicolumn{7}{@{}c@{}}{\small\tablename~\thetable. Targeted and untargeted moments (continued)}\\[6pt]
\toprule
& & \multicolumn{3}{c}{Simulated histories} & & \\
\cmidrule(lr){3-5}
Moment & Data & Median & 5th pct. & 95th pct. & Targeted & Parameters \\
\midrule
\endhead
\midrule
\multicolumn{7}{r}{\textit{Continued on the next page}}\\
\endfoot
\bottomrule
\endlastfoot
\multicolumn{7}{@{}l}{\textit{Panel A. Portugal, 17 decisions, 2008--2024}} \\*
\addlinespace[2pt]
Share of decisions with a raise & 0.35 & 0.41 & 0.29 & 0.53 & Yes & $\phi_F$, $K_F$ \\
Share with no change & 0.41 & 0.35 & 0.24 & 0.47 & Yes & $K_F$, $\chi$ \\
Share with a cut & 0.24 & 0.24 & 0.18 & 0.24 & Yes & $\gamma_F$, $K^-$ \\
Mean raise, log points & 2.15 & 1.98 & 1.47 & 2.78 & Yes & $\phi_F$ \\*
 & (0.55) & & & & & \\
Mean cut, log points & 7.64 & 8.33 & 5.91 & 10.56 & Yes & $\gamma_F$, $\lambda_L$ \\*
 & (2.48) & & & & & \\
No change after no change & 0.57 & 0.57 & 0.33 & 0.75 & No & $K_F$, $\chi$ \\
Real pay as paid, 2012 (2007 = 0) & $-24.7$ & $-23.7$ & $-28.2$ & $-19.1$ & No & $\gamma_F$, $\rho_S$ \\
Real pay as paid, 2024 (2007 = 0) & $-15.9$ & $-14.5$ & $-16.2$ & $-13.0$ & No & $\chi$, $\phi_F$ \\
\midrule
\multicolumn{7}{@{}l}{\textit{Panel B. Bangladesh, 1993--2024, year-by-year estimates}} \\*
\addlinespace[2pt]
New scales & 4 & 3 & 2 & 5 & No & $K_I$ \\
Mean new scale, nominal log points & 41.0 & 47.8 & 36.0 & 67.0 & No & $K_I$, $\phi_C$ \\
Years between scales & 6.0 & 8.0 & 6.0 & 12.0 & No & $K_I$, $s_w$ \\
Interim allowances & 4 & 4 & 1 & 8 & No & $K_F$, $\phi_F$ \\
Real pay, 2024 (1992 = 0) & 15.2 & $-10.4$ & $-38.9$ & 22.5 & No & $s_w$ \\
\midrule
\multicolumn{7}{@{}l}{\textit{Panel C. India, 1998--2024}} \\*
\addlinespace[2pt]
Commission awards & 2 & 2 & 1 & 3 & No & $K_I$ \\
Mean award, nominal log points & 25.6 & 29.8 & 21.7 & 38.3 & No & $K_I$, $\phi_C$ \\
Real pay, 2024 (1997 = 0) & 59.1 & 71.0 & 51.9 & 82.0 & No & $s_w$ \\
\midrule
\multicolumn{7}{@{}l}{\textit{Panel D. All nine annual countries: documented value within the model's 5th--95th percentiles}} \\*
\addlinespace[2pt]
Share with no change & \multicolumn{4}{c}{8 of 9 countries, 7 of 7 that pass} & Yes & \\
Share with a cut & \multicolumn{4}{c}{9 of 9 countries, 7 of 7 that pass} & Yes & \\
Mean raise & \multicolumn{4}{c}{6 of 9 countries, 5 of 7 that pass} & Yes & \\
No change after no change & \multicolumn{4}{c}{7 of 8 countries, 6 of 6 that pass} & No & \\
Real pay at the end of the window & \multicolumn{4}{c}{6 of 9 countries, 6 of 7 that pass} & No & \\
\end{longtable}
\par\endgroup\nopagebreak
{\footnotesize\setstretch{1}\setlength{\parindent}{0pt}\vspace{3pt}Notes: Each documented statistic is compared with the same statistic computed, with the same definitions, dates and normalization, on 500 histories simulated from the estimated model along the documented macroeconomic and fiscal history. In each history the government's decisions are drawn from the model at the gap its own earlier decisions produced, while implementation years, legislated restorations, automatic indexation and the Court's calendar follow the documents. ``Targeted'' marks the decisions that the estimation criterion fits, each at its documented state. The other statistics depend on the simulated path of pay and are not part of the criterion. Standard errors of the documented means over decisions are in parentheses. Real pay is legal pay as paid less consumer-price inflation, in log points. The sizes of scales and awards are nominal changes of the reference grade, net of the allowances they absorb. In Bangladesh and India the window opens after the scale or award of the base year. The long-run calibration, which targets Bangladesh's interval between scales and the size of a cycle, is not used here. Panel D counts the annual countries, and the seven of them that pass the adequacy test, whose documented statistic lies between the model's 5th and 95th percentiles. The mean cut is not counted, because Italy and the United States made none, and a country is counted only for the statistics it has (the persistence of no change needs a decision of no change followed by another decision). The last column names the parameters that govern the statistic most directly (Table~\ref{tab:params}, and the court's strike probability $\rho_S$ of Section~\ref{sec:component}).\par}
\addvspace{\intextsep}

Bangladesh's simulated scales are larger and rarer than the documented ones, but the documented values lie within the
model's intervals. At the year-by-year estimates, real pay rises by \RsBrBgdJumpMed{} log points [\RsBrBgdJumpLo,
\RsBrBgdJumpHi] in the year of a new scale, against a documented \RsBrBgdJumpData{} (Figure~\ref{fig:bridge}, Panel~A).
The simulated scales come \RsBrBgdYearsMed{} years apart [\RsBrBgdYearsLo, \RsBrBgdYearsHi], against a documented
\RsBrBgdYearsData{} over 1993--2024 (Table~\ref{tab:moments}), at the lower end of the simulated range. The year-by-year
criterion leaves this difference open, because it conditions on the documented timing of the commissions
(Section~\ref{sec:fit_commission}). The long-run calibration targets the interval between scales and the size of a cycle
over the whole documented history. These measures are closely related to the ones compared here, so the calibration
matches them largely by construction and cannot validate them. For that reason the experiment of
Section~\ref{sec:cf_bgd} is reported at both parameter points.

\subsection{Which features generate the patterns}
\label{sec:mechanism}

Each pattern can be traced to a distinct feature of the model. Figure~\ref{fig:mechanism} removes one feature at a time,
holds every other parameter at its estimate, and repeats the simulation of Section~\ref{sec:bridge}. The exercise shows
what each feature does inside the estimated model. Whether a model without the feature could fit the data is a different
question, which would require re-estimating that model on the same decisions. For comparison, the figure reports one
such competing model, the myopic government of Section~\ref{sec:validation}.

\figif{fig_mechanism.pdf}{What generates the patterns}{fig:mechanism}{Panels~A and~B plot Portugal's real legal pay as
paid, relative to 2007, in the data (black) and as the median over 500 simulated histories (Section~\ref{sec:bridge}). The histories come from the estimated model (red), from the model with one feature removed at the estimated parameters (navy, without the Treasury's requirement, without the Court and its ordered restorations, or without the debt overhang), and from the myopic government re-estimated on the same decisions (ochre). The shaded years are years of stress or a program. Panels~C and~D plot, for Bangladesh
over 1993--2024, the number of new scales (solid) and interim allowances (outlined) and the mean size of a new scale, in
the data and as medians over 500 simulated histories of the estimated model and of the model without fixed costs of
acting or without financing costs. Real pay is legal pay less consumer-price inflation.}

\looseness=1 The Treasury's requirement produces the cuts. Without it, the simulated government makes \RsMechNoReqCutsW{} cuts over
Portugal's crisis, and real pay in 2012 is \RsMechNoReqPayTwelve{} log points from its 2007 level instead of
\RsMechEstPayTwelve{} (Panel~A). Inflation alone does not erode pay fast enough to deliver the saving that the Treasury
requires once spreads rise, which is why, inside the estimated model, the crisis cuts of Pattern~2 come from the
requirement.

The Court determines how the cuts translate into pay. Without the Court and the restorations it ordered, the government
makes \RsMechNoCourtCutsW{} cuts instead of four. It no longer proposes cuts for the saving they yield before a court
strikes them (Proposition~\ref{prop:court}), and every cut it makes is permanent. Real pay therefore shows neither the
temporary trough of 2012 nor the restorations of 2015--2016, and in 2018 it is \RsMechNoCourtPayEighteen{} log points
from its 2007 level instead of \RsMechEstPayEighteen{}. What disappears with the Court is the difference between the cut
legislated, the cut paid and the cut that lasts, which is Pattern~3.

The overhang keeps pay low once the crisis is over. Without it, the high debt of the aftermath no longer raises the
political cost of the payroll, and pay recovers years earlier. In 2018 it is \RsMechNoOverPayEighteen{} log points from
its 2007 level instead of \RsMechEstPayEighteen{} (Panel~B), and the share of decisions without a change falls from
\RsMechEstNone{} to \RsMechNoOverNone{}. The myopic government, re-estimated on Portugal's decisions, errs in the same
direction but by less. It raises pay when stress begins in 2010 and restores pay too early in the aftermath, and these
years are where the forward-looking model gains its advantage in fit.

Bangladesh's cycle has the shape of the reset band of Proposition~\ref{prop:band}, but the convex cost that shapes it is
the cost of financing interim allowances. A new scale is cheap to finance, with $\phi_C$ at the lower end of its range,
whereas allowances are not. Without fixed costs of acting, the government revises in small and frequent steps, with
\RsMechBgdNoFixScales{} new scales of \RsMechBgdNoFixSize{} log points on average over 1993--2024 and
\RsMechBgdNoFixAllow{} interim allowances (Panels~C and~D). Without financing costs, it stops appointing commissions
altogether and raises pay through \RsMechBgdNoFinAllow{} allowances. The fixed cost thus makes revisions rare, and the
cost of financing allowances makes a commission's scale, which absorbs the allowances paid since the last one, the
cheaper way to restore pay once the gap is large. With only one of the two costs, the simulated cycle loses its shape,
which is the sense in which the cycle needs both.

\subsection{Validation out of sample}
\label{sec:validation}
\label{sec:fit_common}
\label{sec:fit_timing}
\label{sec:fit_myopic}

\looseness=1 The model's structure transfers across countries, but it does not forecast within them. The first test predicts each
annual country's decisions from the other eight countries' estimates alone and scores the prediction by its log score,
the sum over decisions of the log probability given to the documented one. The model beats constant probabilities in
\RvTransferWinsConst{} of the \RvTransferN{} annual countries and beats every benchmark estimated out of sample in
\RvTransferWinsOut{}. Its total log score, \RvTransferTotModel{}, exceeds that of constant probabilities
(\RvTransferTotConstant{}) and that of the legal default (\RvTransferTotLegal{}), the change that each year's
institutions produce without a new decision. It falls short of a logit on the fiscal state shrunk toward a default prior
\citep{gelman2008}, which scores \RvTransferTotFiscal{}, and Hungary and Lithuania account for the shortfall.

\looseness=1 The second test holds out years within each country. Re-estimated without a block of about four decision years and
scored on it, the model predicts the \RvHeldoutYears{} held-out years worse than the
direction frequencies of each country's other years, \RvHeldoutTotModel{} against \RvHeldoutTotConst{}, and better in
only \RvHeldoutWins{} of \RvHeldoutN{} countries. The losses come from a few years that the re-estimated model all but
rules out. A crisis block contains a country's cuts, and without them the re-estimated requirement does not produce
cuts. Re-estimated without its budgets of 2012--2015, for example, Portugal's model gives the cuts of 2012--2014 almost
no probability.

\looseness=1 A competing model, re-estimated on the same decisions, locates the evidence for forward-looking behavior. A government
that does not look ahead, the limit of the short horizons with which \citet{acharya2013} model sovereign borrowers, has
the same payoff, actions and court and a discount factor of zero. It fits Portugal's decisions \RvMyopicDiffPRT{} points of log likelihood worse than the forward-looking government. It raises pay when stress begins in 2010, misses the size of the cut of 2013 and restores pay too early in the aftermath. It also fits
Spain's decisions worse, by \RvMyopicDiffESP{} points. In every other country it fits as well or better, so the evidence
for anticipation comes from Portugal's history and, more weakly, from Spain's. On the held-out blocks it scores
\RvHeldoutTotMyopic{} in total, better than the forward-looking model's \RvHeldoutTotModel{} but still below the
\RvHeldoutTotConst{} of constant probabilities.

These results determine how the model can be used. It accounts for the documented histories that it fits and supports
counterfactuals conditional on its estimates, but it does not forecast when governments act.

\section{Aggregate implications and institutional design}
\label{sec:design}

This section uses the estimated model to evaluate two institutions that govern the pay decision, the judicial review of
cuts in Portugal and the indexation of pay between commissions in Bangladesh. It first describes how a revision reaches
the economy and how the social cost of a path of pay is measured. Every number in Sections~\ref{sec:cf_prt}
and~\ref{sec:cf_bgd} is a simulated outcome of the estimated model, in which only the institution changes and the
estimated government reoptimizes under the new rules. Because a decision rule depends on the institutions under which it
is chosen, a rule estimated under one institution cannot be carried unchanged to another
\citep{lucas1976,hansen1980}. In dynamic discrete choice models some counterfactuals are identified even when the
model's primitives are not, and others are not \citep{kalouptsidi2021}. A conclusion is therefore reported only if it
holds at every optimum that fits the documented decisions about as well as the best one, and under the sensitivities
fixed in advance. The rule addresses the uncertainty outside the model, about its parameters, as distinct from the taste
shocks that the government faces inside it \citep{hansen2014}. Apart from the calibrated court, the courts and the financed rules
compared below were specified after the estimates and tests, but before they were computed.

\subsection{Transmission and the social objective}
\label{sec:transmission}
\label{sec:objective}

The experiments rank institutions by the social cost of the paths of pay they produce, and part of that cost is what a
path of pay does to output. A revision reaches output through the recipients' spending, through the Treasury's financing
and, if public pay pulls up private wages, through firms' costs (equation~\eqref{eq:formula}). The block that carries
these channels is calibrated rather than estimated, because the panel of legislated acts does not identify the
aggregate response (Section~\ref{sec:limits_patterns}). Micro evidence, however, disciplines two of its inputs, the
empirical foundation that \citet{hansen1996} ask of calibrated parameters. The first
is the share of recipients who are liquidity constrained, who spend a raise when it is paid (Lemma~\ref{lem:mpc}). In
Bangladesh's rural household panel \citep{ifpribihs2020}, \RbihsPreLow{} percent of public-job households held liquid
savings below a month of consumption before the 2015 scale was announced. Similarly, in the 2016 expenditure survey
\citep{bbshies2016}, \RhtmPubTwentySixteen{} percent of households with a public salaried member held financial and
other non-land assets worth less than a month of gross salary. The calibration's constrained share, 0.45, lies below
both.

The second input is the direct effect of public pay on private wages, $\psi$. By Proposition~\ref{prop:psi}(i), $\psi$
is zero where applications for public posts exceed the posts on offer, because a raise then lengthens the queue without
drawing workers away from private firms. Applications for federal jobs in the United States rise with federal pay
relative to private pay \citep{krueger1988}, and the queues studied here are long. Bangladesh's general civil-service examination drew \RqBgdPre{} applicants per advertised post in the
\RqBgdNPreW{} examinations before the 2015 scale and \RqBgdPost{} in the \RqBgdNPostW{} after it, while the number of
posts on offer barely changed (\RqBgdPostsPre{} and \RqBgdPostsPost{} on average) \citep{bpsc2025}. Likewise, a single
recruitment of Indian railway laborers in 2018 drew \RqLowRailApps{} million applications for \RqLowRailPosts{} posts
\citep{pib2018}. The baseline therefore sets $\psi$ to zero. The raw comparison around the 2015 scale does not isolate
the response to an award. On a protocol that treats both countries' examinations alike, pools each country's awards and
allows for a linear trend in the examination date, applications per post in the first three examinations after an award
changed by \RqHBgdQ{} log points \sd{\RqHBgdQSE} in Bangladesh and \RqHIndQ{} \sd{\RqHIndQSE} in India
\citep{bpsc2025,upsc2023}.

The model is closed through the budget but not through output. Pay moves the deficit and debt, and these move the
spread, the onset of stress and the aftermath under which the next decision is taken. The transmission block maps the
resulting path of pay into output, but output feeds back neither into the pay decision nor into the fiscal state
(Assumption~\ref{ass:separation}), and the experiments below are conditional on that separation. In them the output
effects of the alternative paths of pay are small. Across the \RsCtNEnv{} environments of the Portuguese experiment, the
stabilization term of the loss ranges only from \RsCtStabMin{} to \RsCtStabMax{} a year, against \RsCtFinMin{} to
\RsCtFinMax{} for the financing term and \RsCtTreasMin{} to \RsCtTreasMax{} for the Treasury's penalty (the notes to
Appendix Table~\ref{tab:court_terms}). The next step is a completed general equilibrium model, in which revenue, the GDP
denominator of debt and the onset of stress respond to the output that a revision creates.

The social cost of a path of pay is the flow loss
\begin{equation}
 \ell_t=\tfrac a2x_t^2+\tfrac{w_Y}2\big(\tilde y_t+\Delta y_t\big)^2+\eta_T\big[Q_t-S(x_t)\big]^+
 +s_c\,z_0\big[z_S\,\ind^{\mathrm{cr}}_t\,x_t^++\tfrac{z_2}{200}x_t^2\big]+\phi_t\,R_t^2+\lambda_K K_t .
 \label{eq:socialloss}
\end{equation}
Here $x_t$ is the gap after the year's decision, including the part of a struck cut in force, $\tilde y_t$ the cycle's
output gap, $\Delta y_t$ the output path of the transmission block in response to the path of pay, $Q_t$ the financing
requirement and $\ind^{\mathrm{cr}}_t$ the indicator of a crisis year, of market stress or a program. Moreover, $R_t$ is
the payroll increase made by discrete acts, in percent of the payroll at the reference, $\phi_t$ the estimated financing
coefficient of the instrument that makes it, and $K_t$ the fixed cost incurred. The six terms are the misalignment of
pay from the reference, the stabilization loss, the Treasury's penalty on a saving short of the requirement, the
insurance cost borne by the share $s_c$ of employees who consume their pay, the convex cost of financing increases at
once, and a share $\lambda_K$ of the estimated fixed costs. The weights are calibrated. The stabilization weight is
$w_Y=1$, and the insurance term has the scale $z_0=0.45$, a premium $z_S=0.5$ on pay withheld in a crisis and a
curvature $z_2=2$ around the reference, a relative risk aversion of two.

The financing cost applies to increases only, including the restorations that a legislature or a court orders, whereas a
cut is a saving and carries none. A restoration that the government did not decide carries no cost in its own problem,
so the social loss counts a financing cost that the government does not weigh when it proposes a cut. Portugal's
decisions support this. When the government's problem charges court-ordered restorations at a share of the cost of
financing its own raises, the share that fits best is zero, the 95 percent set ends at \RwUpperRound{}, and a government
that charged them in full is rejected (Appendix~\ref{app:estimation}). The government's political payoffs, namely the
preferred gap $x^\star$, the overhang and the value of a raise in an election year, are excluded. The objective values
employees' protection through the pay they receive, and it attaches no separate value to the legal guarantees that a
court enforces. The evidence does not say how much of a fixed cost of acting is a resource cost, so results are reported
both with fixed costs counted ($\lambda_K=1$) and with them excluded ($\lambda_K=0$). Losses are discounted at 0.96 and
reported per year in normalized units. One unit is the misalignment loss of holding pay ten log points from its
reference for a year, which the payoff counts as one percent of a year's payroll at the reference. These units are
neither percentages of GDP nor consumption-equivalent welfare. Appendix~\ref{app:model} derives the objective and states
its weights.

\subsection{Judicial review in Portugal}
\label{sec:cf_prt}

\looseness=1 A court that reverses cuts shortens how long cuts last, but it raises the crisis's social cost, and the timing of its
review determines how much of that cost it imposes. Proposition~\ref{prop:court} says how the government responds to a
court, and the experiment measures what the responses cost. Table~\ref{tab:courts} compares hypothetical courts in
Portugal over the budgets of 2010--2024. Each history starts from the documented state of the 2010 budget and keeps the
documented macroeconomic history and external fiscal shocks, while the deficit, debt, the spread's schedule and the
fiscal state all respond to the simulated path of pay. Each court strikes, with a probability the government knows, a
cut that takes the outstanding cuts above the admissible burden in stress or a program, and the government re-solves its
problem under each court. None of these courts is a prediction of how Portugal's Court would have ruled.

\begin{table}[t]\centering
\begin{threeparttable}
\caption{Judicial review in Portugal, 2010--2024}
\label{tab:courts}
\footnotesize\setlength{\tabcolsep}{2pt}
\begin{tabular*}{\textwidth}{@{\extracolsep{\fill}}l*{7}{c}}
\toprule
& \multicolumn{3}{c}{Treasury} & \multicolumn{2}{c}{Employees} & & \\
\cmidrule(lr){2-4}\cmidrule(lr){5-6}
& \makecell{Temporary\\savings} & \makecell{Durable\\cuts} & \makecell{Restoration\\payments} & \makecell{Cumulative\\cut in force} & \makecell{Pay in\\crisis years} & \makecell{Debt,\\2024} & \makecell{Social\\loss} \\
Court & (1) & (2) & (3) & (4) & (5) & (6) & (7) \\
\midrule
No court & 0.0 & 9.0 & 0.0 & 121 & 85.9 & 92.1 & 25.83 \\
Calibrated court ($\rho_S=0.9$) & 12.9 & 8.5 & 20.0 & 62 & 83.8 & 93.1 & 37.04 \\
Review before effect ($\rho_S=0.9$) & 0.0 & 7.4 & 6.2 & 41 & 86.8 & 94.5 & 29.06 \\
Certain review ($\rho_S=1$) & 19.5 & 5.0 & 23.8 & 54 & 83.8 & 93.2 & 38.11 \\
Certain review before effect ($\rho_S=1$) & 0.0 & 5.0 & 4.3 & 30 & 87.7 & 95.4 & 29.17 \\
\bottomrule
\end{tabular*}
\begin{tablenotes}[flushleft]\footnotesize
\item Notes: Entries are simulated outcomes of the estimated model for Portugal under hypothetical courts, over the budgets of 2010--2024, from the documented state of the 2010 budget with the documented macroeconomic history and external fiscal shocks. A court strikes a cut that takes the outstanding cuts above the admissible burden in stress or a program with the stated probability, which the government knows, and the government re-solves its problem under each court. A cut struck after it is in force stays in force for the share of the year of the calibrated court, whereas review before the cut takes effect leaves no part in force. Entries are means over 400 histories of the strike draws along the policy without taste shocks, or one history where striking is certain or there is no court. Column 1 is the saving from struck cuts while in force, and column 3 the pay increases that undo struck cuts and restore standing ones, which must be financed at once, both in percent of the payroll at the reference summed over 2010--2024. Column 2 sums the cuts that stand, in log points. Column 4 sums over 2010--2024 the cuts in force in each year, standing or struck, in log points of pay. Column 5 is mean pay in crisis years in percent of the reference, column 6 debt at the end of 2024 in percent of GDP, and column 7 the social loss of Section~\ref{sec:objective} with fixed costs counted, the discounted average over 2010--2024 in normalized loss units a year.
\end{tablenotes}
\end{threeparttable}
\end{table}

\paragraph{The Court shortens how long cuts last.} \looseness=1 Without a court, every cut the government makes lasts, and summed
over 2010--2024 the cuts in force amount to \RsCtNoneCumIn{} log points of pay. The calibrated court strikes a cut beyond
the admissible burden with probability 0.9 after the cut has been in force. Facing it, the government also proposes cuts
that the court strikes, because each has a chance of standing and saves pay while it is in force
(Proposition~\ref{prop:court}). The cuts that stand, however, are reversed after the crisis, so the cuts in force sum to
\RsCtEstCumIn{}. This is not protection during the crisis itself. Mean pay in crisis years is \RsCtEstPayCrisis{}
percent of the reference under the court, against \RsCtNonePayCrisis{} without one, and the part of the cuts in force
above the admissible burden in crisis years sums to \RsCtEstCumEx{} log points under the court against \RsCtNoneCumEx{}
without it. The measure fixed in advance for the experiment, which counts a standing cut only in the year it was
imposed, points the same way (Appendix~\ref{app:experiments}).

\paragraph{Reversal is costly.} \looseness=1 Under the calibrated court the social loss is \RsCtEstLossOne{} a year, against
\RsCtNoneLossOne{} without a court, and the difference comes from the restorations. A struck cut is a temporary saving
that is undone when the court rules. Over 2010--2024 the court's struck cuts save \RsCtEstTemp{} percent of a year's
payroll while in force, whereas the pay increases that undo them and restore the cuts that stood add up to
\RsCtEstRest{} percent, which must be financed at once. The financing term of the loss is accordingly
\RsCtEstTermFinancing{} under the court against \RsCtNoneTermFinancing{} without it, while the Treasury's penalty is
lower, \RsCtEstTermTreasury{} against \RsCtNoneTermTreasury{} (Appendix Table~\ref{tab:court_terms}). The comparison
fixed in advance points the same way. Against a Court that reversed nothing, Portugal's Court as documented raised the
loss by \RvCFsensBaseObsMinusNoRev{} a year, and the ranking holds under every sensitivity fixed in advance
(Appendix~\ref{app:experiments}).

\paragraph{Review before a cut takes effect protects more at less cost.} If the court rules before a cut takes effect, a
struck cut saves nothing. The temporary savings therefore disappear, and the restorations shrink to the reversal of the
cuts that stood (\RsCtBeforeRest{} percent of a year's payroll against \RsCtEstRest{}). The cuts in force sum to
\RsCtBeforeCumIn{}, mean pay in crisis years rises to \RsCtBeforePayCrisis{} percent of the reference, and the loss falls
to \RsCtBeforeLossOne{}. At the calibrated strike probability the government still proposes cuts beyond the admissible
burden, because each has a one-in-ten chance of standing. Only review that is also certain removes that incentive, as
Proposition~\ref{prop:court}(ii) predicts. The government then proposes no cut beyond the burden, the cuts in force sum
to \RsCtCertainBeforeCumIn{}, and pay in crisis years rises to \RsCtCertainBeforePayCrisis{} percent of the reference.
The price is debt of \RsCtCertainBeforeDebt{} percent of GDP at the end of 2024 instead of \RsCtEstDebt{}. That review
must also be certain rests on the estimate that the government does not weigh the cost of financing the restorations a
court orders. A government that weighed it at a twentieth of the cost of its own raises, a weight inside the 95 percent
set, would expect to pay for reversing a cut that stands. Review before a cut takes effect at the calibrated probability
would then already remove the incentive (Appendix~\ref{app:experiments}). These results concern Portugal's crisis. In
the model's own world, where the fiscal and macroeconomic states are drawn from the model instead of the documents and
crises occupy about six percent of years, the courts hardly differ, and neither faster nor
more certain review reduces the number of struck cuts (Appendix~\ref{app:experiments}).

\paragraph{Every court costs more than none.} Across the twenty courts of the experiment, which combine four strike
probabilities with five delays of review, every loss exceeds the loss without a court by at least \RsCtMinAboveNone{} a
year (the notes to Appendix Table~\ref{tab:court_terms}). The channel depends on when the court rules. A court that rules
after a cut has taken effect raises the loss through the financing of the restorations. A court that rules before raises
it through the Treasury's penalty, because the admissible burden limits the adjustment the Treasury requires. The
penalty is \RsCtCertainBeforeTermTreasury{} under certain review before a cut takes effect, against
\RsCtNoneTermTreasury{} with no court. The comparison is conditional on Portugal's estimate, whose overhang the
decisions place only within a wide set (Section~\ref{sec:criterion}), on the separation of the decision from output, and
on an objective that values a court only through what it does to pay, the budget and output
(Section~\ref{sec:objective}). The comparison also holds when the government weighs the cost of the restorations at
\RwGridIn{}, the largest weight of the grid inside the 95 percent set. Every court then still costs at least
\RwEdgeMinAboveNone{} a year more than none, although the calibrated court's extra cost falls from \RwZeroCalMinusNone{}
to \RwEdgeCalMinusNone{} (Appendix~\ref{app:experiments}).

\subsection{Financed indexation in Bangladesh}
\label{sec:cf_bgd}

Indexation between commissions protects employees against deep erosion and requires fewer revisions, and it pays as
long as its automatic increases are cheap to finance. Mean pay, however, is lower under every rule. If price indexation
cost nothing to finance, Bangladesh's stationary loss would fall from \RvCFbgdFitDocstatlossOne{} to
\RvCFbgdFitIdxstatlossOne{} a year at the year-by-year estimates, and from \RvCFbgdLRDocstatlossOne{} to
\RvCFbgdLRIdxstatlossOne{} at the long-run calibration (Table~\ref{tab:index}). New scales become rarer, because pay no
longer falls behind prices between them, and pay rarely falls far behind its reference. At the estimates, the 90th
percentile of the gap falls from \RsIdxFitDocPninety{} to \RsIdxFitFullFreePninety{} log points.

\begin{table}[t]\centering
\begin{threeparttable}
\caption{Financed indexation in Bangladesh}
\label{tab:index}
\footnotesize\setlength{\tabcolsep}{2pt}
\begin{tabular*}{\textwidth}{@{\extracolsep{\fill}}l*{7}{c}}
\toprule
& \multicolumn{2}{c}{Employee pay} & Revisions & Financing & \multicolumn{2}{c}{Social loss} & Government \\
\cmidrule(lr){2-3}\cmidrule(lr){4-4}\cmidrule(lr){5-5}\cmidrule(lr){6-7}\cmidrule(lr){8-8}
& \makecell{Mean,\\\% of\\reference} & \makecell{Gap,\\90th\\percentile} & \makecell{Scales\\per\\century} & \makecell{Automatic\\increases} & \makecell{Fixed\\costs\\excluded} & \makecell{Fixed\\costs\\counted} & \makecell{Payoff\\gain} \\
& (1) & (2) & (3) & (4) & (5) & (6) & (7) \\
\midrule
\multicolumn{8}{@{}l}{\textit{Panel A. Year-by-year estimates}} \\
\addlinespace[2pt]
Documented regime & 94.4 & 33.6 & 11.1 & 0.00 & 5.89 & 17.22 & -- \\
Full indexation, financing free & 88.1 & 24.3 & 2.2 & 0.00 & 2.86 & 5.14 & 13.12 \\
Full indexation, financed at $\hat\phi_F$ & 85.7 & 27.4 & 2.2 & 7.72 & 11.58 & 13.76 & 3.02 \\
Half indexation, financed at $\hat\phi_F$ & 92.1 & 27.5 & 8.0 & 2.32 & 5.91 & 14.06 & 4.17 \\
\midrule
\multicolumn{8}{@{}l}{\textit{Panel B. Long-run calibration}} \\
\addlinespace[2pt]
Documented regime & 94.6 & 34.9 & 17.1 & 0.00 & 6.13 & 16.58 & -- \\
Full indexation, financing free & 93.3 & 20.1 & 9.5 & 0.00 & 1.93 & 7.62 & 11.63 \\
Full indexation, financed at $\hat\phi_F$ & 90.0 & 23.5 & 9.6 & 8.49 & 11.14 & 16.91 & 1.15 \\
Half indexation, financed at $\hat\phi_F$ & 93.4 & 26.6 & 14.5 & 2.38 & 5.85 & 14.59 & 3.33 \\
\bottomrule
\end{tabular*}
\begin{tablenotes}[flushleft]\footnotesize
\item Notes: Entries are stationary outcomes of the estimated model for Bangladesh over 20,000 simulated years, under the documented regime and under rules that pay the share $\kappa_P$ of each year's inflation automatically, with the automatic increases financed at the coefficient $\phi_A$. The estimated coefficient of an interim allowance is $\hat\phi_F=0.271$. The government re-solves its problem under every rule. Column 1 is mean pay in percent of the reference schedule, and column 2 the 90th percentile of the gap in log points, so pay falls at least this far behind its reference one year in ten. Column 3 counts new pay scales. Column 4 is the cost of financing the automatic increases, and columns 5 and 6 the social loss of Section~\ref{sec:objective}, all in normalized loss units a year, where one unit is the misalignment loss of holding pay ten log points from its reference for a year. Column 7 is the government's own payoff relative to the documented regime, in the same units, and a positive value means that the government is better off. Panel A uses the year-by-year estimates and Panel B the long-run calibration, whose reference share and cost of appointment approximate the documented interval between scales and the size of a cycle.
\end{tablenotes}
\end{threeparttable}
\end{table}

Once the automatic increases have to be financed, the government's behavior changes, and with fixed costs excluded so
does the ranking of full indexation. The estimated financing coefficients belong to the instruments that Bangladesh uses,
an interim allowance ($\hat\phi_F=\RvphiFBGD$) and a scale, and an automatic increase is neither. The experiment
therefore lets a rule pay a share $\kappa_P$ of each year's inflation automatically and charges each automatic step at a
coefficient $\phi_A$ of its own, in the government's objective as well as in the social loss, and the government
re-solves its problem. It responds by letting pay fall further behind between scales, because a lower level of pay makes
each automatic increase cheaper to finance. Under full indexation financed at $\hat\phi_F$, mean pay is
\RsIdxFitFullFinPay{} percent of the reference, against \RsIdxFitFullFreePay{} under the policy solved without the
charge, and the 90th percentile of the gap is \RsIdxFitFullFinPninety{} log points instead of \RsIdxFitFullFreePninety{}
(Table~\ref{tab:index}). Financed at $\hat\phi_F$ and with fixed costs excluded, full indexation then raises the loss
well above the documented regime's (\RsIdxFitFullFinLossZero{} against \RsIdxFitDocLossZero{}), and half indexation
leaves it almost unchanged (\RsIdxFitHalfFinLossZero{}). With fixed costs counted, both financed rules beat the
documented regime at the estimates (\RsIdxFitFullFinLossOne{} and \RsIdxFitHalfFinLossOne{} against
\RsIdxFitDocLossOne{}), whereas at the long-run calibration only half indexation does (\RsIdxLRHalfFinLossOne{} against
\RsIdxLRDocLossOne{}, and \RsIdxLRFullFinLossOne{} for full indexation). The
reoptimization matters more for pay than for the welfare boundary. Holding the government's choices fixed would misstate
the rule's effect on pay, but it would move the financing that half indexation can bear only from \RvRoneReqHalf{} to
\RvRoneReqFixHalf{} of $\hat\phi_F$ (Appendix~\ref{app:experiments}).

\figif{fig_r1_bgd.pdf}{Welfare boundaries of financed indexation in Bangladesh}{fig:r1_bgd}{Panel~A plots the stationary
social loss without fixed costs at the year-by-year estimates against the coefficient $\phi_A$ at which the automatic
increases are financed, in multiples of the estimated allowance coefficient $\hat\phi_F$, for rules that index the share
$\kappa_P$ of inflation (navy), and the red line is the documented regime. Panel~B plots, for each share, the largest
$\phi_A$ at which the rule's loss without fixed costs is below the documented regime's, at the estimates (red circles) and
at the long-run calibration (ochre diamonds). The whiskers span that coefficient over the 95 percent profile region of the
reference share and the cost of appointment, and its 5th to 95th percentiles over 1,000 redraws of the calibration's
moments. The dotted line marks $\hat\phi_F$. The government re-solves its problem under every rule.}

Partial indexation survives where full indexation does not (Figure~\ref{fig:r1_bgd}). Because the charge is quadratic in
the automatic increase, halving the indexed share at a given gap keeps half of the protection against prices and cuts
the charge to a quarter. Without fixed costs, full indexation lowers the loss below the documented regime's only if its
automatic increases cost less than \RvRoneFitFullstatZero{} of $\hat\phi_F$ to finance at the estimates and
\RvRoneLRFullstatZero{} at the calibration, while half indexation does so up to \RvRoneFitHalfstatZero{} and
\RvRoneLRHalfstatZero{}. Across the profile confidence region of the reference share and the cost of appointment, and
across 1,000 redraws of the calibration's moments, half indexation lowers the loss without fixed costs at every point if its automatic increases cost less
than \RvRoneReqRobustHalf{} of $\hat\phi_F$, whereas full indexation needs less than \RvRoneReqRobustFull{}. By its own
objective the government, too, is better off under half indexation than under the documented regime unless $\phi_A$
exceeds \RvRoneGovFitHalf{} times $\hat\phi_F$ at the estimates and \RvRoneGovLRHalf{} times at the calibration, because
pay stays closer to the level it prefers and it appoints fewer commissions. At both parameter points, wherever half indexation lowers the social
loss, it also leaves the government better off, so the rule does not ask the government to act against its own
interest.

These thresholds are welfare boundaries, conditional on the estimated model and the government's reoptimized policy.
They do not measure what an automatic increase costs to finance. Instead, they say which financing would make a rule
worth adopting. The two parameter points rest on different evidence, since the long-run calibration targets the interval between scales and the size of a cycle, which the year-by-year decisions reject (Section~\ref{sec:fit_commission}).
Moreover, Bangladesh's decisions do not favor the forward-looking commission model over a myopic government
(Section~\ref{sec:validation}), so the gain is a property of the estimated model rather than of a model that the data
single out. Finally, an escape clause or a funded reserve would make the rule conditional on the fiscal state, which
Bangladesh's model does not have.

\makeatletter\let\savedsecFB\@fb@secFB\let\@fb@secFB\relax\makeatother
\section{Conclusion}
\makeatletter\let\@fb@secFB\savedsecFB\makeatother
\label{sec:conclusion}

Public pay is a nominal price that governments set by law, and like other prices that are costly to change, it moves in
steps where it is set at discretion. Because it is one of the largest items of public spending, these steps shape how
inflation, the business cycle and sovereign risk reach the payroll. This paper has argued that two costs shape those steps and that they pull in opposite
directions. The fixed cost of acting makes revisions large and rare, whereas the convex cost of financing an increase at
once makes them small and frequent. Through the budget, each revision also moves the deficit, debt and the risk of
fiscal stress under which the next one is made, and under stress the same decision can turn into a cut.

The documented decisions of eleven countries show three regularities, one for each part of this account. Legal pay set
at discretion erodes against prices and private earnings between acts and is restored in discrete steps, and across
countries the steps are larger where they are rarer. Raises come outside crises, whereas every documented cut came
during market stress or an adjustment program, after large deficits. Finally, where a court can strike cuts, legal
reversals create their own paths of pay and of the Treasury's cash, which can differ from the legislated schedule for
years.

\looseness=-1 A single dynamic program accounts for these regularities. Estimated country by country and simulated along the
documented history, the model reproduces the documented paths of real pay in most of the countries whose decisions it
describes, with the same definitions applied to data and simulations. In Bangladesh, however, the year-by-year estimates
imply larger and rarer scales than the documented ones. Within the estimated model, each regularity can be traced to a
distinct feature. The Treasury's requirement, which carries the spread and the excess deficit into the pay decision, produces the cuts, and the Court produces the difference between the cuts
legislated, paid and lasting. Portugal's debt overhang keeps pay low after its crisis, and the fixed cost of a commission,
together with the cost of financing the allowances paid between commissions, produces the erosion-and-reset cycle of
commission pay.

The same features shape the lessons for the fiscal institutions that govern pay. A court that strikes cuts after they take effect leaves the
government a temporary saving to exploit. It therefore shortens how long cuts last without raising pay during a crisis,
and it raises the crisis's social cost, mainly because the restorations it orders must be financed at once. Review
before a cut takes effect removes the temporary savings and most of the restorations. As long as the government does
not weigh the cost of financing the restorations a court orders, as Portugal's decisions indicate, only review that is
also certain removes the incentive to propose cuts that the court will strike. Indexation between commissions protects
employees against deep erosion and requires fewer revisions while its automatic increases are cheap to finance. Once
their financing enters the government's problem, the government lets pay fall further behind, and a rule that indexes
only part of inflation survives a much wider range of financing costs than full indexation.

\looseness=-1 The natural next step is to close the model through output. Here the model is closed through the budget, but the output
that a revision creates does not feed back into revenue, into the GDP denominator of debt or into the onset of stress,
and the transmission block is calibrated, because the panel of legislated acts cannot identify the aggregate response.
Within these bounds the estimated model is an account of the histories it fits. It does not forecast held-out years
better than constant probabilities, and outside \RsForwardBetterList{} a government that does not look ahead fits the
decisions as well. Two extensions would address these bounds most directly, a general equilibrium closure and decision
histories long enough to hold out a crisis while still estimating the requirement that produces cuts.
\label{text:end}

\singlespacing

\clearpage
\onehalfspacing\advance\baselineskip by 0pt plus 1pt\relax
\appendix
\renewcommand{\figif}{\figifat{!tbp}}
\setlength{\textfloatsep}{16pt plus 22pt minus 4pt}
\makeatletter\setlength{\@fpsep}{14pt}\setlength{\@fpbot}{0pt plus 1fil}\makeatother
\numberwithin{equation}{section}
\numberwithin{table}{section}
\numberwithin{figure}{section}
\part*{Appendix}
\addcontentsline{toc}{part}{Appendix}
\noindent Equation, result, table and figure numbers without a letter refer to the main text, and lettered numbers
refer to this appendix. Appendix~\ref{app:proofs} proves the results of Section~\ref{sec:model}, and
Appendix~\ref{app:model} derives the government's payoff and the Treasury's requirement and describes the fiscal,
court and commission components, the transmission block and the numerical solution. Appendix~\ref{app:data} defines the
data behind the three patterns and the decision histories. Appendix~\ref{app:estimation} reports the fiscal block,
every country's estimates and the inference for Portugal, and Appendix~\ref{app:experiments} documents the two
institutional experiments, their sensitivities and their uncertainty.

\etocsetnexttocdepth{subsection}
\localtableofcontents
\clearpage

\section{Proofs and derivations}
\label{app:proofs}

The results are proved in the order in which they appear in Section~\ref{sec:model}.

\paragraph{Proposition~\ref{prop:band}.}
(i) Fix a policy that only raises pay and keeps the gap in a bounded set. Let the $i$th spell between raises last $T_i$
years and start at the gap $x_i$, and let the raise that ends it have size $J_i$. Without uncertainty the gap rises by
$\mu T_i$ over the spell, and the misalignment loss of the spell is
\[
 \frac a2\int_0^{T_i}(x_i+\mu t-x^\star)^2\,\dd t\ \ge\ \frac{a}{24}\,\mu ^2T_i^3 ,
\]
with equality if and only if the spell is centered on $x^\star$, $x_i=x^\star-\mu T_i/2$. The losses and costs of the
first $N$ spells are therefore at least $\sum_{i\le N}\big(K^++\phi_FJ_i^2+\tfrac a{24}\mu ^2T_i^3\big)$. Over these
spells the gap changes by $\mu \sum_{i\le N}T_i-\sum_{i\le N}J_i$, which is bounded, so the average raise $\bar J_N$ and
the average spell $\bar T_N$ satisfy $\bar J_N=\mu \bar T_N+O(1/N)$. By Jensen's inequality
$\sum_{i\le N}J_i^2\ge N\bar J_N^2$ and $\sum_{i\le N}T_i^3\ge N\bar T_N^3$, so the average loss per year over the first
$N$ spells is at least $C(\mu \bar T_N)+O(1/N)$, where $C$ is the average loss of a stationary cycle
in~\eqref{eq:avgloss},
\[
 C(J)=\frac{K^+\mu }{J}+\phi_F\mu J+\frac{a}{24}J^2 .
\]
$C$ is strictly convex on $J>0$, with $C''(J)=2K^+\mu /J^3+a/12$, and tends to infinity as $J\to0$ and as $J\to\infty$,
so it has a unique minimizer $J^\ast$, the root of $C'(J)=0$. Multiplying $C'(J)=0$ by $J^2$ gives the cubic of the
proposition, whose left side increases from zero on $J>0$, so the positive root is unique. The stationary policy that
raises pay from $x^\star+J^\ast/2$ to $x^\star-J^\ast/2$ has an average loss of exactly $C(J^\ast)$, so no bounded
policy of raises does better, and its spells last $J^\ast/\mu $ years.

(ii) Where waiting is optimal, the first term of \eqref{eq:qvi} is zero,
\[
 \tfrac{\sigma_d^2}{2}V''(x)+\mu V'(x)-\delta_IV(x)=\tfrac a2(x-x^\star)^2 .
\]
\looseness=1 The quadratic $v_0+v_1x+v_2x^2$ with $v_2=-a/(2\delta_I)$, $v_1=(ax^\star+2\mu v_2)/\delta_I$ and
$v_0=(\sigma_d^2v_2+\mu v_1-\tfrac a2x^{\star2})/\delta_I$ solves it, and the homogeneous equation is solved by
$e^{u_1x}$ and $e^{u_2x}$, where $u_1>0>u_2$ are the roots of $\tfrac{\sigma_d^2}{2}u^2+\mu u=\delta_I$. Far below the
trigger the government waits for erosion to bring the gap back to the band, and the expected discounted loss of waiting
is bounded by a quadratic in $x$. Because $e^{u_2x}$ grows faster than any quadratic as $x\to-\infty$, its coefficient is
zero, which leaves the stated form with $v_3$ the coefficient on $e^{u_1x}$. At the trigger the government is
indifferent between waiting and raising, which is value matching. The gain from raising,
$\max_{x'}\big[V(x')-K^+-\phi_F(x-x')^2\big]$, has derivative $-2\phi_F(x-\hat x)$ in $x$ by the envelope theorem, and
an optimal trigger requires the value of waiting to meet it with the same slope, which is smooth pasting
\citep{stokey2009}. The first-order condition for the reset point is $V'(\hat x)=-2\phi_F(\bar x-\hat x)$. After a raise
the gap starts at $\hat x$, and the next raise comes when a Brownian motion with drift $\mu >0$ and variance
$\sigma_d^2$ first rises by $\bar x-\hat x$. This first-passage time is inverse Gaussian with mean $(\bar x-\hat x)/\mu $
and shape $(\bar x-\hat x)^2/\sigma_d^2$, so its variance is $(\bar x-\hat x)\sigma_d^2/\mu ^3$. The three conditions
are necessary. A band that satisfies them is optimal if its value also satisfies \eqref{eq:qvi} at every gap, which a
numerical solution of~\eqref{eq:qvi} by value iteration verifies at the estimated costs of the two commission
countries. \qed

\paragraph{Corollary~\ref{cor:costs}.}
Let $F(J)=\tfrac a{12}J^3+\phi_F\mu J^2-K^+\mu $, and write $u=\tfrac a{12}J^{\ast3}>0$ and $w=\phi_F\mu J^{\ast2}\ge0$.
At the root, $K^+\mu =u+w$, so $s_F=\phi_FJ^{\ast2}/K^+=w/(u+w)\in[0,1)$, and $J^\ast F'(J^\ast)=3u+2w$. By the implicit
function theorem the elasticity of $J^\ast$ with respect to a parameter $p$ is $-p\,(\partial F/\partial p)/(J^\ast
F'(J^\ast))$. With respect to $K^+$ it is $(u+w)/(3u+2w)=1/(3-s_F)$, and with respect to $\phi_F$ it is
$-w/(3u+2w)=-s_F/(3-s_F)$. With respect to $\mu $ it is $(K^+\mu -\phi_F\mu J^{\ast2})/(3u+2w)=u/(3u+2w)=(1-s_F)/(3-s_F)$.
Because $T^\ast=J^\ast/\mu $, the interval has the same elasticities with respect to $K^+$ and $\phi_F$, and the
elasticity $(1-s_F)/(3-s_F)-1=-2/(3-s_F)$ with respect to $\mu $. Parts (a) and (b) follow from the signs, with
$s_F\in[0,1)$. For (c), solving $e=(1-s_F)/(3-s_F)$ for $s_F$ gives $s_F=(1-3e)/(1-e)$, while size and interval together
determine only $\mu =J^\ast/T^\ast$ and $J^\ast$, which is one equation in the two costs. \qed

\paragraph{Lemma~\ref{lem:nesting}.} Under the restrictions the fiscal state is constant and there are no outstanding
cuts. The election phase carries no payoff, since $\eta_W$ is constant and $v_E=0$, the conflict term and the
Treasury's penalty vanish, and next year's gap is $x+\mu $ with certainty. The flow payoff \eqref{eq:support} is then
$u(x)=-\tfrac a2(x-x^\star)^2$ up to a constant. With interim allowances and cuts unavailable, the actions are to wait or
to appoint a commission. With $\rho_R=1$ the new scale follows in the year of the appointment, so appointing at $x$ and
setting the scale $x'\le x$ costs $K_I+\phi_C(x-x')^2$ and leads to the post-decision value $u(x')+\beta_IV(x'+\mu )$.
Leaving the scale unchanged costs $K_I$ and is dominated by waiting.

The problem is deterministic and stationary, so the optimal policy is a stationary map from gaps to actions, and its path
from any reset is eventually periodic. The period need not consist of cycles of one length, because the target of a reset
depends on the gap at which it is made, so the optimum can alternate between lengths. Consider the policies whose cycles
all have one length $T$. After a reset to $\hat x$ the gap takes the values $\hat x,\hat x+\mu ,\dots,\hat x+(T-1)\mu $
at the post-decision dates and is reset by $J=\mu T$. Without discounting, the average loss per year of such a cycle is
\[
 \frac1T\Big[\tfrac a2\sum_{k=0}^{T-1}(\hat x+k\mu -x^\star)^2+K_I+\phi_C(\mu T)^2\Big].
\]
The sum is smallest when the $T$ gaps are centered on $x^\star$, $\hat x=x^\star-\tfrac{(T-1)\mu }2$. Since
$\sum_{k=0}^{T-1}\big(k-\tfrac{T-1}2\big)^2=T(T^2-1)/12$, the average loss is then
\[
 \tfrac a{24}\mu ^2(T^2-1)+\frac{K_I}T+\phi_C\mu ^2T=\Big[\tfrac a{24}J^2+\frac{K_I\mu }J+\phi_C\mu J\Big]-\tfrac a{24}\mu ^2,\qquad J=\mu T.
\]
The bracket is the average loss of the continuous-time problem of Proposition~\ref{prop:band}(i) with $K^+=K_I$ and
$\phi_F=\phi_C$, whose minimizer satisfies $\tfrac a{12}J^3+\phi_C\mu J^2=K_I\mu $. The two objectives differ by a
constant, so over the reals they are minimized at the same $J$. Because the bracket is convex in $J$, the best cycle of one
length has one of the two whole numbers of years around $T^\ast$, and the optimum over all policies may alternate
between the two. The estimated problem discounts the future, and the numerical check below finds the same cycles at
discount factors of 0.90 and 0.99.
\qed

\paragraph{A numerical check of the lemma.} The solver used in the estimation, restricted to the commission problem, was
solved under the lemma's restrictions for erosion speeds of 4, 6 and 9 log points a year, two pairs of costs
$(K_I,\phi_C)$ and discount factors of 0.90 and 0.99. In all \RvNestCases{} cases its revision region and its targets
coincide with those of an independent value iteration of the reset problem on the same grid, and its values differ from
it by at most \RvNestMaxDiff{} up to a constant. In every case the solver's cycle has the whole number of years nearest
the continuous interval of Proposition~\ref{prop:band}(i), or alternates between the two whole numbers around it.

\paragraph{Proposition~\ref{prop:court}.} Let the cut take the outstanding cuts from $L$ to $L+c>\bar B$ at the gap $x$
in a crisis year. If it stands, which happens with probability $1-\rho_S$, it is in force for the whole year, the gap
after the decision is $x+c$ and the outstanding cuts are $L+c$, a position worth $\mathcal W(x+c,L+c,0;s)$ in the
notation of~\eqref{eq:postvalue}. If it is struck, which happens with probability $\rho_S$, it is in
force for the share $\tau$ of the year, so the year's flow payoff is evaluated at the gap $x+\tau c$, while the
outstanding cuts and the continuation are those of waiting, which is worth $\mathcal W(x,L,0;s)$. Undoing the struck
cut is not the government's decision and carries no cost in its problem. The value of proposing the cut is
therefore that of~\eqref{eq:values_court},
\begin{align*}
 &(1-\rho_S)\,\mathcal W(x+c,L+c,0;s)+\rho_S\big[\mathcal W(x,L,0;s)+u(x+\tau c)-u(x)\big]\\
 &\qquad-K^--\lambda_L\big[(1-\rho_S)+\rho_S\tau\big]c ,
\end{align*}
with $u$ evaluated at the year's $(b,E,F)$.
The expected pay in force during the year, and so the expected pay that employees lose, is
$(1-\rho_S)c+\rho_S\tau c=\xi(\rho_S,\tau)\,c$, which is part~(i). With $\rho_S=1$ and $\tau=0$ the value is
$\mathcal W(x,L,0;s)-K^-$, the value of waiting less the fixed cost, which is part~(ii). For part~(iii), $\xi(\rho_S,\tau)=1-\rho_S(1-\tau)$ is zero on $[0,1]^2$ only at $\rho_S=1$ and
$\tau=0$. Its derivatives are $-(1-\tau)$ and $\rho_S$, and its cross-partial is one, so an earlier review raises the
effect of a more certain one, and conversely. \qed

\paragraph{Lemma~\ref{lem:mpc}.}
(i) With the borrowing limit binding before and after the increment, consumption equals income plus new borrowing less
debt service, $c=y+\tilde b_0+\theta_C\hat y-Rb$, where $b$ is inherited debt and $R$ its gross rate. Inherited debt is given on impact, so $\partial c/\partial y=1+\theta_C\,\partial\hat
y/\partial y$ then. If the limit does not respond to the increment, borrowing does not change, inherited debt stays on
its path, and consumption rises by the increment in every period. (ii) Without constraints, income risk or future binding limits, and with $\beta R^D=1$, the Euler equation
makes the consumption increment constant, at the annuity value of the present value of the income increment,
$\sum_{j\ge0}(\beta\rho_y)^j\Delta y=\Delta y/(1-\beta\rho_y)$. The increment is therefore $(1-\beta)\Delta
y/(1-\beta\rho_y)$ in every period, including periods after the income increment has decayed. (iii) With
$\partial\hat y/\partial y=0$, part (i) gives a propensity of one for constrained recipients. Aggregating $m_i\Delta y_i$
over recipients and dividing by the aggregate increment gives the average, weighted by each recipient's share of the
increment. With increments proportional to pay, $s_c$ is therefore the constrained recipients' share of the payroll,
which equals their share of recipients when the two groups are paid alike. The two-type consumption path of a persistent increment is therefore $\Delta C^G_t=s_c\rho_y^t\Delta
Y_0+(1-s_c)m^u\Delta Y_0$, which the transmission block implements with an annuity state fixed at $m^u$ times the
increment at the date of the news (Appendix~\ref{app:model}). \qed

\paragraph{Proposition~\ref{prop:captive}.}
\looseness=1 Under the assumptions, equity and foreign funding do not change within the year, and required reserves change by
$\Delta\mathrm{Res}=r_D\,\Delta D$. Deposits change by the recycled payroll less the deposit-funded part of the
Treasury's non-bank financing, $\Delta D=(s_D-\theta_T-s_R\theta_R)\Omega$, and $\Delta B^b=\theta_B\Omega$. The
balance-sheet identity gives $\Delta L^p=(1-r_D)\Delta D-\Delta B^b=-[\theta_B+(1-r_D)(\theta_T+s_R\theta_R-s_D)]\Omega$.
\looseness=1 With an administered lending rate the change is a rationed quantity. The flow accounting holds within the year, and it
does not describe the stock of securities or retail bonds that the Treasury's placements leave on balance sheets in
later years. \qed

\paragraph{Proposition~\ref{prop:psi}.}
\looseness=1 (i) When applications exceed posts at skill $z$, hiring equals the number of posts whatever the number of applications.
A higher $W$ raises the value of a public job relative to the outside option, so more households apply and the
probability of being hired, the ratio of posts to applications, falls. Because applicants queue while employed and the
composition of applicants does not change, the mass of households supplying labor to formal firms at skill $z$ is
unchanged. Holding private labor demand fixed, so is $w_p(z)$, and $\psi(z)=0$. (ii) When posts exceed applications
every applicant is hired, and a higher $W$ draws applicants from formal employment, unemployment and informality. Formal
supply falls weakly and $w_p$ rises weakly along labor demand, with no change when all hires come from an informal margin
that is elastic at a reservation wage that does not respond to $W$. Both statements hold private labor demand fixed.
In general equilibrium it moves with output, and the total elasticity adds $\psi_Y\,\dd\ln Y/\dd\ln W$. \qed

\paragraph{Equation~\eqref{eq:formula}.}
Linearize goods-market clearing on impact with the monetary response, the exchange rate and expectations held fixed.
With $\psi=0$ the dose transfers $\Omega$ to recipients, who spend $\bar m\Omega$ on impact (Lemma~\ref{lem:mpc}), and
takes $\theta_T\Omega$ from taxpayers, who reduce spending by $m_T\theta_T\Omega$. The domestic share $(1-\alpha_M)$ of
the net change is multiplied by $\kappa_D$. Investment falls by $\eta_K$ times the credit displacement of
Proposition~\ref{prop:captive}. With $\psi>0$ the private wage rises by $\psi\,\dd\ln W=\psi\Omega/s_g$ and lowers output
with elasticity $\varepsilon_W$. A services term, by which better-paid officials raise productivity, would add to the
bracket with the lag at which public services enter production. It is zero in the quantitative results because neither
of its factors is measured. \qed

\section{The model in detail}
\label{app:model}
\label{app:micro}
\label{app:system}

This appendix supports Section~\ref{sec:model}. It derives the payoff and the requirement from the problems of the
agents involved, states the fiscal, court and commission components, describes the transmission block and the numerical
solution, and collects the notation.

\subsection{Microfoundations of the payoff and the requirement}

Every flow is in percent of the payroll at the reference. Public services are worth $100\,v_Sq(x)$, where the quality of
the workforce $q$ is quadratic and concave in the gap, with $q(0)=1$. Quality depends on legal pay relative to the
reference $W^\ast_t=P_t^{1-s_w}(W^p_t)^{s_w}$, where $P_t$ is the price level and $W^p_t$ the private wage, because
effort and morale depend on what pay buys while recruitment and retention depend on pay relative to the private sector
\citep{krueger1988,dalbo2013,finan2017}. The reference is efficient pay, the benchmark of \citet{gomes2015}. A unit of revenue costs
taxpayers $1+\lambda_F$, and at $x=0$ the marginal value of quality equals the marginal excess burden $\lambda_F$ of the
revenue that pays for it. Relative to a planner, the government counts each unit of payroll an extra
$\eta_F(b)$ against it, for the taxpayers who finance it, and an extra $\eta_W(E)$ in its favor, for the employees who receive it. Under probabilistic voting these are the excess densities of swing voters in each group
\citep{lindbeck1987,persson2000}. A visible raise before an election is worth $v_E$ per log point
\citep{rogoff1990,brender2005}, and the government discounts the future at $\beta_I$, below a planner's discount factor
if it may lose office \citep{alesina1990}. The union tolerates gaps up to $x_s(b)$, and beyond its tolerance a dispute
costs $c_U$ per log point of excess, the resistance curve of \citet{ashenfelter1969} run in reverse.

In stress or a program, creditors or the program require the primary balance to improve by the extra interest at the
year's spread, $D^F=\zeta b/100$, and by the deficit above three percent of GDP, $D^D=(d^{\mathrm{ref}}-3)^+$. The
Treasury meets requirement $j$ with contributions $d_{ij}$ of instruments $i$, the payroll among them, at the cost
$\tfrac12C_{ij}d_{ij}^2$, and it finances whatever the payroll fails to contribute at the cost $\eta_T$ per unit of
payroll, the premium on unplanned borrowing in a crisis.

\setcounter{proposition}{0}\renewcommand{\theproposition}{B.\arabic{proposition}}%
\providecommand{\theHproposition}{}\renewcommand{\theHproposition}{B.\arabic{proposition}}%
\begin{proposition}[Flow payoff and requirement]
\label{prop:micro}
Under these assumptions the government's flow objective is the payoff~\eqref{eq:support} up to a term that does not
depend on the year's gap, with
$a=-100\,v_Sq''(0)+(\lambda_F+\eta_F-\eta_W)/100$, and the payroll's required saving is~\eqref{eq:requirement} with
$\gamma_j=C_{Pj}^{-1}/\sum_iC_{ij}^{-1}$. A raise from $x$ to $x'<x$ financed within its year by borrowing whose marginal
cost rises by $\phi_B$ per point of GDP costs $\phi_F[S(x)-S(x')]^2$, with $\phi_F=\tfrac12\phi_Bs_g/100$, and a cut costs
$\lambda_L=(1+\eta_W)\lambda_N$ per log point of pay lost if employees bear a fairness loss $\lambda_N$ per log point
\citep{kahneman1986,bewley1999,kaur2019}.
\end{proposition}

\begin{proof}
Employees receive $100e^{-x/100}$, which costs taxpayers $100(1+\lambda_F)e^{-x/100}$. With the political weights the
payroll term is $-100[\lambda_F+\eta_F(b)-\eta_W(E)]e^{-x/100}=-100[\cdot]+[\cdot]S(x)$. The first part does not
depend on the gap. It varies with debt in the aftermath through $\eta_F(b)$, and the payoff~\eqref{eq:support} leaves it
out, so the overhang acts through the value of the saving $S(x)$ and the union's tolerance alone. Because $q$ is quadratic,
$100v_Sq(x)=100v_S-\lambda_Fx-\tfrac{a_S}2x^2$ with $a_S=-100v_Sq''(0)$, which gives~\eqref{eq:support}. The Treasury's
first-order conditions equate $C_{ij}d_{ij}$ across instruments, so $d_{Pj}=\gamma_jD^j$. A payroll at the gap $x$ saves
$s_gS(x)/100$ percent of GDP, which gives~\eqref{eq:requirement}. A raise adds $s_g[S(x)-S(x')]/100$ percent of GDP to
the year's borrowing at the cost $\tfrac12\phi_B$ times its square, and from the next year the higher payroll is in the
level term. A cut of $c$ log points costs employees $\lambda_Nc$, which the government weighs at $1+\eta_W$.
\end{proof}

The social objective~\eqref{eq:socialloss} is the planner's version of this objective. It centers misalignment at the
reference and excludes the political weights $\eta_F$, $\eta_W$ and $v_E$, and it counts the Treasury's penalty and the
in-year financing of discrete increases, both of which are resource costs here. It keeps the government's curvature $a$,
which exceeds the planner's by $(\eta_F-\eta_W)/100$, a difference of $0.15/100$ at the estimates. The stabilization
weight $w_Y=1$ and the insurance weights $z_S=0.5$ and $z_2=2$ (a relative risk aversion of two) were fixed in the plan.
The insurance term's scale, $z_0=0.45$, is the taxpayers' weight $\eta_F=\eta_W+a\,x^\star$ at the employees' weight
$\eta_W=0.3$ and the starting preferred gap of 7.5 log points. It is held fixed, so that the estimated political weights
do not move the social objective. The
misalignment weight $a=0.02$ normalizes the units, so that a loss of one unit is the misalignment loss of holding
pay ten log points from its reference for a year.

\subsection{Fiscal states, the court and the commission}
\label{app:components}

After its first budget, stress continues with probability 0.6, escalates into a program with probability 0.2 and
otherwise resolves. In later budgets it continues with probability one half, escalates with probability 0.4 and resolves
with probability 0.1. A program persists with probability two thirds. The aftermath lasts while debt exceeds 85 percent of GDP. In the model's transition law the probability that
a year is an aftermath year rises logistically with debt around 85, with a scale of 2.5 points, because a sharp
threshold would make the continuation value discontinuous in debt. Normal and aftermath years are split by whether an
excessive deficit procedure is open, in which case the requirement consists of the deficit term alone. A procedure
opens with probability 0.09 a year, closes with probability 0.26 and is open at the end of a crisis with probability
0.35, and the requirement is capped at 60 percent of the payroll. The onset of
stress follows~\eqref{eq:onset}, the spread~\eqref{eq:spread}, and the primary deficit excluding the payroll its median
in each fiscal state (Table~\ref{tab:fiscal_block}).

The drift~\eqref{eq:drift} and nominal growth in~\eqref{eq:deficit} take inflation and real growth from the country's
Markov chain, estimated on its normal years (Assumption~\ref{ass:separation}). The state's $m_t$ holds the values of the
year before budget year $t$, and the next draw $m_{t+1}$ holds those of budget year $t$, which set the year's drift,
nominal growth and, in~\eqref{eq:onset}, real growth. This is the timing of the estimated onset relation, which predicts
the onset from the previous year's deficit and growth (Table~\ref{tab:fiscal_block}). In the European countries the draw
is shifted by the documented mean differences in inflation and real growth between crisis and normal years when the next
fiscal state is a second or later budget of stress or a program. Such a state follows a crisis year, so the shift stands
in for that of a crisis year. The two differ in two cases. A year that ends a crisis carries no shift, and in Hungary
and Romania, whose crises begin directly with a program, the year before the program carries one. Real
private-wage growth follows its relation to real growth in normal years, plus its mean residual in crisis years and in
aftermath years, keyed in the same way to the next fiscal state, so the reference grows by $s_w$ times this rate.

In Portugal and Italy, a cut proposed in stress or a program that takes the outstanding cuts above the admissible burden
$\bar B=5$ log points is struck with probability $\rho_S=0.9$. The burden is that of Portugal's 2011 reductions, which
the Court upheld (5.1 log points, a cut of 5 on the model's grid of whole log points). A struck cut stays in force for
the whole year in the budgets of stress and the first budget of a program, as Portugal's 2012 suspension did, and for a
fifth of the year in later program years ($\tau=1$ and $\tau=0.2$). The fifth is the average of the 2013 suspension,
struck before it was due, and the 2014 reductions, struck after five months. In later
program years the excess of the outstanding cuts over $\bar B$ is reversed with probability 0.9, and in the aftermath
all outstanding cuts are reversed with yearly probability $r_A=0.5$. A reversal ordered by a court or legislated is not a
decision of the government, and its own problem attaches no cost to it. The social loss of Section~\ref{sec:objective}
counts it as a discrete increase that must be financed at once. Portugal's decisions put the weight that the
government's problem would attach to financing a court-ordered restoration at zero (Appendix~\ref{app:estimation}). In the documented histories, a legislated reversal applies to the cuts
standing at the start of its year and, where the law ended a single measure, to that measure alone. In 2013, for example, Spain's
suppression of the December 2012 extra payment lapsed while its 2010 cut stayed in force, and in 2010 Romania's unpaid
leave of 2009 lapsed while the cut of July 2010 took effect.

In the commission countries a sitting commission reports with the yearly probability $\rho_R$ of the documented
commissions. After each documented scale the gap is set to the model's return point, the gap that maximizes the
post-decision value with no commission sitting, so that the level of the gap does not drift over decades. Interim
measures of undocumented size count for their direction only, and the gap after them is unknown until the next scale.

\paragraph{Other calibrated constants.} The employees' weight is $\eta_W=0.3$ and the union's tolerance $x_s=5$ log
points. The payroll at the reference, $s_g$, is the country's mean compensation of employees over 2000--2024 in the
European countries, the General Schedule payroll in the United States, and the payroll that the scales cover in
Bangladesh and India, 2.0 and 1.6 percent of GDP. The effective rate at a zero spread, $i_c$, is the country's mean
effective rate net of the spread's pass-through outside programs, and pre-crisis debt $\bar b$ is debt in 2007. The
notes to Tables~\ref{tab:params} and~\ref{tab:v2_estimates} give the remaining constants and ratios.

\subsection{The transmission block}
\label{app:transmission}

The block is a log-linear small open economy, solved by the generalized Schur decomposition \citep{klein2000}. Public
employees are constrained or unconstrained (Lemma~\ref{lem:mpc}), and private households are hand-to-mouth or Ricardian.
A firm in the nontraded or export sector maximizes the present value of dividends subject to quadratic price-adjustment
costs, convex capital-adjustment costs, a collateral constraint and nonnegative dividends. This gives labor demand at
the private wage, a Phillips curve in which real marginal cost is proportional to the private wage, and an investment
equation with an accelerator and, where credit is rationed, a credit term with coefficient $\eta_K$. With an administered
lending rate, loans are rationed and the credit equation is the decomposition of Proposition~\ref{prop:captive} plus a
credit-demand term. A tax rule on lagged debt keeps debt stationary, a debt-elastic country premium of the kind that
\citet{schmittgrohe2003} use to close small open economies rises with public debt, and the policy rate equals the foreign rate under a peg and follows a Taylor rule under a float.

\paragraph{The system.} Let $k_t$ be the percent by which a revision raises real public pay above its path without the
revision, so that the dose $\Omega_t$ is $k_t$ times the share of the covered payroll in GDP. Every other variable is a
deviation from the path without the revision. Output $y_t$, the consumption of Ricardian households $c^R_t$, inflation
$\pi_t$, the real exchange rate $e_t$ and the private wage $w^p_t$ are in percent. The consumption of public employees
$c^G_t$ and of private hand-to-mouth households $c^H_t$, taxes $T_t$, public debt $b_t$, private credit $L^p_t$ and
investment $I_t$ are in points of GDP. The block solves
\begingroup\allowdisplaybreaks
\begin{align*}
 \Omega_{t+1}&=\rho_y\,\Omega_t, && \text{pay increment}\\
 c^G_t&=s_c\,\Omega_t+(1-s_c)\,m^u\,\Omega_0, && \text{public employees}\\
 c^H_t&=\varsigma_H\big[\varsigma_L\,(y_t+w^p_t)-T_t\big], && \text{hand-to-mouth households}\\
 y_t&=(1-\alpha_M)\big[c^G_t+c^H_t+(1-\varsigma_H)\,\varsigma_C\,c^R_t\big]+I_t+\varsigma_X\,\varepsilon_X\,e_t, && \text{goods market}\\
 I_t&=\varsigma_I\,y_t+\eta_K\,L^p_t, && \text{investment}\\
 L^p_t&=-\kappa_L\,\Omega_t+\nu\,y_t, && \text{private credit}\\
 w^p_t&=\psi\,k_t+\psi_Y\,y_t, && \text{private wage}\\
 \pi_t&=\beta\,\E_t\pi_{t+1}+\kappa_\pi\,w^p_t, && \text{Phillips curve}\\
 e_t&=c^R_t+\varrho_e\,b_t, && \text{risk sharing, premium}\\
 T_t&=\theta_T\,\Omega_t+\varrho_T\,b_{t-1}, && \text{tax rule}\\
 b_t&=b_{t-1}/\beta+\Omega_t-T_t, && \text{public debt}\\
 e_t&=e_{t-1}-\pi_t . && \text{fixed exchange rate}
\end{align*}
\endgroup
The last equation closes the block for the two countries of the experiments, which both have a fixed exchange rate,
Portugal inside the monetary union. In equation~\eqref{eq:formula}, taxpayers' propensity $m_T$ is the share
$\varsigma_H$ of private households that are hand-to-mouth, and the system implies the multiplier $\kappa_D$ and the
elasticity $\varepsilon_W$.

\paragraph{Calibration.} The parameters common to the two countries are a discount factor $\beta=0.96$, a unit
elasticity of intertemporal substitution, a persistence of the pay increment $\rho_y=0.85$, a labor share
$\varsigma_L=0.50$, an investment share $\varsigma_I=0.25$, an export elasticity $\varepsilon_X=0.50$, a cyclicality
$\nu=0.50$ of credit and $\psi_Y=0.50$ of the private wage, a slope $\kappa_\pi=0.20$ of the Phillips curve, a premium
$\varrho_e=0.10$ and a tax response $\varrho_T=0.10$. They imply $m^u=(1-\beta)/(1-\beta\rho_y)=0.22$.
Table~\ref{tab:calibration} gives the parameters that differ between Portugal and Bangladesh. In Bangladesh the block
uses the disposable increment. Income tax withheld at source takes 2.5 percent of the gross increment and lower
provident-fund subscriptions add 6.3 percent, so recipients receive 103.8 percent of it in cash, which is also the cash
that the Treasury must find.

\begin{table}[!tbp]\centering
\begin{threeparttable}
\caption{Calibration of the transmission block}
\label{tab:calibration}
\footnotesize\setlength{\tabcolsep}{2pt}
\begin{tabular*}{\textwidth}{@{\extracolsep{\fill}}llcc}
\toprule
Parameter & Symbol & Portugal & Bangladesh \\
\midrule
\multicolumn{4}{@{}l}{\textit{Panel A. Shares}} \\
\addlinespace[2pt]
Payroll covered by the revision, percent of GDP & & 11.0 & 2.0 \\
Constrained share of public employees & $s_c$ & 0.45 & 0.45 \\
Hand-to-mouth share of private households & $\varsigma_H$ & 0.30 & 0.55 \\
Consumption, share of GDP & $\varsigma_C$ & 0.64 & 0.68 \\
Exports, share of GDP & $\varsigma_X$ & 0.40 & 0.15 \\
Import content of consumption & $\alpha_M$ & 0.38 & 0.25 \\
\midrule
\multicolumn{4}{@{}l}{\textit{Panel B. Financing of the dose}} \\
\addlinespace[2pt]
Taxes & $\theta_T$ & 0.40 & 0.10 \\
Bank credit & $\theta_B$ & 0.00 & 0.00 \\
Retail bonds & $\theta_R$ & 0.00 & 0.90 \\
Market or foreign borrowing & $\theta_F$ & 0.60 & 0.00 \\
Deposit drawdown per retail bond & $s_R$ & -- & 0.80 \\
Administered lending rate & & No & Yes \\
Credit displacement per unit of dose & $\kappa_L$ & 0.00 & 0.72 \\
\midrule
\multicolumn{4}{@{}l}{\textit{Panel C. Transmission}} \\
\addlinespace[2pt]
Direct effect of public pay on private wages & $\psi$ & 0.00 & 0.00 \\
Investment per point of credit & $\eta_K$ & 0.20 & 0.35 \\
Exchange rate & & Monetary union & Fixed \\
\bottomrule
\end{tabular*}
\begin{tablenotes}[flushleft]\footnotesize
\item Notes: Entries are the calibrated parameters of the transmission block that differ between the two countries of the experiments, and the text lists the common ones. The payroll shares come from the government finance statistics and national budgets, and the financing shares from the fiscal accounts, with the IMF's Article IV consultation for Bangladesh. The constrained share of public employees is disciplined by Bangladesh's household surveys (Section~\ref{sec:objective}) and assumed for Portugal. Where the lending rate is administered, the credit displacement is that of Proposition~\ref{prop:captive} with a zero reserve ratio and recycled deposits equal to the tax share, $\kappa_L=\theta_B+s_R\theta_R$, and it is zero elsewhere. The direct effect of public pay on private wages is set to zero (Section~\ref{sec:objective}).
\end{tablenotes}
\end{threeparttable}
\end{table}

\paragraph{The output path.} The path $\Delta y_t$ of the social loss~\eqref{eq:socialloss} is the block's output
response to the path of real pay under an institution. In Portugal's histories of 2010--2024 that path is measured
relative to the documented path of pay, whose effect the documented output gap already contains, and in the model's own
world it is measured relative to the reference. The response is built by superposition from the block's responses to the
yearly innovations of the path, so that the block's pay increment follows the path exactly. An increase of one percent in
real pay raises output on impact by 0.05 percent in Portugal and by 0.01 percent in Bangladesh, which is 0.43 and
0.42 percent per point of GDP of the dose, and the response fades within about five years. The output path enters the
social loss but does not feed back into the decision problem.

\subsection{Numerical solution}
\label{app:numerics}

The government's problem is solved by value iteration with Anderson acceleration to a tolerance of $10^{-6}$ on a grid
of gaps, outstanding cuts and debt, with cubic Lagrange interpolation on four debt nodes. Sizes are listed on a grid whose
widths enter the nested logit, so that the inclusive values do not depend on how finely the sizes are listed. The solver
reproduces the reset problem of Lemma~\ref{lem:nesting} (Appendix~\ref{app:proofs}), and a value iteration
of~\eqref{eq:qvi} confirms the band of Proposition~\ref{prop:band}(ii) at the estimated costs of the commission
countries.

\condbreak{12\baselineskip}
\begingroup\setstretch{1}\footnotesize\setlength{\tabcolsep}{3pt}
\begin{xltabular}{\textwidth}{l>{\RaggedRight\arraybackslash}X}
\caption{Notation}\label{tab:notation} \\
\toprule
\endfirsthead
\caption[]{Notation (continued)} \\
\toprule
\endhead
\midrule
\multicolumn{2}{r}{\textit{Continued on the next page}} \\
\endfoot
\bottomrule
\endlastfoot
\multicolumn{2}{@{}l}{\textit{The pay decision}}\\
$W_t,\ W^\ast_t,\ x_t$ & legal pay of a grade step; the reference schedule; the gap $x_t=100(\ln W^\ast_t-\ln W_t)$, log points \\
$\pi_t,\ \omega_t,\ s_w,\ \kappa_P$ & inflation; growth of nominal private wages; share of real private-wage growth in the reference; share of inflation indexed by rule \\
$a,\ x^\star,\ \eta_F,\ \eta_W,\ \lambda_F$ & misalignment weight; preferred gap $x^\star=(\eta_F-\eta_W)/a$; extra political weights of taxpayers and employees; excess burden of revenue \\
$S(x),\ a_S$ & saving on the payroll at the gap $x$, $S(x)=100(1-e^{-x/100})$; curvature of the value of services \\
$K^+,\ \mu,\ \sigma_d,\ J^\ast,\ T^\ast,\ s_F$ & benchmark: fixed cost; drift and volatility of the gap; size and interval of a reset; $s_F=\phi_FJ^{\ast2}/K^+$ \\
$\bar x,\ \hat x,\ e$ & benchmark: trigger and reset point of the band; elasticity of the size of a reset to the drift \\
$\beta_I,\ \delta_I$ & the government's discount factor and rate \\
$F_t\in\{N,M,C,A\},\ \ind^{\mathrm{cr}}_t$ & fiscal state: normal, market stress, adjustment program under conditionality, aftermath; indicator of a crisis year (market stress or a program) \\
$E_t,\ m_t,\ r_t,\ \ind^C_t$ & election phase; macroeconomic state (inflation and real growth of the year before the budget year); whether last year's change was a raise; whether a commission is sitting \\
$b_t,\ d_t,\ p(F),\ s_g,\ n_t$ & debt and deficit, percent of GDP; primary deficit excluding the payroll; payroll at the reference, percent of GDP; nominal growth \\
$\zeta(b,F),\ \zeta_0,\ \zeta_1,\ i(b,F),\ i_c,\ \iota$ & spread over Germany; its level at 60 percent of GDP and slope above it; effective interest rate; effective rate at a zero spread; pass-through of the spread \\
$c_0,\ c_g,\ c_d,\ \Lambda$ & onset of stress: constant, real growth, deficit; the logistic function \\
$Q,\ d^{\mathrm{ref}},\ \gamma_F,\ \gamma_D,\ \eta_T$ & required saving on the payroll; deficit at the reference payroll; the requirement's shares of interest pressure and of the excess deficit; the Treasury's penalty per point of shortfall \\
$\chi,\ \bar b,\ x_s,\ c_U$ & debt overhang per point of debt above its pre-crisis level; the union's tolerance; conflict cost per log point beyond it \\
$\phi_F,\ K_F,\ K^-,\ \lambda_L,\ v_E$ & financing cost of a raise; restart cost; fixed cost of a cut; cost per log point of pay cut; value of a visible raise in an election year \\
$L_t,\ \bar B,\ \rho_S,\ \tau,\ r_A$ & outstanding cuts; admissible burden; probability of striking a cut beyond it; share of the year a struck cut stays in force; yearly probability of reversal in the aftermath \\
$\xi(\rho_S,\tau)$ & expected share of a cut beyond the admissible burden that is in force in its year, $1-\rho_S(1-\tau)$ \\
$\lambda_R$ & share of the financing cost $\phi_F$ at which the government's problem would charge the restorations a court orders; zero at the estimate (Appendix~\ref{app:estimation}) \\
$K_A,\ K_I,\ \phi_C,\ \rho_R$ & fixed cost of an interim allowance; cost of appointing a commission; financing cost of its scale; yearly probability of its report \\
$k,\ j,\ h_j,\ v,\ I_k,\ V$ & direction and size; width of a size; value of an action; inclusive value of a direction; value function \\
$u,\ c,\ g_t$ & flow payoff; size of a cut; real growth \\
$\mathcal W,\ \mathcal A(s),\ \varepsilon_{kj},\ V^C$ & post-decision value; actions available in the state $s$; taste shocks; value of a year in which a commission reports \\
$\ind^E_t,\ L^{\mathrm{rev}}_t$ & indicator of an election year; outstanding cuts reversed at the start of the year \\
$\sigma,\ \sigma_J$ & scales of the taste shocks to the directions and to the sizes within a direction \\
$v_S,\ q(x),\ P_t,\ W^p_t,\ \lambda_N,\ \phi_B$ & microfoundation (Appendix~\ref{app:model}): value of services; quality of the workforce; price level; private wage; employees' fairness loss per log point of pay cut; slope of the marginal cost of borrowing \\
\addlinespace
\multicolumn{2}{@{}l}{\textit{Transmission and the social objective}}\\
$\Omega_t,\ \rho_y$ & the dose, points of GDP; persistence of the income increment \\
$\beta,\ R^D,\ \hat y,\ \tilde b_0$ & households' discount factor; gross deposit rate; documented income; intercept of the borrowing limit \\
$s_c,\ m^u,\ \bar m,\ m_T,\ \theta_C$ & constrained share; annuity propensity; impact propensity; taxpayers' propensity; borrowing limit per unit of documented income \\
$\theta_T,\theta_B,\theta_R,\theta_F,\ s_R,\ s_D,\ r_D$ & financing shares; deposit drawdown per retail bond; payroll recycled to deposits; reserve ratio \\
$L^p,\ \kappa_L,\ \eta_K,\ \kappa_D,\ \alpha_M$ & private loans; credit displacement per unit of dose; investment per point of credit; nontraded multiplier; import content of spending \\
$w_p(z),\ c_a,\ \psi,\ \psi_Y,\ \varepsilon_W$ & private wage at skill $z$; cost of applying for a public post; elasticities of the private wage to public pay and to output; output elasticity to the private wage \\
$k_t,\ c^G_t,\ c^H_t,\ c^R_t,\ e_t,\ T_t,\ I_t$ & transmission block (Appendix~\ref{app:transmission}): percent increase of real public pay; consumption of public employees, of private hand-to-mouth households and of Ricardian households; real exchange rate; taxes; investment \\
$\varsigma_H,\ \varsigma_L,\ \varsigma_C,\ \varsigma_X,\ \varsigma_I$ & transmission block: hand-to-mouth share of private households; labor share; shares of consumption, exports and investment in GDP \\
$\varepsilon_X,\ \nu,\ \kappa_\pi,\ \varrho_e,\ \varrho_T$ & transmission block: export elasticity; cyclicality of credit; slope of the Phillips curve; debt-elastic premium; response of taxes to lagged debt \\
$\ell_t,\ \tilde y_t,\ \Delta y_t,\ w_Y$ & social flow loss; output gap; output path of the transmission block; stabilization weight \\
$z_0,\ z_S,\ z_2,\ R_t,\ \phi_t,\ K_t,\ \lambda_K$ & scale, crisis premium and curvature of the insurance term; payroll increase made by discrete acts; estimated financing coefficient of the instrument that makes it; fixed costs incurred; share of fixed costs counted \\
$\phi_A$ & financing coefficient of the automatic increases of an indexation rule (Section~\ref{sec:cf_bgd}) \\
\end{xltabular}
\par\vspace{2pt}\noindent{\footnotesize Notes: Gaps and sizes are in log points, fiscal quantities in percent of GDP, and costs and losses in the normalized units of Section~\ref{sec:objective}.\par}
\endgroup

\section{Data and measurement}
\label{app:data}
\label{app:provenance}

This appendix supports Section~\ref{sec:patterns}. It defines the series behind the three patterns and the decision
histories on which the model is estimated, and it reports the evidence that the panel of legislated acts cannot identify
the aggregate response to a revision. Units and sources for every variable are recorded in the replication archive.

\subsection{Legal pay, prices and earnings}

\paragraph{Legal pay as paid.} The yearly change of legal pay is the documented general change of the schedule, in log
points. In a decision year it is the change the government decided for the pay year, and in an implementation or measured year the documented change. Automatic indexation paid by rule (India's dearness allowance, Cyprus's
cost-of-living adjustment) is added. A one-off cut that lapses by design adds its reversal in the following year
(Spain's 2012 suppression of the December extra payment, Romania's unpaid days of 2009), and a court-ordered restoration
adds the restored share (Lithuania's coefficients and class allowances, from October 2013). Portugal's changes are annual
averages of the schedules in force, computed day by day from the dated acts (Section~\ref{sec:pattern_payment}), so that a
cut struck by the Court counts for the part of the year it was in force. Back pay for earlier years, such as Spain's
refund of the 2012 payment and Lithuania's compensation of 2016--2020, is not a change in the level of pay.

\paragraph{Prices and earnings.} \looseness=1 Consumer-price inflation is 100 times the log change of the price index, from the World Development Indicators
\citep{wdi2026} and, for the European countries, Eurostat's harmonized index \citep{eurostat2026}. Private earnings are
the hourly wages and salaries of the business economy from Eurostat's Labour Cost Index for the European countries
\citep{eurostat2026}, and the employment cost index of private wages and salaries for the United States
\citep{blseci2026}, dated as the pay formula dates it. Bangladesh and India have no
long series of private earnings, and their nominal GDP series break where the accounts were rebased. In Bangladesh,
nominal GDP rises by 22 log points in fiscal 1996, the first year of the 1995/96-base accounts, while real GDP grows by
4.5 percent. The benchmark of Figure~\ref{fig:pay_erosion} for the two countries is therefore income per head,
constructed by cumulating the growth of real GDP per head and consumer-price inflation. The model's reference for them
grows with nominal GDP per head, whose growth in Bangladesh's fiscal 1996 is replaced by the same construction.

\paragraph{The size and spacing of raises across countries.} \looseness=1 The rank correlation of Section~\ref{sec:pattern_erosion}
uses the ledger's acts and the general raises that the audit found, excluding court rulings and restorations, over
spells whose years are all documented, for the countries with at least three raises, two documented spells and two
raises of known size.

\subsection{The documented decision histories}

\looseness=1 The decision of a budget cycle is the general change that the government proposes for the coming year, net of the
year's other decisions. Deductions levied only on public employees count as cuts, and general taxes do not. Sizes are
the decided change of average pay, converted to log points. Graduated schedules without a published average are
evaluated at a representative gross pay, stated with the sources. The change in the United States is the General
Schedule increase, base plus the average locality increase. Every year except 2012 is a decision, because the statutory formula is never applied in full. In each year an alternative plan of the President or an act of Congress
sets the locality increase, and in \RsUsaBaseBelowW{} of the \RsUsaFormulaYearsW{} decision years it also holds the
base increase below the formula, which the base matches in the other \RsUsaBaseAtW{}. The freeze of 2012 was fixed by a
law of December 2010, so that year is an implementation year. The scales of Bangladesh and India are the documented changes of the reference grade at
each commission's award, net of the allowances it absorbed. Hungary's general pay base stayed frozen from 2013
while its increases were sector-specific, and Romania's pay has moved by targeted equalization steps since 2017, so in
both countries the general decisions cover only part of how pay is set. Both remain in every table.

\subsection{Fiscal data and states}

\begin{table}[!tbp]\centering
\begin{threeparttable}
\caption{Raise episodes on documented histories}
\label{tab:lpdid}
\footnotesize\setlength{\tabcolsep}{2pt}
\begin{tabular*}{\textwidth}{@{\extracolsep{\fill}}lcccccccc}
\toprule
& \multicolumn{4}{c}{Output} & \multicolumn{4}{c}{Consumption} \\
\cmidrule(lr){2-5}\cmidrule(lr){6-9}
Horizon & Raw & Adjusted & Raw & Adjusted & Raw & Adjusted & Raw & Adjusted \\
& (1) & (2) & (3) & (4) & (5) & (6) & (7) & (8) \\
\midrule
$h=-4$ & $-1.02$ &  & $-0.48$ &  & 0.53 &  & 0.31 &  \\
& (1.54) &  & (1.81) &  & (1.92) &  & (1.79) &  \\
$h=-3$ & $-1.46$ &  & $-0.78$ &  & $-0.58$ &  & $-1.30$ &  \\
& (1.08) &  & (1.07) &  & (2.19) &  & (1.12) &  \\
$h=-2$ & $-0.39$ &  & $-0.72$ &  & 1.69 &  & 0.48 &  \\
& (0.57) &  & (0.53) &  & (1.64) &  & (0.72) &  \\
\addlinespace[3pt]
$h=0$ & $-0.85$ & $-1.33$ & 0.08 & $-0.37$ & 0.82 & 1.05 & $-0.27$ & $-0.38$ \\
& (1.17) & (2.70) & (0.73) & (0.96) & (1.01) & (10.07) & (0.96) & (1.76) \\
$h=1$ & 0.29 & $-0.68$ & 1.00 & 0.11 & 0.14 & 0.59 & 0.32 & 0.10 \\
& (1.79) & (19.62) & (1.88) & (2.46) & (1.75) & (22.82) & (2.03) & (3.20) \\
$h=2$ & 3.30 & 1.84 & 0.49 & $-0.86$ & 2.29 & 2.97 & $-0.02$ & $-0.36$ \\
& (2.38) & (6.95) & (3.29) & (7.54) & (2.65) & (6.46) & (2.47) & (6.18) \\
$h=3$ & 3.97 & 2.02 & 0.76 & $-1.03$ & 4.83\sym{**} & 5.73 & 0.61 & 0.16 \\
& (2.94) & (8.98) & (2.28) & (11.01) & (1.80) & (29.02) & (3.12) & (23.54) \\
\midrule
Joint $p$-value, $h=-4,-3,-2$ & \multicolumn{2}{c}{0.960} & \multicolumn{2}{c}{0.704} & \multicolumn{2}{c}{0.905} & \multicolumn{2}{c}{0.522} \\
Raise episodes, $h=0$ & \multicolumn{2}{c}{18} & \multicolumn{2}{c}{26} & \multicolumn{2}{c}{18} & \multicolumn{2}{c}{26} \\
Clean controls, $h=0$ & \multicolumn{2}{c}{54} & \multicolumn{2}{c}{55} & \multicolumn{2}{c}{49} & \multicolumn{2}{c}{50} \\
Countries, $h=0$ & \multicolumn{2}{c}{13} & \multicolumn{2}{c}{22} & \multicolumn{2}{c}{12} & \multicolumn{2}{c}{21} \\
Country fixed effects & Yes & Yes & No & No & Yes & Yes & No & No \\
Year fixed effects & Yes & Yes & Yes & Yes & Yes & Yes & Yes & Yes \\
\bottomrule
\end{tabular*}
\begin{tablenotes}[flushleft]\footnotesize
\item Notes: The dependent variable is 100 $\times$ the change in log real GDP (columns 1--4) or in log real household consumption (columns 5--8) from $t-1$ to $t+h$. Standard errors are in parentheses. Those of the raw estimates are clustered by country, while those of the adjusted estimates, and the joint $p$-value of the pre-treatment differentials, come from a joint bootstrap of the whole path with 499 replications that resample countries. The bootstrap errors are wide because few countries connect treated episodes to controls and resampled panels often lose them. Stars use $p$-values from t tests with one fewer degree of freedom than countries for the raw estimates, and normal tests with the bootstrap standard errors for the adjusted ones. With so few connecting countries, both tests are only approximate. The events are every documented discretionary general pay act, that is, the ledger's acts and the general revisions the year-by-year audit found. Adjustments under a standing formula, such as the United States' General Schedule, are not events. Only country-years whose status the audit documents, or in which the ledger records an act, are covered. A raise episode starts in $t$ with a raise first paid in $t$ after three documented years without an act of either sign. Clean controls are documented and free of acts over $[t-3,t+h]$, so treated episodes and controls share the same three-year history. Lagged growth and inflation are controls, dated before the window for $h<0$. Country fixed effects absorb each country's trend growth in the long difference, and the specification with country and year fixed effects is the preferred one. Adjusted estimates subtract the linear extrapolation of the pre-treatment differential. Beyond $h=3$ fewer than seven countries supply clean controls, and the estimates are not reported. No cut is preceded by three documented years without an act, so the documented design cannot estimate the response to cuts (0 starts). \mbox{\sym{*} $p<0.10$}, \mbox{\sym{**} $p<0.05$}, \mbox{\sym{***} $p<0.01$}.
\end{tablenotes}
\end{threeparttable}
\end{table}

For the 27 EU countries the fiscal panel uses Eurostat's general government finance statistics in percent of GDP, namely gross debt, net lending, interest payable and compensation of employees \citep{eurostat2026}. For the United States,
Bangladesh and India it uses the IMF's Fiscal Monitor and World Economic Outlook \citep{imf2026}. The spread is the October
spread of the ten-year yield over Germany's in the year before the budget. The deficit is minus net lending, and the
primary deficit excluding the payroll is the deficit less interest and compensation. Market stress is dated when market
access was lost or threatened without a program, and a program when the country was under IMF or EU conditionality. The documented states exist for the eight European countries of the estimation and for Latvia. Under an objective
definition of stress, a spread of four points applied to every euro member, the deficit's coefficient in the onset logit is \RvOnObjD{} (Table~\ref{tab:fiscal_block}, column 4).

\subsection{The comparison group of the panel}

The ledger records an act when a statute, decree, court ruling, collective agreement with statutory force or program
conditionality changes the pay of a broad class of public employees by a discrete amount. Freezes, progression along a
scale and automatic indexation under an existing rule are not acts. By this definition the January adjustments of the
United States are changes under a standing formula and not acts, whereas the decision history above treats each of them
as a decision, because an alternative plan or an act of Congress set it. The episode design needs acts that depart from
a country's routine, and the decision model needs every year in which the government chose. A year without a recorded act is evidence of no
change only if someone looked, so a year-by-year audit applied the ledger's definition, under a written protocol blind
to macroeconomic outcomes, to every country-year that enters a comparison as a control or pre-episode year. Before
recording a documented absence, it required a source whose coverage includes the year's general pay measures, such as the budget law, an official pay table or an official pay history. The absence of news is not documentation. Across
\RprovNctry{} countries and \RprovNyears{} years, the audit finds \RprovNewGen{} general revisions that the ledger
omits, many of them routine adjustments decided in each year's budget (in Ghana, Hong Kong, Russia, Georgia, Spain, Egypt
and Japan among others), and documents the absence of a general act in \RprovNoEvent{} years. The other years are
\RprovLedger{} acts already in the ledger, 6 general acts whose net sign is not established, 20 years with only a one-off
payment or a minimum-wage change, \RprovSector{} with only sector-specific acts, \RprovIndex{} documented January
adjustments of the United States, and \RprovUnknown{} years that remain unknown. In a random sample of \RauditN{} years
outside the windows researched for the ledger, the audit finds a general revision in \RauditGen{}.

On the documented histories that remain, the response of output to the start of a raise episode is estimated by a local
projection \citep{jorda2005} with clean controls \citep{dube2025},
\begin{equation}
 y_{i,t+h}-y_{i,t-1}=\beta_h\,D_{it}+\gamma_h'X_{i,t-1}+\alpha_{i,h}+\delta_{t,h}+e_{i,t,h},
 \label{eq:lpdid}
\end{equation}
where an episode starts, $D_{it}=1$, when an act is first paid in $t$ after three documented years without one, and a
clean control is a year in which every year from $t-3$ to $t+h$ is documented and free of acts. The regressions control
for lagged growth and inflation and include country and year effects, with standard errors clustered by country. Only
\RdocNstartRaise{} raise episodes start after three documented years without an act. The regressions with country
effects use the 18 of them in countries that also supply clean controls, of which \RdocGdpRaiseNttwo{} remain at two
years, and those without country effects use 26 (Table~\ref{tab:lpdid}). The estimates,
\RdocGdpRaiseHzero{} percent on impact and \RdocGdpRaiseHtwo{} percent after two years, are therefore uninformative
about the size of the response, and no documented cut episode follows three documented years without an act. This is
why Section~\ref{sec:transmission} calibrates the transmission block rather than estimating it.

\subsection{The response of raises to inflation}
\label{app:size_erosion}

Table~\ref{tab:size_erosion} reports the regressions behind the two elasticities of Section~\ref{sec:pattern_erosion},
on the spells of its size-and-spacing statistic. The speed of erosion of a spell is mean consumer-price inflation from
the year of the previous raise to the year before the raise, the erosion accumulated when the government decides. The
median spell lasts one year, the median raise is \RsEroMedRaise{} log points and median inflation \RsEroMedInfl{}
log points a year. With country fixed effects the elasticity of the size of a raise is close to a third (column 1). It
remains within one standard error of a third with decade fixed effects (column 2) and when inflation is measured over
the three years before the spell, which uses no information from the spell itself (column 3), and leaving out one
country at a time moves it between \RsEroLooLo{} and \RsEroLooHi{}. Country by country the response differs widely, from
\RsEroCtryLo{} in Egypt to \RsEroCtryHi{} in Mauritius among the nine countries with at least eight spells. Across countries the elasticity is \RsEroBetween{} and not distinguishable from one
(column 6), as it must be if real pay is not to drift without bound. In levels the relation appears only on the
one-year spells with inflation below 30 log points, where a log point more inflation goes with a raise larger by
\RsEroPassOne{} log points \sd{\RsEroPassOneSE}. Over all spells the slope on accumulated inflation is \RsEroLevel{}
\sd{\RsEroLevelSE}.

\begin{table}[!tbp]\centering
\begin{threeparttable}
\caption{The response of the size and spacing of raises to inflation}
\label{tab:size_erosion}
\footnotesize\setlength{\tabcolsep}{2pt}
\begin{tabular*}{\textwidth}{@{\extracolsep{\fill}}l*{6}{c}}
\toprule
& (1) & (2) & (3) & (4) & (5) & (6) \\
\midrule
\multicolumn{7}{@{}l}{\textit{Panel A. Log size of the raise}} \\
\addlinespace[2pt]
Log inflation & 0.35\sym{***} & 0.32\sym{***} & 0.26\sym{**} & 0.20 & 0.57\sym{***} & 0.80\sym{***} \\
 & (0.06) & (0.05) & (0.10) & (0.25) & (0.10) & (0.22) \\
$p$-value, coefficient $=1/3$ & 0.730 & 0.831 & 0.448 & 0.602 & 0.030 & 0.038 \\
\midrule
\multicolumn{7}{@{}l}{\textit{Panel B. Log years since the previous raise}} \\
\addlinespace[2pt]
Log inflation & $-0.11$ & $-0.11$ & 0.00 & $-0.27$ & $-0.09$ & $-0.02$ \\
 & (0.09) & (0.09) & (0.06) & (0.20) & (0.05) & (0.06) \\
$p$-value, coefficient $=-2/3$ & $<0.001$ & $<0.001$ & $<0.001$ & 0.078 & $<0.001$ & $<0.001$ \\
\midrule
Country fixed effects & Yes & Yes & Yes & Yes & No & No \\
Decade fixed effects & No & Yes & No & No & No & No \\
Inflation measured before the spell & No & No & Yes & No & No & No \\
Spells of two years or more only & No & No & No & Yes & No & No \\
Country means & No & No & No & No & No & Yes \\
Countries & 30 & 30 & 30 & 9 & 32 & 32 \\
Observations & 185 & 185 & 186 & 24 & 187 & 32 \\
\bottomrule
\end{tabular*}
\begin{tablenotes}[flushleft]\footnotesize
\item Notes: Each column regresses the log of the size of a raise (Panel A) or the log of the years since the previous raise (Panel B) on the log of mean consumer-price inflation since the previous raise, so the coefficients are elasticities. They are associations across spells and not effects of inflation, because a government chooses when a spell ends. Inflation is the mean over the years from the previous raise to the year before the raise, in log points a year, and in column 3 the mean over the three years that end with the previous raise. Standard errors are in parentheses, clustered by country in columns 1--5 and robust in column 6. Stars use $p$-values from $t$ tests with one fewer degree of freedom than countries, and with the residual degrees of freedom in column 6. The $p$-value rows test the coefficient against the value that Corollary~\ref{cor:costs} gives under a fixed cost alone. An observation is a spell between consecutive raises whose years are all documented, in the countries of Section~\ref{sec:pattern_erosion}, that ends in a raise of known size and has positive mean inflation. In column 6 it is a country's mean over its spells. Raises are the ledger's acts and the general raises that the audit found, with the acts of a year summed and court rulings and restorations excluded. Country fixed effects leave out the two countries with a single spell. \mbox{\sym{*} $p<0.10$}, \mbox{\sym{**} $p<0.05$}, \mbox{\sym{***} $p<0.01$}.
\end{tablenotes}
\end{threeparttable}
\end{table}

The interval does not respond as the reset problem requires. Its elasticity is small in every column, and the value of
$-2/3$ that Corollary~\ref{cor:costs} gives under a fixed cost alone is rejected in every column but the fourth, whose
\RsEroLongN{} spells of two years or more, where the timing of a raise is a choice, are too few
to tell. Read through Corollary~\ref{cor:costs}(c), the estimate of column 1 would
put the financing share $s_F$ at zero and no higher than \RsEroSFmax{} at the 95 percent level. That reading requires
a government that can shorten the interval, which the spells of a single year rule out. For a government that revises
every year, an elasticity below one means that pay falls behind when inflation is high and catches up later. A convex
cost of financing an increase at once produces this smoothing, but so would raises that follow trend inflation or
erosion that consumer prices measure with error, and the regressions cannot tell these~apart.

\makeatletter\let\savedsecFB\@fb@secFB\let\@fb@secFB\relax\makeatother %
\section{Estimation, identification checks and inference}
\makeatletter\let\@fb@secFB\savedsecFB\makeatother
\label{app:estimation}

This appendix supports Section~\ref{sec:quant}. It reports the fiscal relations estimated without pay decisions, every
country's estimates, the fit and validation of the model country by country, the recovery of known parameters from random
starts, the confidence sets for Portugal's parameters and the weight that Portugal's government places on financing the
restorations a court orders. It first records the plan that was fixed in advance, with its departures, and two corrections
made after the tests had run.

\paragraph{The plan and its departures.} The specification of Section~\ref{sec:model}, the criterion and the estimator
of Section~\ref{sec:criterion}, the adequacy test, the tests of Section~\ref{sec:validation} and the environments of the
two experiments were written down in a plan before any result reported in this paper was computed. The plan replaced
one written for an earlier specification, which had been estimated on the same decisions and in which debt and the risk
of stress did not respond to pay. Four departures from the plan were made, each recorded before the computation that it
affects. First, the stabilization term of the social loss uses the transmission block's output path, which the plan
required and the first version of the programs had omitted. Second, the sensitivities to the excess burden, which the
plan listed, were added to the programs. Third, the experiments report their Monte Carlo errors. Fourth, every search
runs until its step falls below its convergence threshold, because the plan's caps on the number of rounds had stopped
most searches before they converged. Everything added after the tests had run is identified where it is reported, namely
the comparison of Section~\ref{sec:bridge}, the mechanism exercise of Section~\ref{sec:mechanism}, the courts of
different speed and predictability and the financed rules of Section~\ref{sec:design}, the weight on court-ordered
restorations below and the two corrections that follow.

\paragraph{The fiscal block.} \looseness=1 Table~\ref{tab:fiscal_block} reports the relations~\eqref{eq:spread} and~\eqref{eq:onset}
and the primary deficits by fiscal state, all estimated on fiscal and market data alone. The deficit, not debt, predicts
the onset of stress. Its coefficient is \RvOnD{} on the documented states of the eight European countries
and Latvia, \RvOnGreeceD{} with Greece included and \RvOnObjD{} under an objective definition of stress
(a spread of four points, applied to every euro member), while debt above 60 percent of GDP, put in place of the deficit,
enters with \RvOnHinge{}. Because only \RsOnEventsW{} onsets identify the deficit's coefficient, the model carries it as a fixed input. The relations do not rest on Hungary and Romania. Without them the deficit's
coefficient is \RvRobOnD{} \sd{\RvRobOnDse} on \RvRobOnEvents{} onsets. The pass-through is the exception. Without the two
countries it is \RvRobMu{} \sd{\RvRobMuSe}, about half the estimate the model uses, because spreads reach the effective
rate faster outside the euro area.

\begin{table}[!tbp]\centering
\begin{threeparttable}
\caption{The fiscal block}
\label{tab:fiscal_block}
\footnotesize\setlength{\tabcolsep}{2pt}
\begin{tabular*}{\textwidth}{@{\extracolsep{\fill}}lcccc}
\toprule
 & (1) & (2) & (3) & (4) \\
\midrule
\multicolumn{5}{@{}l}{\textit{Panel A. The October spread over Germany (percentage points)}} \\
\addlinespace[2pt]
Debt above 60 percent of GDP, $(b-60)^+$ & 0.017\sym{***} & 0.016\sym{***} & 0.031\sym{***} & \\
 & (0.002) & (0.002) & (0.007) & \\
Stress or program & 2.429\sym{***} & 1.965\sym{***} &  & \\
 & (0.664) & (0.571) &  & \\
\quad $\times\,(b-60)^+$ & 0.010 & 0.035\sym{***} &  & \\
 & (0.016) & (0.010) &  & \\
Outside the euro area & 2.421\sym{***} & 2.614\sym{***} & 1.473\sym{***} & \\
 & (0.606) & (0.568) & (0.459) & \\
\quad $\times\,(b-60)^+$ & 0.051 & 0.016 & 0.037 & \\
 & (0.035) & (0.040) & (0.048) & \\
Constant & 0.300\sym{**} & 0.258\sym{**} & 0.315\sym{*} & \\
 & (0.123) & (0.112) & (0.164) & \\
Country-years & 211 & 236 & 593 & \\
$R^2$ & 0.55 & & & \\
\midrule
\multicolumn{5}{@{}l}{\textit{Panel B. The onset of stress or a program (logit)}} \\
\addlinespace[2pt]
Deficit, previous year (percent of GDP) & 0.329\sym{**} &  & 0.310\sym{***} & 0.254\sym{***} \\
 & (0.135) &  & (0.104) & (0.056) \\
Real growth, previous year & $-0.089$ & $-0.206$\sym{*} & $-0.092$ & $-0.311$\sym{***} \\
 & (0.117) & (0.107) & (0.109) & (0.080) \\
Debt above 60 percent of GDP, $(b-60)^+$ &  & $-0.021$\sym{**} &  &  \\
 &  & (0.010) &  &  \\
Constant & $-4.141$\sym{***} & $-2.244$\sym{***} & $-4.194$\sym{***} & $-6.056$\sym{***} \\
 & (0.810) & (0.113) & (0.713) & (0.764) \\
Onsets & 9 & 9 & 10 & 6 \\
Country-years at risk & 181 & 181 & 196 & 330 \\
\midrule
\multicolumn{5}{@{}l}{\textit{Panel C. Interest and the primary deficit}} \\
\addlinespace[2pt]
Pass-through of the spread to the effective rate & 0.092\sym{*} & 0.042 & & \\
 & (0.046) & (0.051) & & \\
Primary deficit excluding the payroll: stress & 5.70\sym{***} & 6.23\sym{**} & & \\
 & (0.75) & (2.23) & & \\
\quad program & 2.20\sym{***} & 2.32\sym{***} & & \\
 & (0.67) & (0.64) & & \\
\quad aftermath & 1.70\sym{***} & 2.01\sym{*} & & \\
 & (0.58) & (0.88) & & \\
\bottomrule
\end{tabular*}
\begin{tablenotes}[flushleft]\footnotesize
\item Notes: The nine documented countries are the eight European countries of the estimation and Latvia. Panel A reports OLS regressions of the October spread over Germany before the budget year, in percentage points, on debt at the start of the budget year. Panel B reports logits of entering stress or a program from a normal or aftermath year. Panel C reports the regression of the change in the effective interest rate on the change in the spread, outside programs (1) and in all years (2), and the primary deficit excluding the payroll relative to normal years, in percent of GDP, from country fixed-effects median (1) and mean (2) regressions. Standard errors, in parentheses, are clustered by country, except in the median regression (default standard errors). Stars use t tests with one fewer degree of freedom than countries (clustered OLS), the normal distribution (logits) and t tests with the residual degrees of freedom (median regression). In Panel A, column (1) is the model's schedule on the nine documented countries, 2000--2024, column (2) adds Greece, and column (3) uses the 26 EU countries other than Germany with one schedule, because the other countries' fiscal states are not documented. In Panel B, column (1) is the model's onset on the nine documented countries, column (2) puts the debt hinge in place of the deficit, column (3) adds Greece, and column (4) covers every euro member, with stress defined by an October spread of at least four points. \mbox{\sym{*} $p<0.10$}, \mbox{\sym{**} $p<0.05$}, \mbox{\sym{***} $p<0.01$}.
\end{tablenotes}
\end{threeparttable}
\end{table}

\paragraph{Two corrections after the tests.} \looseness=1 After every test and counterfactual had been computed, two coding errors
were found while the comparison of Section~\ref{sec:bridge} was being built, and both were corrected. The first was in the documented pay paths, where the coding of a legislated reversal reversed every cut outstanding after the year's decision.
In Spain in 2013 it therefore reversed the 2010 cut, which stayed in force, together with the suppression of the December 2012 extra payment, which lapsed. In Romania in 2010 it reversed the year's own new cut of July 2010 together with the
unpaid leave of 2009, so the documented path never contained Romania's 2010 cut. A legislated reversal now applies to the
cuts standing at the start of its year and, where the law ended one measure only, to that measure. The second was in an input series. Bangladesh's growth of nominal income per head in 1996, which joins two national-accounts bases without a
splice, is set to the growth of real GDP per head plus consumer-price inflation, and Bangladesh's macroeconomic process
is re-estimated with it. At the earlier estimates, the corrections change the criterion of Spain, Romania and Bangladesh
and leave every other country's criterion unchanged to six decimals. The three countries were re-estimated with the same
search and starts, the pooled round was repeated with every country, and every test and counterfactual that involves them
was recomputed. Portugal's inference and the experiments with its courts use Portugal's likelihood and histories alone,
which the corrections do not change. The pooled round again left the two common ratios at their starting values. The corrections change one verdict of the adequacy test. Romania, which failed at the earlier estimates, passes once its
documented path contains the cut of 2010. Spain's overhang, estimated at 0.16 before, is now zero, the forward-looking
government now fits Spain's decisions better than the myopic one, and Spain's documented real pay at the end of its
window now lies just above the model's interval. In Bangladesh the documented size and spacing of the scales now lie
within the range of the simulated histories, the profile region of the reference share no longer reaches its upper
end, and half indexation financed at the estimated allowance coefficient leaves the loss without fixed costs almost
unchanged instead of lowering it. Elsewhere the numbers change but the conclusions do not.

\paragraph{The estimates.} Table~\ref{tab:v2_estimates} reports each country's estimates, its fit and the adequacy test.
Several estimates lie at a bound of the search range, among them the restart cost and the cost of a cut in countries
that rarely cut. In those countries the decisions bound the cost from below but not from above, and the counterfactuals
do not depend on it. The requirement's coefficients $\gamma_F$ and $\gamma_D$ are searched between 0 and 1.5. Under the
derivation of Appendix~\ref{app:model} they are shares, which cannot exceed one, and the interest coefficient is below
one in Portugal, Cyprus, Italy and Hungary. It is at the upper end of its range in Ireland, Romania and Lithuania and
close to it in Spain. The crisis cuts of these countries were larger than a share of the extra interest bill can account
for, so there the coefficient scales the requirement and is not a share. Steps along \RvAttainDirections{} random
directions from each country's estimate raise the criterion only in \RvAttainImprovingList, by \RvAttainGainIRL{} and
\RvAttainGainUSA{} points of log likelihood. The two common ratios that are not searched are profiled. At half and at twice the starting value of
the value of a visible raise in an election year, the countries' total log likelihood changes by \RvProfileBElo{} and
\RvProfileBEhi{}, so that value is not identified. At twice the starting value of the cost of conflict with the union the
total changes by \RvProfileKhi{}, whereas at half of it the problem has no solution for \RvProfileKloInfeasible{} and the
other countries' total rises by \RvProfileKloRest{}, so the cost of conflict is bounded above but not below.

\begin{table}[!tbp]\centering
\begin{threeparttable}
\caption{Estimates of the decision problem}
\label{tab:v2_estimates}
\footnotesize\setlength{\tabcolsep}{2pt}
\begin{tabular*}{\textwidth}{@{\extracolsep{\fill}}lccccccccccc}
\toprule
 & PRT & IRL & ESP & CYP & ITA & ROU & HUN & LTU & USA & BGD & IND \\
\midrule
$\gamma_F$ & 0.388 & 1.500 & 1.489 & 0.331 & 0.499 & 1.500 & 0.635 & 1.500 &  &  &  \\
$\gamma_D$ & 0.030 & 0.963 & 0.324 & 0.000 & 0.041 & 1.463 & 0.923 & 0.549 &  &  &  \\
$\chi$ & 0.34 & 0.88 & 0.00 & 1.16 & 0.00 & 0.00 & 0.00 & 0.00 &  &  &  \\
$s_w$ & 0.00 & 0.16 & 0.64 & 0.12 & 0.38 & 0.30 & 0.07 & 0.00 & 0.59 & 0.22 & 0.81 \\
$\lambda_L$ & 3.15 & 11.74 & 12.44 & 5.31 & 15.00 & 0.00 & 0.00 & 15.00 & 15.00 & 15.00 & 8.81 \\
$\phi_F$ & 1.285 & 0.812 & 0.246 & 0.035 & 0.151 & 0.488 & 2.000 & 0.022 & 2.000 & 0.271 & 0.277 \\
$K^-$ & 0.22 & 0.10 & 7.20 & 22.17 & 50.00 & 28.74 & 29.44 & 50.00 & 50.00 & 50.00 & 50.00 \\
$K_F$ ($K_A$) & 1.9 & 31.9 & 10.7 & 7.4 & 15.4 & 0.0 & 13.5 & 60.0 & 53.7 & 60.0 & 16.6 \\
$K_I$ &  &  &  &  &  &  &  &  &  & 101.2 & 26.8 \\
$\phi_C$ &  &  &  &  &  &  &  &  &  & 0.0002 & 0.0008 \\
$\sigma$ & 0.46 & 4.02 & 2.09 & 7.38 & 6.25 & 50.00 & 27.81 & 43.18 & 16.66 & 14.42 & 3.69 \\
$\sigma_J/\sigma$ & 1.00 & 0.91 & 0.47 & 0.94 & 0.11 & 0.34 & 0.05 & 0.71 & 0.19 & 0.94 & 0.12 \\
\midrule
Log likelihood & $-13.70$ & $-19.53$ & $-17.26$ & $-21.56$ & $-21.03$ & $-36.37$ & $-8.95$ & $-42.61$ & $-45.55$ & $-36.45$ & $-15.20$ \\
Decision years & 17 & 11 & 14 & 11 & 17 & 12 & 6 & 17 & 29 & 26 & 34 \\
Direction match & 1.00 & 0.91 & 0.79 & 0.82 & 0.82 & 0.67 & 0.33 & 0.59 & 0.93 & 0.77 & 0.91 \\
Adequacy test & Passes & Passes & Passes & Passes & Passes & Passes & Fails & Fails & Passes & Passes & Passes \\
Optima within 1 point & 1 & 1 & 1 & 1 & 1 & 1 & 1 & 1 & 1 & 1 & 1 \\
\bottomrule
\end{tabular*}
\begin{tablenotes}[flushleft]\footnotesize
\item Notes: The columns are Portugal (PRT), Ireland (IRL), Spain (ESP), Cyprus (CYP), Italy (ITA), Romania (ROU), Hungary (HUN), Lithuania (LTU), the United States (USA), Bangladesh (BGD) and India (IND). Each column is the country's best estimate after the search over the common ratios, evaluated again on the grid and at the tolerance used for reporting. Costs are in the normalized units of Section~\ref{sec:objective}, the payoff's percent of a year's payroll at the reference, and a blank cell means that the country does not have the component. The common ratios are $x^\star=7.50$ and $\eta_T/a=300$, and $v_E/a=2$ and $c_U/a=45$ are profiled. The direction match is the share of decision years in which the model's most likely direction is the documented one. The adequacy test is that of Section~\ref{sec:criterion}. The last row counts the distinct optima within one point of log likelihood of the best. The search ranges are $[0,1.5]$ for $\gamma_F$ and $\gamma_D$, $[0,3]$ for $\chi$, $[0,1]$ for $s_w$, $[0,15]$ for $\lambda_L$, $[0.001,2]$ for $\phi_F$, $[0.1,50]$ for $K^-$, $[0,60]$ for $K_F$, $[0,300]$ for $K_I$, $[0.0002,0.2]$ for $\phi_C$, $[0.02,50]$ for $\sigma$ and $[0.05,1]$ for $\sigma_J/\sigma$, and an estimate equal to an end of its range is at a bound.
\end{tablenotes}
\end{threeparttable}
\end{table}

\paragraph{Fit and validation by country.} Table~\ref{tab:country} collects the evidence of Sections~\ref{sec:criterion}
to~\ref{sec:validation} country by country. The countries that pass the adequacy test also match most of the statistics
of the data--model comparison. Of the two that fail it, Lithuania matches one of five, and Hungary, whose six decision
years leave wide intervals, matches four. Out of sample, the two tests differ. Predicted from the other countries' estimates, the model beats constant probabilities in most annual countries, but re-estimated without a block of years
it predicts that block worse than the country's own direction frequencies in most countries. The last column shows that
the myopic government fits as well as the forward-looking one everywhere except in \RsForwardBetterList.

\begin{table}[!tbp]\centering
\begin{threeparttable}
\caption{Fit and validation by country}
\label{tab:country}
\footnotesize\setlength{\tabcolsep}{2pt}
\begin{tabular*}{\textwidth}{@{\extracolsep{\fill}}l*{7}{c}}
\toprule
& \multicolumn{3}{c}{Fit of the decisions} & Data and model & \multicolumn{3}{c}{Out of sample and a competing model} \\
\cmidrule(lr){2-4}\cmidrule(lr){5-5}\cmidrule(lr){6-8}
& Decision & Direction & Adequacy & Statistics & Transfer & Held-out & Forward \\
& years & match & test & in band & & years & less myopic \\
& (1) & (2) & (3) & (4) & (5) & (6) & (7) \\
\midrule
\multicolumn{8}{@{}l}{\textit{Annual pay setting}} \\
Portugal & 17 & 1.00 & Passes & 5 of 5 & 9.5 & $-11.6$ & 6.9 \\
Ireland & 11 & 0.91 & Passes & 4 of 4 & 7.8 & $-1.4$ & $-0.3$ \\
Spain & 14 & 0.79 & Passes & 4 of 5 & 5.2 & $-4.0$ & 1.2 \\
Cyprus & 11 & 0.82 & Passes & 4 of 5 & 2.4 & 2.7 & $-0.5$ \\
Italy & 17 & 0.82 & Passes & 5 of 5 & 1.3 & $-3.7$ & $-0.6$ \\
United States & 29 & 0.93 & Passes & 5 of 5 & 11.3 & $-1.6$ & $-0.4$ \\
Romania & 12 & 0.67 & Passes & 4 of 5 & $-1.5$ & $-2.1$ & $-1.1$ \\
Hungary & 6 & 0.33 & Fails & 4 of 5 & $-3.7$ & $-22.3$ & $-0.5$ \\
Lithuania & 17 & 0.59 & Fails & 1 of 5 & $-5.8$ & 3.5 & $-7.6$ \\
\multicolumn{8}{@{}l}{\textit{Pay commissions}} \\
Bangladesh & 26 & 0.77 & Passes & 5 of 5 & -- & $-1.6$ & $-4.2$ \\
India & 34 & 0.91 & Passes & 3 of 3 & -- & 6.3 & $-1.3$ \\
\bottomrule
\end{tabular*}
\begin{tablenotes}[flushleft]\footnotesize
\item Notes: Columns 1--3 give the inputs and the verdict of the adequacy test of Section~\ref{sec:criterion}, namely the number of decision years, the share of them in which the model's most likely direction is the documented one, and whether the country passes. Column 4 counts the statistics of Table~\ref{tab:moments} whose documented value lies between the 5th and 95th percentiles of the same statistic over 500 histories simulated from the estimated model. For the annual countries these are the five statistics of Panel~D, and for Bangladesh and India those of Panels~B and~C. Column 5 is the log score of the documented directions predicted from the other annual countries' estimates alone, less the score of constant probabilities estimated on the same countries (Section~\ref{sec:validation}), and it is computed for the annual countries only. Column 6 is the log score of the held-out blocks under the model re-estimated without them, less the score of the country's own direction frequencies. Column 7 is the log likelihood of the forward-looking model less that of the myopic government, each estimated on the country's decisions at the calibrated common ratios, so a positive value favors the forward-looking model. Columns 5--7 are in points of log probability.
\end{tablenotes}
\end{threeparttable}
\end{table}

\paragraph{Recovery from random starts.} Histories are simulated from Portugal's estimate, from a perturbed version of
it, and from the estimates of Ireland, Spain and the United States, and each is re-estimated by the full search from
random starts, none of them at the true parameters. In \RvRecoveryReached{} of the \RvRecoveryN{} histories the search
attains at least the log likelihood of the true parameters, including both of Portugal's, which was the condition fixed
in advance for reporting inference. It falls short in both histories of the perturbed Portugal ($\gamma_F$ ten percent
higher and $\chi$ twenty percent lower), where it settles on much larger taste shocks. Attaining the likelihood does not pin down each parameter. In Portugal's two histories $\gamma_F$ is recovered at \RvRecPRTgamFA{} and \RvRecPRTgamFB{}
against a true value of \RvRecPRTgamFTrue{}, and $\chi$ at \RvRecPRTchiA{} and \RvRecPRTchiB{} against \RvRecPRTchiTrue{}.

\paragraph{Inference for Portugal.} The confidence sets for $\gamma_F$, $\chi$ and $x^\star$ invert likelihood-ratio
statistics whose critical values come from a parametric bootstrap of \RvVFourB{} histories at Portugal's estimate, each
re-estimated by the full search with the parameter free and with it held at its true value (Table~\ref{tab:v2_v4}). The
critical values for $\gamma_F$ and $\chi$, \RvVFourCutGamF{} and \RvVFourCutChi{}, lie far above the 3.84 of the
$\chi^2(1)$ distribution, although their median statistics are only \RvVFourMedGamF{} and \RvVFourMedChi{}. That for $x^\star$, \RvVFourCutX{}, lies below it, with a median of \RvVFourMedX{}. Part of the long upper tail comes from the search
itself. In the replicates with large statistics the restricted estimate has much larger taste shocks than the best fit,
so the bootstrap calibrates the statistic as the search computes it. At the $\chi^2(1)$ critical value the sets would be
$[\RvVFourLoCsqGamF, \RvVFourHiCsqGamF]$ for $\gamma_F$ and $[\RvVFourLoCsqChi, \RvVFourHiCsqChi]$ for $\chi$. The sets reported are inner bounds, since at 11 of the 16 profile points the restricted search stopped at its maximum number of rounds, and a longer
search could only raise the likelihood there and widen the sets. For $\gamma_F$ the capped points lie far outside the
set, but for $\chi$ the lowest points are close to its critical value, so the set may extend to zero.

\begin{table}[!tbp]\centering
\begin{threeparttable}
\caption{Inference for Portugal: bootstrap critical values and confidence sets}
\label{tab:v2_v4}
\footnotesize\setlength{\tabcolsep}{2pt}
\begin{tabular*}{\textwidth}{@{\extracolsep{\fill}}lccccc}
\toprule
& & \multicolumn{2}{c}{Bootstrap statistic} & \multicolumn{2}{c}{95 percent confidence set} \\
\cmidrule(lr){3-4}\cmidrule(lr){5-6}
Parameter & Estimate & Critical value & Median & Bootstrap & $\chi^2(1)$ \\
\midrule
$\gamma_F$ & 0.388 & 8.59 & 1.10 & [0.306, 0.519] & [0.351, 0.447] \\
$\chi$ & 0.34 & 14.61 & 0.08 & [0.17, 0.86$^{a}$] & [0.24, 0.63] \\
$x^\star$ & 7.50 & 2.57 & 0.24 & [5.00$^{a}$, 10.00$^{a}$] & [5.00$^{a}$, 10.00$^{a}$] \\
\bottomrule
\end{tabular*}
\begin{tablenotes}[flushleft]\footnotesize
\item Notes: The bootstrap simulates 19 histories, on the coarser of the search's two grids, at Portugal's estimate and re-estimates each without restriction and with the parameter held at its true value. The critical value is the largest of the 19 replicates' likelihood-ratio statistics, the 95 percent critical value, and the median is that of the replicates' statistics. For $x^\star$, the statistic compares the best of 5, 6.5, 8.5 and 10 with the true 7.5. In every replicate the search attains at least the log likelihood of the true parameters. A 95 percent confidence set contains the values whose profile likelihood-ratio statistic on the reporting grid, against the best of the estimate and the profile points, is at most the bootstrap or the $\chi^2(1)$ critical value, with the ends interpolated linearly between profile points. The profile points of $\gamma_F$ and $\chi$ are at 0, 0.25, 0.5, 0.75, 1.5 and 2.5 times the estimate. A superscript $a$ marks an end of a set that reaches the end of the profiled range. At 11 of the 16 profile points the search stopped at its maximum number of rounds before converging. A longer search could only raise the likelihood at these points, so the sets could only widen.
\end{tablenotes}
\end{threeparttable}
\end{table}

\paragraph{Transfer and held-out years.} \looseness=1 Each annual country's decisions are predicted from the distribution of the
other eight countries' estimates and scored against benchmarks estimated on the same eight countries, namely constant probabilities, the legal default of each year's institutions, a logit on the fiscal state, and a logit with all of the
model's state information, the last two shrunk toward a default prior \citep{gelman2008}. The shrunk logits score
\RvTransferTotFiscal{} and \RvTransferTotState{}, against the model's \RvTransferTotModel{}. Hungary and Lithuania
account for \RvTransferShortHL{} points of the model's shortfall in log score against the first of them, and in the other
seven countries the model leads it by \RvTransferLeadRest{}. For the held-out years, the model is re-estimated from new
random starts without each block of about four decision years, or without each spell between scales in the commission
countries, and scored on that block. Of the \RvHeldoutYears{} held-out years, \RvHeldoutMiss{} receive a probability below five percent.

\begin{table}[!tbp]\centering
\begin{threeparttable}
\caption{Portugal's profile likelihood in the weight on court-ordered restorations}
\label{tab:restoration_weight}
\footnotesize\setlength{\tabcolsep}{2pt}
\begin{tabular*}{\textwidth}{@{\extracolsep{\fill}}l*{8}{c}}
\toprule
& & & \multicolumn{6}{c}{Estimates} \\
\cmidrule(lr){4-9}
\makecell{Weight\\$\lambda_R$} & \makecell{Log\\likelihood} & \makecell{Likelihood-ratio\\statistic} & $\phi_F$ & $K_F$ & $\lambda_L$ & $K^-$ & $\sigma$ & $\chi$ \\
& (1) & (2) & (3) & (4) & (5) & (6) & (7) & (8) \\
\midrule
0 (estimate) & $-13.70$ & 0.00 & 1.28 & 1.87 & 3.15 & 0.22 & 0.46 & 0.34 \\
0.05 & $-15.51$ & 3.61 & 1.30 & 2.41 & 2.35 & 1.22 & 0.61 & 0.33 \\
0.1 & $-15.86$ & 4.31 & 0.87 & 0.00 & 1.89 & 0.65 & 0.43 & 0.63 \\
0.25 & $-16.25$ & 5.09 & 0.34 & 0.00 & 1.81 & 0.90 & 0.56 & 0.83 \\
1 & $-18.21$ & 9.02 & 0.04 & 0.00 & 2.52 & 0.35 & 0.45 & 0.66 \\
\bottomrule
\end{tabular*}
\begin{tablenotes}[flushleft]\footnotesize
\item Notes: Entries are Portugal's log likelihood and estimates when the government's problem charges the restorations a court orders at the share $\lambda_R$ of the financing cost $\phi_F$ of its own raises, with Portugal's ten parameters re-estimated at each weight by the search that produced the estimates, from the same starts and from the estimate. The row for $\lambda_R=0$ is the estimate of Table~\ref{tab:v2_estimates}. Column 2 is twice the difference between the largest log likelihood and the row's, and the weights whose statistic is at most 3.84 form the 95 percent set. Columns 3 to 8 are the financing cost of a raise, the restart cost, the cost per log point of pay cut, the fixed cost of a cut, the scale of the taste shocks to the directions and the debt overhang.
\end{tablenotes}
\end{threeparttable}
\end{table}

\paragraph{The weight on court-ordered restorations.} The government's problem charges the cost of financing an
increase at once only on the increases that the government decides. A restoration that a court orders is also an
increase, and the social loss charges it, but in the government's own problem it costs nothing. Whether a government
weighs the cost of increases it did not choose is a question about its behavior, so Portugal's decisions are asked to
answer it. The government's problem is generalized to charge court-ordered restorations at the share $\lambda_R$ of the
financing cost $\phi_F$ of its own raises. Two restorations are charged: the undoing, at the start of the next year, of
the part of a struck cut that was in force, and the reversal of cuts that stood. At $\lambda_R=0$ the problem is the one estimated in the paper, and at $\lambda_R=1$ the government pays for a restoration it is ordered to make exactly as it pays
for a raise it chooses. Portugal's ten parameters are re-estimated at $\lambda_R\in\{0.05,0.1,0.25,1\}$ by the search
that produced the estimates, from the same starts and from the estimate itself (Table~\ref{tab:restoration_weight}).

\looseness=1 The decisions choose $\lambda_R=0$. The log likelihood falls at every positive weight, the 95 percent set ends at
\RwUpper{}, interpolated between \RwGridIn{} and \RwGridOut{}, and a government that charged itself the full cost is
rejected, with a likelihood-ratio statistic of \RwLROne{} ($p=\RwPOne$). Two checks guard the comparison. A search may
stop short of the best fit at its weight, which would favor some weights over others. Evaluating at every weight each of the
\RwCrossNWord{} optima that the searches found leaves every point of the profile unchanged, however, and none of them fits better than
the estimate when restorations are free. The grid also skips the weights between 0 and 0.05, where the profile could rise. It does not, because at the estimate the log likelihood already falls by \RwSlopeDrop{} at a weight of
\RwSlopeW{}, and by the envelope theorem the profile's slope at zero is the slope of the estimate's criterion. The
weight cannot be negative, so its estimate lies on the boundary of the parameter space, where the critical value of 3.84
errs toward a wider set \citep{andrews2001}.

\looseness=1 What identifies the weight is the cut of 2012, the largest cut that Portugal's government proposed and one that the
Court struck. A government that must pay to restore a struck cut gains much less from proposing it, so at the estimate a
weight of 0.05 alone lowers the log likelihood of the 2012 decision by \RwTwelveLossFive{}. Re-estimation recovers most
of that fit, but only at a cost elsewhere. At small weights it moves the cost of a cut from its size to its fixed part,
raising the fixed cost from \RwKdnZero{} to \RwKdnFive{} and lowering the cost per log point from \RwLamLZero{} to
\RwLamLFive{}, and the cut of 2013 then fits worse. At large weights it makes financing almost free, with $\phi_F$ at
\RwPhiOne{} and no restart cost, so that restorations cost little. The raises of 2008 and 2022--2024 then fit worse,
by \RwRaisesLossOne{} in log likelihood. The raises pin down the financing cost, and given that cost the cuts of 2012 and
2013 pin down the weight. Inside the set the financing cost and the overhang hardly move (\RwPhiFive{} and \RwChiFive{}
at a weight of 0.05, against \RwPhiZero{} and \RwChiZero{} at the estimate), while the split of a cut's cost between its
fixed part and its size moves as just described. Appendix~\ref{app:experiments} recomputes the court experiments at a
weight of 0.05.

\section{The experiments: design, sensitivity and uncertainty}
\label{app:experiments}
\label{app:est_cf}
\label{app:est_rules}

This appendix supports Section~\ref{sec:design}. It explains how the two experiments are simulated, reports the
comparison of Portugal's legal environments that was fixed in advance together with its sensitivities, records the
originally specified measure of the burden on employees and its limitation, recomputes the courts for a government that
weighs the cost of the restorations they order, and reports the boundaries of financed indexation under parameter
uncertainty. The courts of different speed and predictability, their recomputation for such a government, and the
financed rules were specified after the paper's estimates and tests, but before they were computed.

\paragraph{Portugal's histories.} Each environment is simulated over the budgets of 2010--2024 from the documented state
of the 2010 budget. The documented macroeconomic and electoral history and the external shocks of the fiscal accounts
are held fixed, while the payroll, interest, the deficit and debt, the spread's schedule component, the requirement and
the fiscal state respond to the path of pay. Along the documented path of pay, the simulation reproduces the documented
fiscal states. After 2024 the loss is the expected loss in the model's own world from the end-2024 state. In that world
Portugal's primary balance in normal years is 2.7 points of GDP tighter than in its documented normal years, so after 2024
debt falls faster than it did in those years.

\paragraph{The legal environments fixed in advance.} The plan compared three environments at Portugal's estimate, namely the Court as documented, striking in the years in which it struck and with the documented restorations, no reversal, and strict reversal, in which any cut beyond the admissible burden is struck with certainty in every crisis year. Counting
the future, the Court's reversal of cuts raised the social loss by \RvCFsensBaseObsMinusNoRev{} a year against a Court
that reversed nothing (\RvCFprtObslosstotalOne{} against \RvCFprtNoRevlosstotalOne{}), through the financing of the
restorations, whose term is \RvCFprtObstotalfinancing{} with reversal against \RvCFprtNoRevtotalfinancing{} without it.
These losses count the expected loss after 2024, whereas those of Table~\ref{tab:courts} average over 2010--2024 alone,
so the two are not comparable in level.
The ranking would turn only if the financing term were weighted at less than \RvCFprtRevObsNoRevOne{} of its value, and
every sensitivity fixed in advance leaves it in place, with reversal raising the loss by roughly four to six units a year.
With every country re-estimated at an excess burden of 0 or 1 instead of 0.3, reversal raises the loss by \RvCFsensLamZeroObsMinusNoRev{} and \RvCFsensLamOneObsMinusNoRev{} a year. With the share
of a struck cut in force halved or doubled it raises the loss by \RvCFsensInHalfObsMinusNoRev{} and
\RvCFsensInDoubleObsMinusNoRev{}, and with the documented restorations made immediate or delayed by two years, by
\RvCFsensRestNowObsMinusNoRev{} and \RvCFsensRestLateObsMinusNoRev{}. With the union's tolerance held at its normal level
in the aftermath it raises the loss by \RvCFsensTolFixedObsMinusNoRev{}. Portugal's log likelihood then changes by
\RvUdiffPRT{} and its overhang rises from \RvUchiEstPRT{} to \RvUchiPRT{}, so the size of the overhang is not separately
identified from the union's tolerance. The documented and the strict Court differ by less than half a point, too little
to rank them.

\paragraph{Courts of different speed and predictability.} A court strikes a cut that takes the outstanding cuts above the
admissible burden in stress or a program with probability $\rho_S\in\{0.5,0.75,0.9,1\}$, which the government knows. A
struck cut stays in force for the calibrated court's shares $\tau$ of the year scaled by a delay
$f\in\{0,0.25,0.5,1,2\}$, at most for the whole year, so that $f=1$ is the calibrated court and $f=0$ is review before a
cut takes effect. The government re-solves its problem
under each of the twenty courts and under no court, and the histories of 2010--2024 draw the strikes 400 times along the
policy without taste shocks. In the model's own world, where crises occupy about six percent of years, the courts hardly
differ, and the two statements fixed in advance for that world, that faster review and that certain review would each
reduce the number of struck cuts, both fail.

\paragraph{The specified measure of the burden on employees.} The plan measured the burden on employees by the excessive cut imposed, namely the part above the admissible burden of the cuts that stand in crisis years, counted in the year
in which they are imposed, plus the in-force part of struck cuts. That measure counts a standing cut once, although the
cut stays in force until it is reversed. By that measure the calibrated court imposes \RsCtEstImposed{} log points against
\RsCtNoneImposed{} with no court, and lax courts appear to protect employees by letting more cuts stand. The corrected
measure of Table~\ref{tab:courts} sums the cuts in force over every year in which they are in force. Its excessive part,
the cuts in force above the admissible burden in crisis years, is \RsCtEstCumEx{} under the calibrated court,
\RsCtNoneCumEx{} with no court, \RsCtBeforeCumEx{} with review before a cut takes effect and \RsCtCertainBeforeCumEx{} with
certain review before it. The corrected measure is computed from the same simulated histories, and it reproduces every
other outcome of those histories. It was added after the experiment had been computed.

\paragraph{The social loss by term.} Table~\ref{tab:court_terms} splits the social loss of each court of
Table~\ref{tab:courts} into the terms of equation~\eqref{eq:socialloss}. Courts that rule after a cut has taken effect
raise the loss through the financing of the restorations, and courts that rule before raise it through the Treasury's
penalty, while the stabilization term, which carries the transmission block's output response, hardly differs across
courts.

\begin{table}[!tbp]\centering
\begin{threeparttable}
\caption{Judicial review in Portugal: the social loss by term}
\label{tab:court_terms}
\footnotesize\setlength{\tabcolsep}{2pt}
\begin{tabular*}{\textwidth}{@{\extracolsep{\fill}}l*{7}{c}}
\toprule
& \multicolumn{6}{c}{Terms of the social loss} & \\
\cmidrule(lr){2-7}
& Misalignment & Stabilization & \makecell{Treasury's\\penalty} & Insurance & Financing & \makecell{Fixed\\costs} & Total \\
Court & (1) & (2) & (3) & (4) & (5) & (6) & (7) \\
\midrule
No court & 3.34 & 4.70 & 14.39 & 1.29 & 1.92 & 0.18 & 25.83 \\
Calibrated court ($\rho_S=0.9$) & 3.42 & 4.84 & 9.93 & 1.42 & 17.16 & 0.27 & 37.04 \\
Review before effect ($\rho_S=0.9$) & 2.84 & 4.71 & 17.20 & 1.15 & 2.98 & 0.19 & 29.06 \\
Certain review ($\rho_S=1$) & 3.25 & 4.78 & 9.11 & 1.38 & 19.30 & 0.27 & 38.11 \\
Certain review before effect ($\rho_S=1$) & 2.53 & 4.62 & 18.96 & 1.04 & 1.89 & 0.12 & 29.17 \\
\bottomrule
\end{tabular*}
\begin{tablenotes}[flushleft]\footnotesize
\item Notes: The social loss of Table~\ref{tab:courts}, column 7, by the terms of equation~\eqref{eq:socialloss}, in normalized loss units a year. The terms are the misalignment of pay from its reference (1), the stabilization loss from the output gap and the transmission block's response to the path of pay (2), the Treasury's penalty on a saving short of its requirement (3), the insurance cost of employees who consume their pay (4), the cost of financing increases at once, restorations included (5), and the fixed costs (6). The courts, histories and discounting are those of Table~\ref{tab:courts}. Across all 21 environments of the experiment, the 20 courts of Section~\ref{sec:cf_prt} and no court, the stabilization term ranges from 4.62 to 5.04, the financing term from 1.89 to 25.08 and the Treasury's penalty from 7.04 to 18.96, and every court's total exceeds the loss without a court by at least 2.29.
\end{tablenotes}
\end{threeparttable}
\end{table}

\paragraph{Courts when the government weighs the restorations.} The comparisons of Section~\ref{sec:cf_prt} rest on the
estimate that the government does not weigh the cost of financing the restorations a court orders
(Appendix~\ref{app:estimation}). Table~\ref{tab:courts_edge} recomputes the courts of Table~\ref{tab:courts} at a weight
of 0.05, the largest weight of the grid inside the 95 percent set, which ends at \RwUpper{}, with Portugal's parameters
re-estimated at that weight. Of the \RwEdgeNWord{} comparisons on which the section's statements rest, \RwEdgeHoldWord{}
keep their sign. The calibrated court still shortens how long cuts last without raising pay in crisis years, review before a
cut takes effect is still cheaper than review after it, and every court still costs more than none, by at least
\RwEdgeMinAboveNone{} a year. Several differences are smaller. The calibrated court raises the loss by
\RwEdgeCalMinusNone{} instead of \RwZeroCalMinusNone{}, because the cuts it strikes are smaller and their restorations cheaper to finance. Its temporary savings are \RwEdgeTempCal{} percent of a year's payroll against \RwZeroTempCal{}, and
its financing term is \RwEdgeFinCal{} against \RwZeroFinCal{}. Review before a cut takes effect lowers the loss by
\RwEdgeBeforeGain{} instead of \RwZeroBeforeGain{}.

\looseness=1 The comparison that changes is the incentive to propose cuts beyond the burden when review comes before a cut takes
effect. At the estimate the government proposes such cuts at the calibrated strike probability, \RwZeroExcessBefore{} on
average over 2010--2024, because each stands one time in ten. At a weight of 0.05 it proposes none, because a cut that stands is reversed after the crisis and the government now pays for the reversal. Review before a cut takes effect at the
calibrated probability then yields exactly the outcomes of certain review before it. With the same parameters but no
weight on the restorations, the government proposes \RwDecExcessBefore{} such cuts, so the change comes from the
weight, not from the parameters.

\begin{table}[!tbp]\centering
\begin{threeparttable}
\caption{Judicial review in Portugal when the government weighs the restorations}
\label{tab:courts_edge}
\footnotesize\setlength{\tabcolsep}{2pt}
\begin{tabular*}{\textwidth}{@{\extracolsep{\fill}}l*{8}{c}}
\toprule
& \multicolumn{2}{c}{\makecell{Proposals beyond\\the burden}} & \multicolumn{2}{c}{\makecell{Cumulative\\cut in force}} & \multicolumn{2}{c}{\makecell{Pay in\\crisis years}} & \multicolumn{2}{c}{\makecell{Social\\loss}} \\
\cmidrule(lr){2-3}\cmidrule(lr){4-5}\cmidrule(lr){6-7}\cmidrule(lr){8-9}
Court & $\lambda_R=0$ & $0.05$ & $\lambda_R=0$ & $0.05$ & $\lambda_R=0$ & $0.05$ & $\lambda_R=0$ & $0.05$ \\
& (1) & (2) & (3) & (4) & (5) & (6) & (7) & (8) \\
\midrule
No court & 3.0 & 1.0 & 121 & 112 & 85.9 & 85.7 & 25.83 & 25.25 \\
Calibrated court ($\rho_S=0.9$) & 2.7 & 2.7 & 62 & 58 & 83.8 & 84.3 & 37.04 & 33.85 \\
Review before effect ($\rho_S=0.9$) & 2.7 & 0.0 & 41 & 30 & 86.8 & 87.7 & 29.06 & 29.33 \\
Certain review ($\rho_S=1$) & 3.0 & 3.0 & 54 & 52 & 83.8 & 84.1 & 38.11 & 36.05 \\
Certain review before effect ($\rho_S=1$) & 0.0 & 0.0 & 30 & 30 & 87.7 & 87.7 & 29.17 & 29.33 \\
\bottomrule
\end{tabular*}
\begin{tablenotes}[flushleft]\footnotesize
\item Notes: The courts of Table~\ref{tab:courts} with the government's problem at the estimate ($\lambda_R=0$, the columns of Table~\ref{tab:courts}) and at the largest weight of the grid inside the 95 percent set of Table~\ref{tab:restoration_weight} ($\lambda_R=0.05$), with Portugal's parameters re-estimated at that weight. Each history starts from the documented state of the 2010 budget as the model at those parameters reconstructs it, and the histories, draws and objective are those of Table~\ref{tab:courts}. Columns 1 and 2 count the proposals of cuts that take the outstanding cuts above the admissible burden in a crisis year, over 2010--2024, as means over the histories. The other columns are columns 4, 5 and 7 of Table~\ref{tab:courts}. Without a court no restoration is ordered, so the weight has nothing to charge, and the first row differs between the two weights only through the parameters.
\end{tablenotes}
\end{threeparttable}
\end{table}

\begin{table}[!tbp]\centering
\begin{threeparttable}
\caption{Financed indexation in Bangladesh: the boundaries under parameter uncertainty}
\label{tab:r1_bgd_uncertainty}
\footnotesize\setlength{\tabcolsep}{2pt}
\begin{tabular*}{\textwidth}{@{\extracolsep{\fill}}lcccccc}
\toprule
 & \multicolumn{2}{c}{Profile region} & \multicolumn{2}{c}{Calibration draws} & \multicolumn{2}{c}{Preferred gap $x^\star$} \\
\cmidrule(lr){2-3}\cmidrule(lr){4-5}\cmidrule(lr){6-7}
 & $\lambda_K=1$ & $\lambda_K=0$ & $\lambda_K=1$ & $\lambda_K=0$ & 5 & 10 \\
\midrule
$\kappa_P=0.25$ & 3.32--4.00\textsuperscript{a} & 1.51--2.88 & 3.16, 3.54, 4.00 & 1.84, 2.32, 2.42 & $>$4.00, 2.05 & $>$4.00, 1.88 \\
$\kappa_P=0.5$ & 1.84--2.75 & 0.77--1.39 & 1.63, 1.78, 1.98 & 0.89, 1.04, 1.12 & 2.18, 0.89 & 2.22, 0.85 \\
$\kappa_P=0.75$ & 1.36--2.03 & 0.44--0.87 & 1.14, 1.21, 1.40 & 0.54, 0.60, 0.65 & 1.64, 0.51 & 1.67, 0.50 \\
$\kappa_P=1$ & 1.14--1.75 & 0.13--0.60 & 0.93, 0.96, 1.19 & 0.36, 0.43, 0.44 & 1.47, 0.22 & 1.47, 0.22 \\
\bottomrule
\end{tabular*}
\begin{tablenotes}[flushleft]\footnotesize
\item Notes: Each entry describes the financing coefficient $\phi_A$, as a multiple of $\hat\phi_F=0.271$, below which an indexation rule that pays the share $\kappa_P$ of inflation automatically lowers the stationary social loss below the documented regime's, with the government re-solving its problem (Figure~\ref{fig:r1_bgd}). The profile region contains the points of a grid in the reference share $s_w$ and the cost of appointment $K_I$ whose profile likelihood-ratio statistic, with Bangladesh's other parameters re-estimated at each point, is at most the 95 percent point of $\chi^2(2)$. Its entries are the smallest and the largest value over the region. The calibration draws redraw the two documented long-run moments from their sampling distributions and recalibrate $s_w$ and $K_I$, and the entries are the 5th, 50th and 95th percentiles over 1,000 draws. The last two columns re-estimate Bangladesh with the preferred gap $x^\star$ at 5 and at 10 instead of 7.5, and give the entries at $\lambda_K=1$ and 0. \textsuperscript{a}The upper end lies above the grid. These calculations were specified after the paper's estimates and tests, and fixed before they were computed.
\end{tablenotes}
\end{threeparttable}
\end{table}

\paragraph{Financed indexation.} A rule pays the share $\kappa_P\in\{0.25,0.5,0.75,1\}$ of each year's inflation
automatically between commissions, and the payroll increase of each automatic step is charged at $\phi_A$, from zero to
four times the estimated allowance coefficient $\hat\phi_F$, in the government's objective and in the social loss. The
government re-solves its problem under every rule. Holding its policy at the solution without the charge instead would raise the financing that half indexation can bear from \RvRoneReqHalf{} to \RvRoneReqFixHalf{} of $\hat\phi_F$. The boundary of a
share is the largest $\phi_A$ at which the rule's stationary loss lies below the documented regime's, interpolated
between grid points. Its uncertainty is measured in three ways (Table~\ref{tab:r1_bgd_uncertainty}), over the 95 percent
profile region of the reference share and the cost of appointment at the year-by-year estimates, which spans reference
shares from \RvRoneUOneSwLo{} to \RvRoneUOneSwHi{} and costs of appointment from \RvRoneUOneKiLo{} to \RvRoneUOneKiHi{}, over 1,000 redraws of the calibration's moments, each of which re-selects the calibration, and with the preferred gap at
5 or 10 log points.

\end{document}